\documentclass[11pt,a4paper]{article}
\usepackage{jheppub} 
\usepackage{graphicx}
\usepackage{bm}

\usepackage[T1]{fontenc} 

\usepackage[utf8]{inputenc}
\usepackage[english]{babel}

\usepackage[dvipsnames]{xcolor}

\usepackage[skip=3pt]{caption}
\usepackage{framed}
\usepackage{xcolor}
\usepackage{rotating}
\usepackage{empheq}
\usepackage{afterpage}

\usepackage{tabularray}

\usepackage[skip=3pt]{subcaption}
\usepackage{float}
\usepackage{enumerate}
\usepackage{enumitem}

\newcommand{\comment}[1]{[$\clubsuit${\ {\color{red}CP:} {\bf #1}}]}
\newcommand{\cmd}[1]{{\color{red}[{$\spadesuit$ #1}]}}

\newcommand{\lbar}{\lower0.2ex\hbox{$\mathchar'26$}\mkern-10mu \lambda}

\def\Im{\mathrm{Im}}
\def\Re{\mathrm{Re}}

\def\mrmd{{\mathrm{d}}}

\def\exp{\mathrm{exp}}

\def\a{\alpha}
\def\b{\beta}
\def\d{\delta}
\def\D{\Delta}

\def\ve{\varepsilon}
\def\f{\phi}
\def\F{\Phi}

\def\g{\gamma}
\def\G{\Gamma}

\def\n{\nu}
\def\o{\omega}
\def\O{\Omega}

\def\s{\sigma}
\def\S{\Sigma}
\def\t{\theta}

\def\mfrD{\mathfrak{D}}
\def\mfrS{\mathfrak{S}}

\def\mcalB{{\mathcal{B}}}
\def\mcalC{\mathcal{C}}

\def\mcalE{\mathcal{E}}
\def\mcalF{\mathcal{F}}

\def\mcalL{{\mathcal{L}}}
\def\mcalP{\mathcal{P}}
\def\mcalS{\mathcal{S}}
\def\mcalV{{\mathcal{V}}}
\def\mcalY{\mathcal{Y}}

\def\mcalN{\mathcal{N}}

\def\mcalT{{\mathcal{T}}} 
\def\mcalI{\mathcal{I}}
\def\mcalIbar{\overline{\mcalI}}
\def\mcalG{\mathcal{G}}

\def\mcalK{\mathcal{K}}
\def\mcalX{\mathcal{X}}
\def\mcalZ{\mathcal{Z}}

\def\mbbZ{\mathbb{Z}}
\def\mbbR{\mathbb{R}}
\def\mbbN{\mathbb{N}}
\def\mbbC{\mathbb{C}}

\def\msfS{\mathsf{S}}

\def\*{\star}

\def\ttO{\tilde \O}

\def\tPi{\tilde \Pi}

\def\ts{\tilde s}

\def\tPi {\tilde{\Pi}}

\def\mrmPT{\mathrm{PT}}

\def\mrmBT{\mathrm{BT}}

\def\mrmF{\mathrm{F}}

\def\mrmB{\mathrm{B}}

\def\mrmL{\mathrm{L}}
\def\mrmR{\mathrm{R}}
\def\mrmP{\mathrm{P}}

\def\mrmI{\mathrm{I}}
\def\mrmII{\mathrm{II}}
\def\mrmIII{\mathrm{III}}
\def\mrmIV{\mathrm{IV}}

\def\mrmNP{\mathrm{NP}}

\def\mrmITW{\mathrm{ITW}}

\def\med{\mathrm{med}}

\def\<{\langle}
\def\>{\rangle} 
\def\dee{\partial}

\def\Im{\mathrm{Im}}
\def\Re{\mathrm{Re}}
\def\sgn{\mathrm{sign}}

\def\mrmUBr{\mathrm{UBr}}
\def\mrmBr{\mathrm{Br}}
\def\mrmAR{\mathrm{AR}}

\def\Disc{\mathrm{Disc}}
\def\arccot{\mathrm{arccot}}
\def\mrmcr{\mathrm{cr}}
\def\dD{\dot{\Delta}}

\usepackage{slashed}

\definecolor{darkbrown}{rgb}{0.4, 0.26, 0.13}
\definecolor{paleblue}{rgb}{0.69, 0.93, 0.93}
\definecolor{lightskyblue}{rgb}{0.53, 0.81, 0.98}
\definecolor{skyblue}{rgb}{0.53, 0.81, 0.92}
\definecolor{darkred}{rgb}{0.55, 0.0, 0.0}
\definecolor{darkblue}{rgb}{0.0, 0.0, 0.55}
\definecolor{darkpastelgreen}{rgb}{0.01, 0.75, 0.24}
\definecolor{lightgreen}{rgb}{0.56, 0.93, 0.56}

\usepackage{babel}
\usepackage{beramono}
\usepackage[T1]{fontenc}

\definecolor{identifiercolor}{rgb}{.4,.6,.56}
\definecolor{stringcolor}{gray}{0.5}
\definecolor{inactivecolor}{rgb}{0.15,0.15,0.5}

\usepackage{listings}
\title{\boldmath 
Exact WKB analysis of inverted triple-well: resonance, PT-symmetry breaking, and resurgence
}

\author[1]{Syo Kamata,}
\emailAdd{skamata11phys@gmail.com}
\affiliation[1]{Department of Physics, The University of Tokyo, Tokyo 113-0033, Japan}

\author[2]{Tatsuhiro Misumi,}
\emailAdd{misumi@phys.kindai.ac.jp}
\affiliation[2]{Department of Physics, Kindai University, Osaka 577-8502, Japan}

\author[3]{Cihan Pazarba\c{s}\i{},}
\emailAdd{cihan.pazarbasi@gmail.com}
\affiliation[3]{Okinawa Institute of Science and Technology, Okinawa 904-0495, Japan}

\author[4,5]{Hidetoshi Taya}
\emailAdd{h\_taya@keio.jp}
\affiliation[4]{Hiyoshi Department of Physics, Keio University, Hiyoshi, Kanagawa 223-8521, Japan}
\affiliation[5]{iTHEMS, TRIP Headquarters, RIKEN, Wako, Saitama 351-0198, Japan}

\abstract{We study non-Hermitian quantum mechanics of an inverted triple-well potential within the exact WKB framework. For a single classical potential, different Siegert boundary conditions define three distinct quantum problems: the PT-symmetric, resonance, and anti-resonance systems. For each case, we derive the exact quantization condition and construct the associated trans-series solution. By identifying the resurgent structures and cancellations in these non-Hermitian setups, we obtain the median-summed series, clarifying when the spectra are real or complex in accordance with the physical properties of each system.
Establishing explicit links to the semi-classical path integral formalism, we elucidate the roles of bounce and bion configurations in these non-Hermitian systems. This analysis predicts PT-symmetry breaking, which we also verify numerically. Using the median quantization conditions, we prove the existence of this symmetry breaking and establish an exact equation for the exceptional point, which emerges as a remarkably simple algebraic relation between the bounce and bion actions. We further show that the median-summed non-perturbative correction to the spectrum vanishes at the exceptional point, while the resurgent structure survives through a universal minimal trans-series. For the resonance and anti-resonance systems, we find that the exact median-summed spectra are related by complex conjugation, representing time reversal in this setting, are necessarily complex, and do not exhibit an exceptional point. Although their spectra differ significantly from the PT-symmetric case, they share the same minimal trans-series. By maintaining explicit links with the path integral saddles and the formal theory of resurgence, our analysis provides a unified and general perspective on the quantization of non-Hermitian theories.
} 

\begin{document}

\maketitle
\flushbottom	

\section{Introduction}

Non-Hermitian quantum mechanics~\cite{Ashida:2020dkc,Hatano:2026tdm} provides a natural framework for describing quantum systems with gain, loss, decay, and/or external driving; namely, situations that cannot in general be captured within ordinary Hermitian quantum mechanics. In this sense, non-Hermitian quantum mechanics can serve as an effective description of open quantum systems interacting with an environment and can encode intrinsically non-equilibrium phenomena. One of its most well-known characteristics is the appearance of the complex spectrum\footnote{Technically, the price to pay to include the complex spectrum is extending the space of solutions to rigged-Hilbert space \cite{Bohm89,delaMadrid:2012xrc} and allowing the oscillating solutions, for which the normalizabilty is ensured by the Zel'dovich regularization \cite{Zeldovich61}. Provided by this established structure, the quantization procedure can be performed as in the Hermitian quantum mechanics by solving the Schr\"odinger equation with appropriate boundary conditions. Since we will not need the normalization throughout this paper, we will not discuss this mathematical construction.}, in contrast to the real spectra of standard Hermitian systems. The presence of complex eigenvalues indicates the nonequilibrium nature of the system, as the corresponding quantum states, also called \textit{resonance} states, dissipate in time, leading to a non-zero probability flux through the boundaries of the system. Although the lack of probability conservation leads to a non-unitary time evolution, it is known that overall characteristics\footnote{Although it describes an open system that interacts with the environment, non-Hermitian quantum mechanics does not include the effects of backreaction from the environment, which would lead to a constant change in the quantum system itself.} of a resonance system can be captured by the tools of Hermitian quantum mechanics, such as the Wentzel-Kramers-Brillouin (WKB) approximation for which the $\a$ decay stands as the most famous example~\cite{Gamow:1928zz}.




The non-Hermitian character of a Hamiltonian, however, does not necessarily imply a complex spectrum~\cite{Mostafazadeh:2001jk,Mostafazadeh:2008pw}. The appearance or absence of complex eigenvalues is determined by genuinely quantum properties of the system, rather than by the classical Hamiltonian alone. A celebrated class of non-Hermitian systems with real spectra is provided by PT-symmetric quantum mechanics, which can admit a consistent unitary formulation in the PT-unbroken phase, similarly to Hermitian quantum mechanics~\cite{Bender:1998gh,Bender:2002vv}. Originally developed as a mathematical extension of the Hermitian quantum theory, PT-symmetric quantum mechanics has grown into an active research field on the experimental side as well; see~\cite{Bender:2019cwm,Bender:2023cem} for recent comprehensive reviews.




Heuristically, the reality of the spectrum, or the lack of complex eigenvalues in a PT-symmetric system, can be understood in terms of the balance between gain and loss. More specifically, real spectrum can be understood as the conservation of probability in a finite region, leading to unitary behaviors. This balance can break down at a critical point, known as an \textit{exceptional point}, beyond which the PT-symmetric system develops complex eigenvalues. 
This transition, referred to as \textit{PT-symmetry breaking}, signals the onset of genuinely non-Hermitian behaviors, or the breakdown of the equilibrium character of the system. 

In one-dimensional quantum mechanics with a non-Hermitian Hamiltonian, 
the gain and loss due to the probability flux and the corresponding quantum states can be formulated clearly in terms of the Siegert boundary condition~\cite{Siegert:1939zz}. 
In this setting, a boundary condition on the real axis is specified by choosing whether the surviving asymptotic wave is incoming or outgoing at each spatial infinity. This choice determines the pattern of the probability flux, thereby distinguishing resonance, anti-resonance, and PT-symmetric systems.
Choosing outgoing waves on both sides of the spatial infinity defines a resonance system, while choosing incoming ones yields the time reversal of the resonance, called an anti-resonance system. 
By contrast, choosing one incoming and one outgoing wave leads to a PT-symmetric system, which can lead to zero total probability flux. 

Recently, similar spectral problems have also been studied by imposing boundary conditions in the complexified coordinate plane rather than on the real axis~\cite{Bender:2007nj,Ai:2019fri,Kamata:2023opn,Kamata:2025dkk,Garbrecht:2025tos}. In this approach, the relevant complex boundary conditions are typically selected by the analytic continuation of an associated Hermitian problem, so that the two descriptions remain closely related to each other. In the present paper, in contrast, we focus on boundary conditions imposed on the real axis and on the corresponding probability-flow interpretation mentioned above. From this viewpoint, the PT-symmetric, resonant, and anti-resonant characters of the system are determined by the choice of incoming and outgoing waves at the spatial infinity. This suggests the possibility of analyzing these different non-Hermitian quantum problems in one WKB-type connection problem, which is one of the main motivations of the present work.

Although the choice of boundary conditions determines whether the system is PT-symmetric or (anti-)resonant, this does not quantitatively determine the spectral properties. In general, a proper treatment of a non-Hermitian system with a real potential requires non-perturbative inputs. For example, it is well known that the complex eigenvalues of resonance systems are tied to quantum tunneling and are therefore intrinsically non-perturbative. In the PT-symmetric case, in contrast, there is no \textit{a priori} criterion that determines whether the exact spectrum is real or complex, and one must instead analyze the eigenvalue problem explicitly. Moreover, for real potentials, perturbation theory alone is not directly informative in this respect, since it typically yields real formal series only.

Recent developments in a wide range of physical systems have shown that a consistent non-perturbative quantization often requires a proper resurgent framework~\cite{Zinn-Justin:1981qzi,Zinn-Justin:1983yiy,Zinn-Justin:2004vcw,Zinn-Justin:2004qzw, Jentschura:2004jg, Jentschura:2010zza, Jentschura:2011zza, Basar:2013eka, Dunne:2013ada, Dunne:2014bca,Misumi:2014raa, Misumi:2014jua,Misumi:2014rsa,Misumi:2014bsa, Misumi:2016fno,Fujimori:2018kqp,Basar:2015xna, Misumi:2015dua,Dunne:2016qix,Behtash:2015loa,Behtash:2015zha,Kozcaz:2016wvy, Dunne:2016jsr, Fujimori:2016ljw,Dorigoni:2017smz,Basar:2017hpr, Fujimori:2017oab, Fujimori:2017osz,Behtash:2018voa, Sueishi:2019xcj, Dunne:2020gtk, Fujimori:2022lng,Hongo:2018rpy,Fujimori:2018nvz,Fujimori:2019skd,Misumi:2019upg,Fujimori:2020zka,Fujimori:2021oqg,Unsal:2020yeh,Nishimura:2021lno,Pazarbasi:2021ifb,Pazarbasi:2021fey,Cavusoglu:2023bai}. In Hermitian quantum mechanics, resurgence explains how non-perturbative sectors cancel the ambiguities of non-Borel-summable perturbative expansions, thereby restoring a single-valued, real spectrum. In a non-Hermitian system, in contrast, the exact spectrum may well be complex, but the same resurgent interplay remains essential in order to keep it single-valued and physically meaningful. One must therefore distinguish carefully between imaginary parts that arise as artifacts of Borel resummation and those that represent genuine physical features of the quantum system.


The main motivation of the present paper is performing a comprehensive resurgence analysis of non-Hermitian quantum mechanics to analyze the associated non-perturbative, possibly complex, spectra. This task requires a highly capable and comprehensive approach. In one-dimensional quantum theories, the exact WKB (EWKB) formalism~\cite{BPV, Voros1983,Silverstone, AKT1, AKT2, Takei1, DDP2, DP1, Takei2, Kawai1, Takei3, AKT3, AKT4, Schafke1, Iwaki1} is distinguished as such a powerful framework for spectral problems and resurgence theory. Within this framework, physical observables are encoded in the geometry of Stokes graphs, monodromies, and Voros multipliers through a rigorous synthesis of resurgence theory and the global connection problem for the Schr\"odinger equation~\cite{Sueishi:2020rug, Sueishi:2021xti, Kamata:2021jrs,Misumi:2024gtf, Misumi:2025ijd}. Recently, the EWKB formalism has been widely used in the analysis of various theoretical and physical setups: 4d~$\mcalN=2$ gauge theories~\cite{Nekrasov:2009rc,Mironov:2009uv,Kashani-Poor:2015pca,Kashani-Poor:2016edc,Ashok:2016yxz,Yan:2020kkb}, wall-crossing phenomena~\cite{Gaiotto:2012rg,Allegretti:2020dyt}, ODE/IM correspondence~\cite{Dorey:2001uw, Dorey:2007zx,Ito:2018eon,Ito:2019jio,Imaizumi:2020fxf,Ito:2025pfo}, TBA equations~\cite{Emery:2020qqu,Ito:2024nlt,Ito:2025sgq}, topological string theory~\cite{Grassi:2014cla,Grassi:2014zfa,Codesido:2017dns,Codesido:2017jwp,Hollands:2019wbr,Ashok:2019gee,Coman:2020qgf,Iwaki:2023cek}, time-dependent quantum systems~\cite{Taya:2020dco,Enomoto:2020xlf, Enomoto:2022mti,Namba:2025ejw,Fujimori:2025kkc}, black hole quasi-normal modes~\cite{Miyachi:2025ptm,Miyachi:2025dyk}, and see also other applications~\cite{Duan:2018dvj,Imaizumi:2022dgj,vanSpaendonck:2022kit,Bucciotti:2023trp, Ture:2024nbi}.  It has also been utilized in the context of non-Hermitian quantum mechanics to analyze both PT-symmetric quantum mechanics~\cite{DP98:PTsymmetry1,DP98:PTsymmetry2,delabaere2000spectral,Kamata:2023opn,Kamata:2024tyb,Kamata:2025dkk}, and resonance systems \cite{Ito:2023cyz,Morikawa:2025xjq}. 

Despite this technical sophistication, at its core, EWKB is a refined version of the familiar WKB connection problem.~In this sense, it provides a natural framework for analyzing PT-symmetric, resonance, and anti-resonance systems within a unified picture. Therefore, together with its close connection to resurgence theory, the EWKB framework is particularly suited to a comprehensive analysis of one-dimensional non-Hermitian quantum systems.


\noindent \textbf{\underline{Scope of this paper:}} In this paper, we consider an inverted triple-well (ITW) potential with parity symmetry and study its quantization within the EWKB formalism. The ITW potential consists of two perturbatively degenerate wells and three barriers, giving rise to non-perturbative WKB cycles associated with instanton configurations. Owing to the parity symmetry, there are only two independent non-perturbative configurations, namely, a bion and a bounce. Their interplay in the exact QCs leads to a remarkably rich structure, making this system interesting from both physical and mathematical perspectives. 

By employing the Siegert boundary conditions, we will analyze the quantum mechanical ITW system in the EWKB framework and derive the exact quantization conditions (QCs) for the PT-symmetric, resonance, and anti-resonance cases. We will particularly show that each quantization condition is linked to an element of a single two-by-two transition matrix relating the asymptotic regions on the real axis.  This allows us to perform a detailed analysis of the PT-symmetric, resonance, and anti-resonance systems on an equal footing, revealing their precise links with the particular boundary condition choices.

Furthermore, solving the exact QCs for each system, we will uncover the trans-series solutions and non-perturbative spectra. Throughout these discussions, our emphasis will be on how the resurgence structure is arranged according to the physical properties of each system. In order to clarify these arrangements, we will first keep a close connection to semi-classical physics and identify (multi-)bion and (multi-)bounce configurations in the non-perturbative spectrum. While mainly restricted in the semi-classical regime, this approach will also enable us to identify how resurgence theory works in non-Hermitian quantum mechanics in comparison with the well-established Hermitian theories. Later, utilizing the median summation and Alien calculus in the EWKB framework, we will obtain results which are equivalent to the exact solutions\footnote{We note that with the exact solutions, we refer to the exactness in the context of the resurgence theory where the median summations lead to the resurgent functions. Although closed forms for these functions are not always available, they encode the entire information of the associated analytic functions. The trans-series, on the other hand, can be considered as a collection of small parameter expansions of these resurgent functions. Contrary to the simple asymptotic expansions, since a trans-series have a unique well-defined Borel summation, they can be utilized for proper non-perturbative approximations of the resurgent functions. In this paper, the connection to the semi-classical physics is considered at the trans-series level as approximations of ``exact'' solutions which we discuss using median summations and Alien calculus in EWKB framework.}

Let us summarize our main results and outline the organization of the paper:
\begin{itemize}[wide]
    \item At the level of the exact QCs, we show that the resonance and anti-resonance systems are not invariant under time reversal, reflecting their intrinsically irreversible and non-equilibrium characters. Rather, their QCs are related to each other by complex conjugation, so that the two systems are mapped into one another as time-reversed partners. In the PT-symmetric case, in contrast, the time-reversal operation maps the QC to itself, up to an irrelevant overall constant.
    
    We also identify the distinct roles played by the complex conjugation (time-reversal) operator and the Stokes automorphism.
    Although these operations appear closely related in Hermitian settings, they are conceptually different in the present problem. 
    Our analysis explicitly demonstrates that complex conjugation relates QCs corresponding to opposite directions of time flow, whereas the Stokes automorphism connects different analytic continuations of a given QC without changing the direction of time.
    
   
    \item In the PT-symmetric case, the exact QCs admit two distinct trans-series solutions, governing the physical spectra in different parameter regimes. One of these solutions yields a complex spectrum, thereby signaling a PT-symmetry breaking. We show moreover that the trans-series structure undergoes a drastic reorganization across the exceptional point, offering a new perspective on the origin of the PT-symmetry breaking.
    \item In all cases, by identifying the relevant resurgent structures and the associated cancellations, we obtain the median-summed trans-series contributions to the physical spectra. This analysis differs from the more familiar Hermitian cases, since in non-Hermitian systems one must allow genuinely complex spectra. Accordingly, in addition to the resurgence cancellations themselves, the signs of the resulting imaginary contributions become essential in the analysis of the resonance, anti-resonance, and broken PT-symmetric cases. We show that these signs are fully consistent with the physical character of each system. To our knowledge, a detailed resurgence analysis of a non-Hermitian theory from this perspective has not been carried out before.
    \item Utilizing the Stokes automorphism, we obtain the \textit{exact} median-summed non-perturbative spectra from the median QCs for each system. Along with recovering the corresponding trans-series solutions, this approach allows us to analyze the exceptional point directly in the PT-symmetric case. In this way, we derive an explicit equation for the exact locations of the exceptional points and establish exact bounds separating the broken and unbroken PT-symmetric phases. We also prove the reality of the spectrum in the corresponding phases without invoking any trans-series construction or numerical computation. Moreover, we show that the exact median-summed non-perturbative correction to the spectrum vanishes at the exceptional point. We further demonstrate that the non-perturbative splitting exhibits a characteristic square-root behavior near the exceptional point. These results illuminate both the non-perturbative nature of the transition and its continuity.
    \item We uncover a universal minimal trans-series that participates in the resurgence cancellations in the same way for every system and in every parameter regime, including at the exceptional point, where the remaining non-perturbative corrections cancel exactly. By applying Alien calculus to the median QCs, we show that this minimal series is generated from the perturbative sector and thus captures the universal part of the resurgent structure that is forced solely by the non-Borel-summability of perturbation theory. The remaining, sector-dependent non-perturbative contributions then encode the genuine physical distinctions between PT-symmetric and (anti-)resonant boundary conditions.    
\end{itemize}

\noindent \textbf{\underline{Comparison with related literature}:} With its emphasis on the role of the resurgence structure of the trans-series solutions, utilization of the median summation and Alien calculus, as well as its strong link with semi-classical quantization, our discussion is distinguished from the previous Exact WKB analyses of non-Hermitian quantum systems. Among these works, let us also point out several of them that are more closely related to our work: 

Recently, several systems closely related to the resonance phenomenon have been analyzed via the EWKB formalism and the associated Stokes geometry. This includes a study of the exactly solvable Rosen-Morse potential by utilizing the Siegert boundary conditions and Zel’dovich regularization \cite{Morikawa:2025xjq}. A more closely related work to our motivation was performed by considering the Stark effect in the hydrogen atom in \cite{Ito:2023cyz}. In the EWKB-related part of this work, the authors derived the exact quantization using the outgoing boundary conditions and computed the complex energy spectra by identifying their relation to a tunneling WKB cycle.

The analysis of PT-symmetric systems in the EWKB formalism, on the other hand, goes back to the seminal letters by Delabaere and Pham \cite{DP98:PTsymmetry1,DP98:PTsymmetry2}. In the first of these two subsequent works \cite{DP98:PTsymmetry1}, the authors studied the PT-symmetric cubic oscillator briefly and established the reality of the spectrum via median summation in the EWKB framework. In the latter paper \cite{DP98:PTsymmetry2}, considering the quartic oscillator with an imaginary linear term, they established an exact critical point condition for the PT-symmetry breaking, which depends on the strength of the imaginary coupling and a real non-perturbative effect. While the exact PT-symmetry breaking condition resembles our result, the specific mechanism based on the semi-classical configurations that we establish in this paper presents an essential difference from \cite{DP98:PTsymmetry2}. 

Another related PT-symmetry breaking study was performed by Delabaere and Trinh in \cite{delabaere2000spectral}. Concentrating on the complex cubic oscillator, in this paper, the authors established the Riemann-surface structure of the spectrum and the branching associated with the PT-symmetry breaking. They also related the PT-symmetric system to the corresponding resonance through an analytic continuation. While both aspects of \cite{delabaere2000spectral} exist in the present paper, our discussion treats the PT-symmetric and (anti-)resonance systems as distinct boundary value problems encoded in a common exact-WKB connection matrix, and further develops their energy trans-series. Particularly, the semi-classical interpretation of the PT-symmetry breaking in this paper is distinguished from \cite{delabaere2000spectral}.

Finally, an approach similar to that in this paper was presented for the inverted quartic double-well (IDW) and for inverted oscillators without a quadratic (mass) term. These studies were carried out in connection to the Ai-Bender-Sarkar (ABS) conjecture~\cite{Ai:2022csx}\footnote{The ABS conjecture relates the (logarithm of the) partition functions of a resonance theory and an associated PT-symmetric theory. According to the conjecture, this continuation ensures the reality of the PT-symmetric spectrum. We do not directly address the ABS conjecture here, since it requires a different theoretical setup. We note, however, that our analysis reveals no direct link between the resonance and PT-symmetric systems.} in~\cite{Kamata:2023opn}, with the zeta-function sum rules in~\cite{Kamata:2025dkk}, and with a non-perturbative version of the Gutzwiller trace formula in~\cite{Kamata:2024tyb}. By contrast, the ITW potential has a much richer semi-classical structure than the IDW, and this allows us to obtain a more intricate and more general picture of resurgence in non-Hermitian systems. In particular, we provide a more detailed analysis of the resurgent structure and the associated cancellations, which further clarifies the contrast with the better-understood Hermitian cases.

By employing EWKB together with the Siegert boundary condition and analyzing the complex spectra comprehensively in view of resurgence theory and semi-classical quantization, our analysis of the ITW potential encompasses and extends earlier examples in the literature. In this sense, it presents a minimal yet highly non-trivial setting for non-Hermitian quantum mechanics, demonstrating both the effectiveness and the broader applicability of the EWKB framework.

\paragraph{\underline{Outline}:} The organization of this paper is as follows. In Section~\ref{Section: EWKB_review}, we review the EWKB formalism, Stokes automorphisms, the median summation, and Alien calculus that are needed in the later analysis. In Section~\ref{sec:ITW}, we introduce the ITW system, discuss its boundary conditions and numerical solutions, and formulate the EWKB connection problem. In Section~\ref{Section: Trans-seriesDirect}, we derive the trans-series solutions for the PT-symmetric, resonance, and anti-resonance cases and discuss the associated resurgence cancellations, leading to the median-summed spectra. In Section~\ref{Section: ExactSolutions}, we solve the median QCs exactly, determine the PT-symmetry breaking and exceptional-point conditions exactly, and analyze the associated resurgent structures, including the minimal trans-series, by using Alien calculus. Section~\ref{sec:summary} is devoted to discussion and outlook.

\section{Review of exact WKB and resurgence theory} \label{Section: EWKB_review}

In this section, we will briefly introduce resurgence theory and the EWKB formalism. Rather than a comprehensive and pedagogical review of these subjects, we will focus on the aspects which are necessary for the analysis of the ITW system. Before starting the technical discussion, let us summarize the aspects of EWKB and resurgence theory which we utilize in this paper:

In its essence, understanding the quantum spectrum of a physical system in the EWKB framework requires the application of resurgence theory at two different levels. First, the setup of EWKB itself is based on the resurgence of the asymptotic WKB expansion. This leads to resurgent wave functions which are equivalent to the exact solutions of the corresponding differential equations. Although these solutions are valid in certain regions of the complexified physical space, which we call Stokes sectors, they lead to an exact version of the conventional WKB connection formula. Then, imposing boundary conditions leads to exact QCs,  whose exactness is a direct consequence of the exact connection formula. 

The solutions to exact QCs reveal the physical spectrum of the associated quantum system. It turns out, however, that solving the exact QC reveals another layer of resurgence structure encoding the physical spectrum. Although, in our approach, it is essentially rooted in the resurgent wave functions and their analytical properties, a resurgent analysis of the spectrum generally involves a much deeper and richer structure. Conventionally, the resurgence structure is disclosed in a semiclassical treatment via trans-series solutions of the spectrum. This approach leads to physically manifest interpretations of the intricate relationships between perturbative expansions and non-perturbative tunneling configurations. From a practical point of view, however, it only has limited computational power, depending on the complications of the physical system itself. Utilizing the formal resurgence theory through median summations and Alien calculus, on the other hand, can overcome this difficulty by revealing the exact resurgent spectrum at least at the formal level. Recently, in different quantum mechanical settings \cite{vanSpaendonck:2023znn,Kamata:2024tyb,Fujimori:2025kkc}, the practical power of this approach has been presented. In this paper, we will consider both directions to provide a comprehensive analysis of the non-Hermitian ITW system, providing both a strong link to the semi-classical quantization and an exact resolution of the non-perturbative resurgent spectra.


\subsection{A brief review of resurgence theory} \label{Section: Resurgence_review}
Resurgence theory is one of the central mathematical tools in our discussion to obtain an equivalent of the exact quantization of ITW. The need to incorporate resurgence theory into the spectral problems, however, can be observed without considering the exact quantization problem. Probably the most fundamental observation is the factorially divergent behaviour of the perturbative series. Although the first few orders of such a series might lead to good approximations, the divergence  prevents understanding the physical observables to the full extent, which requires uncovering the associated analytical function. To give a proper meaning to a factorially divergent series, it should be handled by the Borel summation method, which we will introduce shortly. If the series is Borel summable, the summation leads to the analytic function under consideration directly. When we have a non-summable series, however, the problem becomes more intricate and it is the point where trans-series representations are necessary and resurgence theory starts playing a significant role.

Let us focus on the spectral problems: The analytical function under consideration is the eigenvalues $E(\hbar)$ and we express their trans-series as
\begin{equation}
	E(\hbar) = \sum_{m=0}^\infty e^{-m\frac{S}{\hbar}} \sum_{n=0}^\infty \mcalE_n^{(m)}\hbar^n \, . \label{Trans-seriesGeneral}
\end{equation}
In general, not only the standard perturbation series, which corresponds to the $m=0$ case, but all series are divergent and non-Borel summable. In this case, the Borel summation of each series leads to \textit{non-perturbatively ambiguous} results and indicates the multi-valued character of $E(\hbar)$ at first sight. This is not acceptable because there should be only one spectrum for the physical system. When the entire trans-series is considered, however, the summed trans-series leads to physically appropriate results via cancellation among different ambiguous terms. In the literature, this resurgence cancellation is known as the Bogomolny-Zinn--Justin (BZJ) mechanism~\cite{Bogomolny77,Bogomolny80,Zinn-Justin:1981qzi,Zinn-Justin:1983yiy} and has been tested in various systems~\cite{Zinn-Justin:2004vcw,Zinn-Justin:2004qzw, Basar:2013eka, Dunne:2013ada, Dunne:2014bca,Misumi:2014raa, Misumi:2014jua,Misumi:2014rsa,Misumi:2014bsa,
	Misumi:2016fno,Fujimori:2018kqp,Basar:2015xna, Misumi:2015dua,Dunne:2016qix,Dunne:2020gtk, Behtash:2015loa,Behtash:2015zha,Kozcaz:2016wvy, Dunne:2016jsr, Fujimori:2016ljw, Basar:2017hpr, Fujimori:2017oab, Fujimori:2017osz,Behtash:2018voa, Sueishi:2019xcj, Fujimori:2022lng, Cavusoglu:2023bai,Misumi:2025ijd}.

The process which leads to the single-valued trans-series is called \textit{median summation}. While ambiguity cancellation is a practically viable and physically transparent method to obtain median summations up to some order $m$ in the trans-series, in order to leverage resurgence theory fully and uncover the complete non-perturbative spectrum, one needs to employ the elements of formal resurgence theory, which we will utilize in the EWKB framework. For this reason, in the following, we will discuss these formal aspects and introduce Borel summation, Stokes automorphisms and Alien derivatives. Note that we will keep our discussion very brief and focus on setting necessary conceptual notions which will help with our purpose. For thorough discussions, we refer to~\cite{DP1,Dorigoni2019}. \\

Let us consider a trans-series consisting of infinitely many factorially divergent series such that 
\begin{equation}
	\F(\hbar) = \sum_{m=0}^\infty e^{-m\frac{A}{\hbar}} \sum_{n=0}^\infty f^{(m)}_n \, \hbar^n\, \quad f^{(m)}_n \sim \left( A e^{i \t_\hbar}\right)^{-n} n!	\, , \label{DivergentSeries}
\end{equation}
where $\t_\hbar=\arg \hbar$ and we assume $A>0$. 

\begin{figure}[t]
	\centering
	\includegraphics[width=0.65\textwidth]{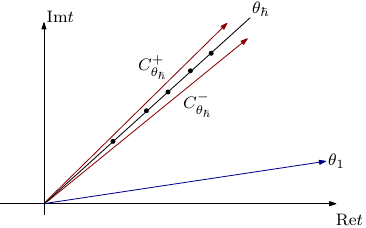}
	\caption{ An example of a Borel plane, where dots in the $\t=\t_\hbar$ direction represent the Borel singularities. Two generic cases of Laplace transform integral contours corresponding to the Borel summations are illustrated with  arrowed lines. {\bf (i)} The contour in the direction $\t=\t_1$ doesn't hit any singularity and corresponds to the Borel summable case. {\bf (ii)} Since the singularities lie on the Stokes ray along the $\t=\t_\hbar$ direction, the Laplace integral cannot directly be taken in this direction. It should be analytically continued and taken along the $C^+_{\t_\hbar}$  or $C^-_{\t_\hbar}$ contours, leading to ambiguous lateral summations. This represents the non-Borel summable case.}
	\label{Figure: BorelPlane}
\end{figure}
\noindent\textbf{\underline{Borel summation}:} The directional Borel summation of each series in \eqref{DivergentSeries} is defined as 
\begin{equation}
	\mrmF_m(\hbar,\t) = \msfS_\t\left[\F_m(\hbar)\right] = \int_{0}^{\infty e^{i\t}} \mrmd t\, e^{-\frac{t}{\hbar}}\, \mathfrak{B}\left[\F_m\right](t)\, ,\label{BorelIntegral_single}
\end{equation}
where $\F_m = \sum f_n^{(m)} \hbar^n$ is the divergent series in sector $m$ in \eqref{DivergentSeries} and $\t$ is a direction on the complex Borel plane of $t$. (See Fig.~\ref{Figure: BorelPlane}.) In \eqref{BorelIntegral_single}, the function $\mathfrak{B}\left[\F_m(\hbar)\right](t)$ is called the Borel transform of $\F_m(\hbar)$ and the Laplace integral acts as an inverse Borel transform that maps back to the original variable space $\hbar$.  

For the asymptotic behaviour in \eqref{DivergentSeries}, the Borel transform $\mcalB[\F_m]$ reads as
\begin{equation}
	\mcalB[\F_m](t)
	= \frac{1}{\hbar}\sum_{n=0}^{\infty}\frac{f_n^{(m)}}{n!}\,t^n. \label{BorelTransform}
\end{equation}
Note that in \eqref{BorelIntegral_single} and \eqref{BorelTransform}, we rescale the Laplace integration variable, leading to the exponential kernel $e^{-\frac{t}{\hbar}}$. We also absorb the overall factor $1/\hbar$ arising from the integration measure into the definition of $\mcalB[\F_m]$.

The factorial $n!$ in the denominator in \eqref{BorelTransform} cancels the original factorial divergent behaviour in \eqref{DivergentSeries} and makes the associated asymptotic behaviour a geometric one. Then, $\mcalB\left[\F_m\right](t)$ defines a function with singularities on the complex $t$-plane along the ray $\t=\t_\hbar$ as illustrated\footnote{Note that the precise singularity structure of the Borel transform $\mcalB[\F_m](t)$ is determined by the exact coefficients $f_n^{(m)}$ of the asymptotic series in \eqref{DivergentSeries}. In general, the singularity structure of the Borel transform of an asymptotic series could be more complicated than that shown in Fig.~\ref{Figure: BorelPlane}. Here, we assume that the singularities lie along a single ray to keep the discussion brief.} in Fig.~\ref{Figure: BorelPlane}. As long as the Laplace integral doesn't hit a singularity of the Borel transform $\mathfrak{B}[\F_m](t)$, as in the direction $\t=\t_1$ in Fig~\ref{Figure: BorelPlane}, the inverse transform is well-defined, leading to a unique resurgent function $\mrmF_m(\hbar,\t)$ and we have a Borel summable case.

When the Laplace integral hits a singularity or singularities on the Stokes ray in the direction $\t=\t_\hbar$ on $t$-plane as in Fig.~\ref{Figure: BorelPlane}, however, the original integral requires analytic continuation. Then, the corresponding lateral summations $\msfS_{\t_\hbar^\pm}=\lim_{\ve\rightarrow 0^+}\msfS_{\t_\hbar \pm \ve}$ are obtained via Laplace integrals on contours $C^\pm_{\t_\hbar}$. Crucially, the lateral summations are not equal to each other, i.e. $\msfS_{\t_\hbar^+}\left[\F_m(\hbar)\right]\neq \msfS_{\t_\hbar^-}\left[\F_m(\hbar)\right]$ and therefore, they lead to a double-valued inverse transform. 

This situation is called \textit{Borel ambiguity} as there is no a priori prescription to determine the analytic continuation direction. From the mathematical point of view, the ambiguity stems from a discontinuity of the lateral summation $\mrmF_m$. The discontinuous jump which is induced while passing through the corresponding singular ray is called the \textit{Stokes phenomenon} and the mathematical object that encodes the discontinuity is called the \textit{Stokes automorphism}.  Combining the Stokes automorphism with the lateral Borel summations leads to a single-valued resurgent function on the Stokes ray at $\t=\t_\hbar$, resolving the ambiguity problem. The process is called \textit{median summation}, which we briefly introduce along with the proper definition of the Stokes automorphism next.

Since we are interested in the resurgent function which is represented by the summed trans-series in \eqref{DivergentSeries}, let us also introduce the representation for its directional Borel summation:
\begin{equation}
	\mrmF(\hbar,\t) =\msfS_{\t}\left[\F(\hbar)\right] = \int_{0}^{\infty e^{i\t}} \mrmd t\,  e^{-\frac{t}{\hbar}} 
	\mathfrak{B}\left[\F\right](t) \, ,
	\label{BorelIntegral_definition}
\end{equation}
where $\mcalB[\F]$ denotes the sum of the Borel transforms of the individual sectors. We note that with the specific choice of the asymptotic behaviour in \eqref{DivergentSeries}, for the $e^{-m\frac{A}{\hbar}}\F_m$ sector in the trans-series, the leading Borel singularity of $\F_m$ produces an ambiguity carrying the exponential factor $e^{-(m+1)\frac{A}{\hbar}}$ when $\t_\hbar=0$. This matches the exponential factor of the next sector in the trans-series, allowing possible ambiguity cancellation. Such cancellations underlie the aforementioned BZJ mechanism. Applied to the full trans-series, the median summation procedure implements these cancellations and provides a prescription for obtaining unambiguous physical quantities using resurgence theory. \\


\noindent\textbf{\underline{Stokes automorphism ${\bm \mfrS_\t}$}:} The Stokes automorphism compares the Borel summations along the two sides of a ray $\t=\t_1$. Then, it is defined as
\begin{equation}
	\msfS_{\t_1^+} = \msfS_{\t_1^-} \circ 
	\mfrS_{\t_1}
	= \msfS_{\t_1^-} \circ \left(\mathrm{Id} - \mathrm{disc}_{\t_1}\right)\, , 
\end{equation}
where $\t^{\pm}_1 = \t_1 \pm 0 $ and $\mathrm{disc}_{\t_1}$ corresponds to the discontinuity of $\mrmF(\hbar,\t)$ at $\t = \t_1$. If the function is continuous at $\t=\t_1$, it holds that $\msfS_{\t_1} = \mathrm{Id}$ and that the two Borel summations are equal to each other as
\begin{equation}
	\msfS_{\t_1^+}\left[\F(\hbar)\right] = \msfS_{\t_1^-}\left[\F(\hbar)\right] = \msfS_{\t_1}\left[\F (\hbar)\right] \, .
\end{equation}
This equality means that the asymptotic series is Borel summable, i.e. $\mrmF(\hbar, \t_1) = \msfS_{\t_1}\left[\F(\hbar)\right]$. 

Let us now consider the direction $\t = \t_\hbar$ where the asymptotic series is not Borel summable. Then, the directional Borel summations $\msfS_{\t_\hbar^\pm}$ are related to each other via the Stokes automorphism as
\begin{equation}
	\msfS_{\t_\hbar^+}\left[\F(\hbar)\right] = \msfS_{\t_\hbar^-}\circ
	\mfrS_{\t_\hbar}\left[\F(\hbar)\right]
	\, .  \label{StokesAuto_Singular}
\end{equation}
Using this relationship we can also analyze the summation exactly on the Stokes ray as well by considering half-Stokes automorphism and defining the summation as

\begin{equation}
	F(\hbar,\t_\hbar) = \mcalS_{\t_\hbar}\left[\F(\hbar)\right] = \msfS_{\t_\hbar^-} \circ \mfrS_{\t_\hbar}^{1/2} \left[\F(\hbar)\right] = \msfS_{\t_\hbar^+} \circ \mfrS_{\t_\hbar}^{-1/2}\left[\F(\hbar)\right]\, ,  \label{BorelSummation_OnStokesRay}
\end{equation}
where the last equality is simply guaranteed by~\eqref{StokesAuto_Singular}. The function $F(\hbar,\t_\hbar)$ is called \textit{median summation} of the trans-series and it corresponds to the formal definition of the resurgent function with the asymptotic trans-series expansion in \eqref{DivergentSeries}.


\noindent \textbf{\underline{Alien Derivative}:} The Stokes automorphism $\mfrS_{\t_\hbar}$ encodes the entire discontinuity along the Stokes ray at $\t=\t_\hbar$. Instead, it is also possible the discontinuity by singling out each singularity on a Stokes ray as well. The operation for this purpose is called the Alien derivative and it is related to the Stokes automorphism as

\begin{equation}
	\mfrS_{\t_\hbar} = \exp\left\{\sum_{\o \in \G_{\t_\hbar}} e^{- \frac{S_\o}{\hbar}} \D_{\o}\right\}  \, , \label{AlienDerivative_Definition}
\end{equation}
where $\G_{\t_\hbar}$ is the set of singularities $\o$ along the singular ray $\t=\t_\hbar$ in Fig.~\ref{Figure: BorelPlane}. Note that each $S_\o$ corresponds to the position of the singularity $\o$ on the Borel plane and the exponential factors $e^{-\frac{S_\o}{\hbar}}$ are linked to the exponentials in the corresponding trans-series. 

It is also convenient to incorporate the exponentials in \eqref{AlienDerivative_Definition} into the Alien derivative by defining $\dD_\o = e^{-\frac{S_\o}{\hbar}}\D_\o$. We can even define an Alien derivative for the entire summation in \eqref{AlienDerivative_Definition} as 
\begin{equation}
	\dD_{\t_\hbar} = \sum_{\o\in\F_{\t_\hbar}} \dD_{\o} = \sum_{\o \in \G_{\t_\hbar}} e^{- \frac{S_\o}{\hbar}} \D_{\o} \label{PointedAlien}
\end{equation}
and express the Stokes automorphism as
\begin{equation}
	\mfrS_{\t_\hbar} = e^{\dD_{\t_\hbar}} \, 
	. \label{AlienDerivative_Dotted}
\end{equation}
In this form, the ``pointed'' Alien derivative $\dD_{\t_\hbar}$ probes the all the singularities along $\t=\t_\hbar$ ray together and it fits for our purposes in this paper. Probing each singularity individually stands as an interesting way to uncover ``exact'' trans-series structure, which was recently discussed in the literature~\cite{vanSpaendonck:2023znn}.

As the name suggests, the Alien derivative possesses the ordinary derivative properties. A useful one, which we will utilize later, is the Leibniz rule: When there is a functional such that $\G(\F(\hbar))$, 
the action of $\dD_{\t_\hbar}$ is decomposed into two parts via the chain rule as
\begin{equation}
	\dD_{\t_\hbar} \left[\G(\F(\hbar),\hbar)\right] = \dD_\f\left[\G(\f,\hbar)\right]\Big\vert_{\f=\F(\hbar)} + \frac{\dee \G}{\dee \F}\Big\vert_{\f=\F(\hbar)} \dD_{\t_\hbar} \F\, . \label{AlienChainRule}
\end{equation}
Note that we differentiate the notation of the two Alien derivatives on the right side of \eqref{AlienChainRule} as $\dD_{\t_\hbar}$ and $\dD_{\f}$ to indicate that $\dD_\f$ acts on $\G(\f)$ operationally. One can consider it as a partial Alien derivative in comparison with the ordinary calculus.

To understand the role of the Alien derivative in relation to the trans-series~\eqref{DivergentSeries}, let us consider the expansion of the exponential in \eqref{AlienDerivative_Dotted}
\begin{equation}
	\mfrS_{\t_\hbar}^\nu  = \sum_{k=0}^\infty \frac{\nu^k}{k!} \dD_{\t_\hbar}^k \, . \label{AlienDerivativeExpansion}
\end{equation}
This is a non-trivial re-arrangement of the entire Stokes discontinuity in a power series $\nu$ where the order corresponds to the number of Alien derivative acted on a given function. For example, considering the median summed function $F(\hbar,\t_\hbar)$ in~\eqref{BorelSummation_OnStokesRay} and using \eqref{AlienDerivativeExpansion}, we express the action of the Stokes automorphism as
\begin{equation}
	\mfrS_{\t_\hbar}^{\nu} \left[F(\hbar,\t_\hbar)\right] = \sum_{k=0}^\infty F^{(k)}(\hbar,\t_\hbar)\, \frac{\nu^k}{k!}\, , \label{AlienExpansion_MedianSummed}
\end{equation}
where $F^{(k)}(\hbar,\t_\hbar) = \dD_{\t_\hbar}^k \left[F(\hbar,\t_\hbar)\right]$. With this formula, we can also return back to the trans-series~\eqref{DivergentSeries}: First, we need to remove the directional Borel summations $\msfS_{\t_\hbar^\pm}$ in~\eqref{BorelSummation_OnStokesRay}, which is equivalent to re-expanding $F(\hbar,\t_\hbar)$ for $\hbar \ll 1$. Then, defining $\F^{(\nu)}(\hbar) = \msfS_{\t_\hbar}^{\nu}[F(\hbar,\t_\hbar)]$, we get the $\hbar\ll 1$ expansion as
\begin{equation}
	\F^{(\nu)}(\hbar) = \sum_{k=0}\frac{\nu^k}{k!} \sum_{n=0}^\infty F^{(k)}_n \hbar^n \, ,  \label{Trans-series_Alien}
\end{equation}
and subsequently setting $\nu = \pm\frac{1}{2}$, we recover the original trans-series \eqref{DivergentSeries}. Note that the choice of the sign of $\nu=\pm\frac{1}{2}$ is \textit{ambiguous}, which is expected as we returned back to the asymptotic trans-series level. 

Despite its equivalence to~\eqref{DivergentSeries}, however, \eqref{Trans-series_Alien} is organized in a different way. The importance of this reorganization reveals itself in~\eqref{AlienExpansion_MedianSummed}: We first observe that at the leading order in~\eqref{AlienExpansion_MedianSummed}, i.e., $k=0$, there is no ambiguity and $F^{(0)} = F(\hbar,\t)$ is the median-summed function. The higher-order terms, on the other hand, only appear in the lateral Borel summations, i.e., $\msfS_{\t_\hbar^\pm}$ for $\nu = \mp \frac{1}{2}$ and canceled out in the median summation. As a result, this organization difference with the original trans-series \eqref{DivergentSeries} becomes convenient to separate the median summed function and its resurgence structure.


This approach stands out as a powerful way to extract the median-summed function along with the associated resurgence structure. In Section~\ref{Section: Spectrum_EWKB}, we will discuss how Stokes automorphism and Alien derivatives are utilized in the EWKB framework, particularly in connection with this paper's context. Before that let us introduce EWKB formalism briefly.

\subsection{A brief introduction to exact WKB} \label{Section: BriefIntro_EWKB}
Throughout this paper, we discuss the non-Hermitian quantum mechanics of ITW in the EWKB formalism and obtain the spectrum focusing on the solutions of the time-independent Schr\"odinger equation of form,
\begin{equation}\label{Schr\"odingerEquation}
	\left[-\frac{\hbar^2}{2}\frac{\mrmd^2}{\mrmd x^2} + Q(x)\right]\f(x) = 0\, , 
\end{equation} 
where $Q(x) =  V(x) - E$. Let us assume $V(x)$ to be a polynomial and briefly discuss the EWKB formalism and present formulas that will be used in the upcoming sections. Rather than providing a thorough explanation of the mathematical construction of the EWKB framework, we concentrate on the general aspects and the necessary details, which we will utilize in the following sections. More detailed explanations can be found in the literature; see e.g., \cite{Kawai1,Sueishi:2020rug,Sueishi:2021xti,Misumi:2024gtf,Misumi:2025ijd}.  

The WKB approach to~\eqref{Schr\"odingerEquation} starts with the ansatz,
\begin{equation}
	\f^{\pm}(x) = \exp\left\{\pm \int_{x_0}^x \mrmd x'\, s(x',\hbar)\right\} \, , \qquad s(x,\hbar) = \sum_{n=-1}^\infty s_n \hbar^n \, .\label{WKB_ansatz}
\end{equation}
Imposing it to \eqref{Schr\"odingerEquation} and solving the resulting Riccati equation separates even and odd terms series expansion in \eqref{WKB_ansatz}, leading to
\begin{align}
	\f^\pm(x) & = \frac{1}{\sqrt{\ts(x,\hbar)}}\exp\left\{\pm \int_{x_0}^x \mrmd x' \ts(x,\hbar) \right\} \\&= \frac{e^{\pm \s(x)}}{\sqrt{\s(x)}}\, \sum_{n=0}^\infty \f_n^\pm\, \hbar^n\, . \label{Asymptotic_WaveFunction}
\end{align} 
where
\begin{equation}
	\ts(x,\hbar) = \sum_{n=0}^\infty \ts_{2n-1}\, \hbar^{2n-1}\, ,  \label{WKBseries_OddTerms}
\end{equation}
is the series of odd terms of \eqref{WKB_ansatz} and
\begin{equation}
	\s(x) = \frac{1}{\hbar} \int_{x_0}^{x} \mrmd z\, \sqrt{Q(x)}\, .\label{StokesIntegral} 
\end{equation}
Note that the lower bound $x_0$ is called the normalization point. In general, $x_0$ could be chosen arbitrarily, however for our purpose, it is chosen as a point satisfying $Q(x_0)=0$ throughout this paper.

\begin{figure}
	\begin{subfigure}[h]{0.48\textwidth}
		\caption{\underline{$\Im \, \s(x) \neq 0$: Borel summable case}}	\label{Figure: WaveFunction_Borel1}
		\vspace{10pt}
		\includegraphics[width=\textwidth]{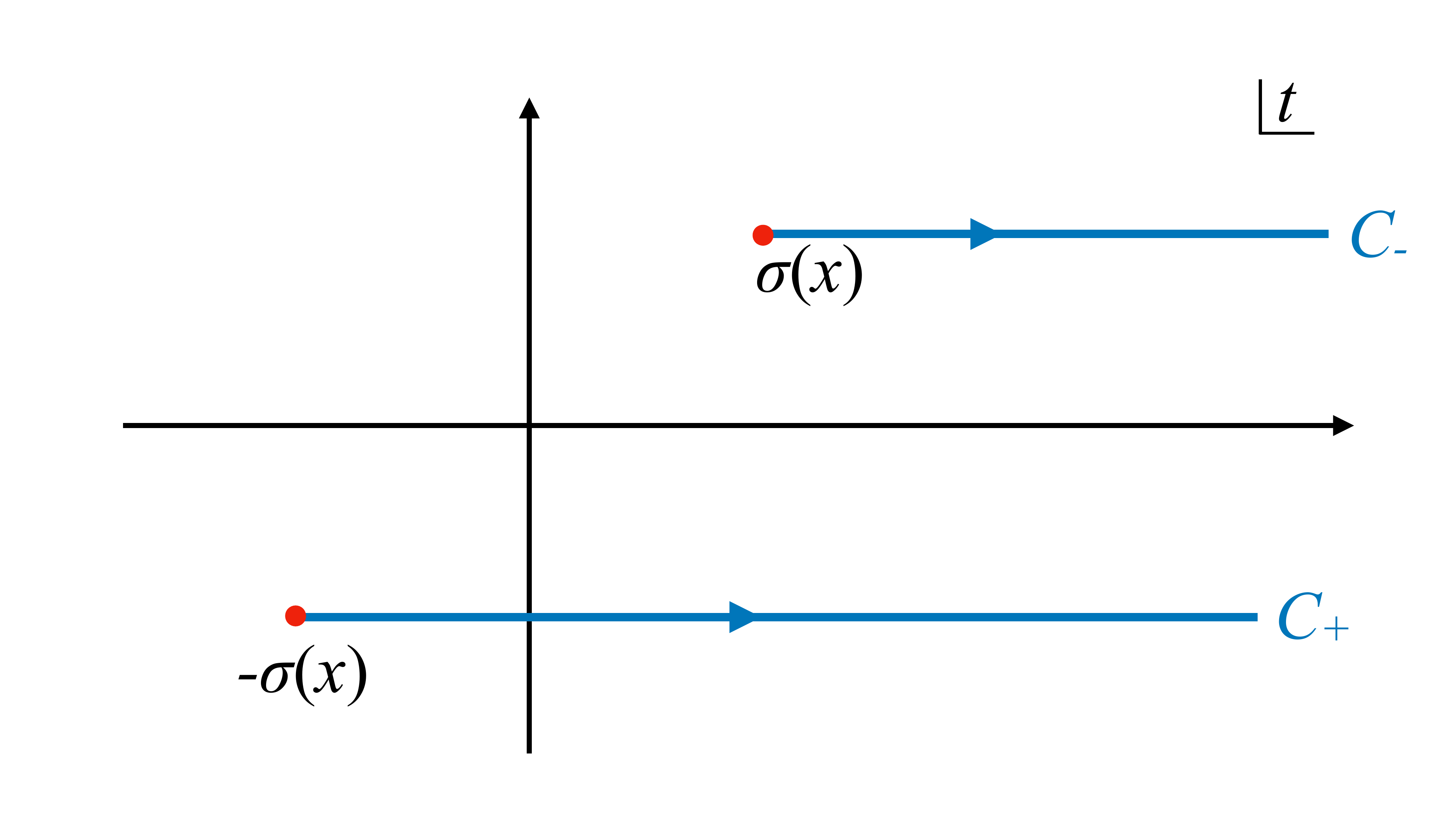}
	\end{subfigure}
	~\hfill
	\begin{subfigure}[h]{0.48\textwidth}
		\caption{\underline{$\Im \, \s(x) = 0$: Singular case for $\psi^{(+)}$}}	\label{Figure: WaveFunction_Borel2}
		\vspace{10pt}
		\includegraphics[width=\textwidth]{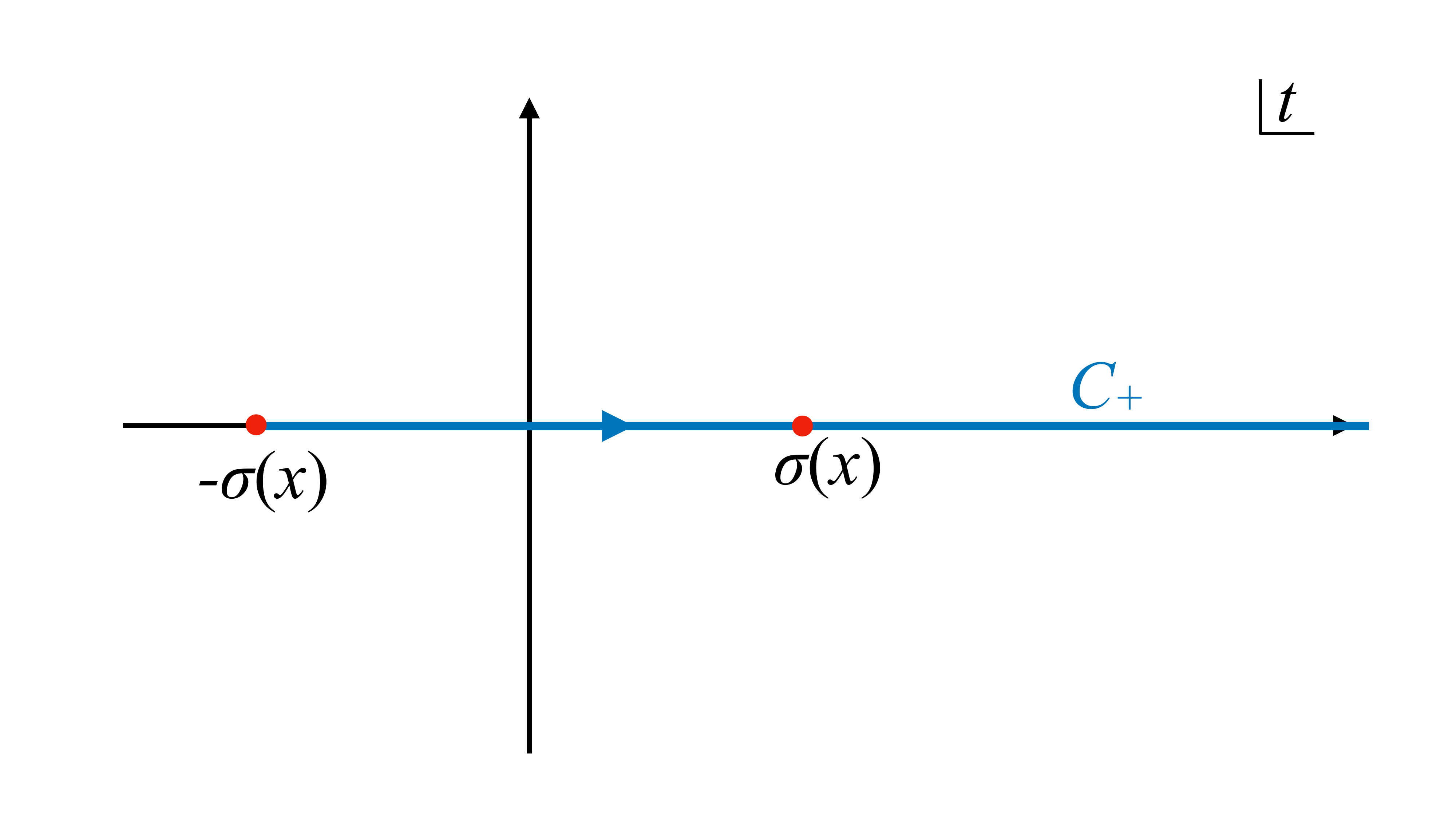}
	\end{subfigure}
	\caption{\textbf{(a)} Borel-summable case with $\Im \, \sigma(x)\neq 0$, where the contours $C_\pm$ avoid the singularities at $t=\pm \sigma(x)$. \textbf{(b)} Singular case with $\Im \, \sigma(x)=0$ and $\Re\, \s(x)>0$, where the contour hits a Borel singularity and $\psi^{(+)}$ becomes discontinuous. It is straightforward to deduce that when $\Re\, \s(x)<0$, the locations of the singularities exchange and $\psi^{(-)}(x)$ becomes singular.} \label{Figure: WaveFunction_Borel}
\end{figure}

It is known that the series~\eqref{Asymptotic_WaveFunction} is factorially divergent~\cite{AKT1}, and, therefore, it should be treated via the Borel-summation procedure. In general, the Borel summation of $\f^{(\pm)}(x)$ is expressed as an integral of the form, 
\begin{equation}
	\psi^{(\pm)}(x) = \frac{1}{\sqrt{\s(x)}}\int_{\mp \s(x)}^{\infty + \Im\left[\mp \s(x)\right]} \mrmd t\, e^{-t} \mathfrak{B}\left[\f^{\pm}\right](t\pm \s(x))\, , \label{Borelsummed_solution}
\end{equation}
where we introduced the rescaling $t\rightarrow \hbar t$ with respect to \eqref{BorelIntegral_single}. It is known that the Borel transform $\mathfrak{B}\left[\f^{(\pm)}\right]$ has singularities at $t=\pm \s(x)$. Then, when $\Im \, \s =0$, for one of the solutions $\psi^{(\pm)}$, the Borel contour $C_\pm = [\mp \s(x),\infty)$ hits the singularity at $\pm\s(x)$, making the corresponding solution $\psi^{(\pm)}$ become discontinuous. See Fig.~\ref{Figure: WaveFunction_Borel} for an illustration of the Borel summable and discontinuous cases associated to \eqref{Borelsummed_solution}.

On the $x$-plane, the singularity corresponds to a \textit{Stokes} curve, which is given by
\begin{equation}
	\Im \, \s(x) 
	= 0 \, .\label{StokesCurve}
\end{equation}
For each curve satisfying~\eqref{StokesCurve}, the particular singular solution is determined by the sign of $\Re \, \s(x)$: If $\Re \, \s(x) > 0$, $\psi^{(+)}$ is singular, while if $\Re \, \s(x)<0$, the singular solution is $\psi^{(-)}$. Note that this can easily be deduced from the Borel contours in~\eqref{Borelsummed_solution} and the locations of the singularities of $\mathfrak{B}\left[\f^{(\pm)}\right]$.

\paragraph{\underline{Airy-type approach}:}For one-dimensional smooth curves, the points $x_0\in \mbbC$ such that $Q(x_0)=0$ are special for the WKB analysis because they correspond to branch points of $\sqrt{Q(x)}$, which defines the Stokes curve~\eqref{StokesCurve}. From the physical perspective, they are turning points of the potential $V(x)$ at the classical energy level $E$. 

\begin{figure}
	\begin{subfigure}[h]{0.36\textwidth}
		\caption{\underline{Stokes diagram for Airy equation}}	\label{Figure: AiryDiagram}
		\vspace{10pt}
		\includegraphics[width=\textwidth]{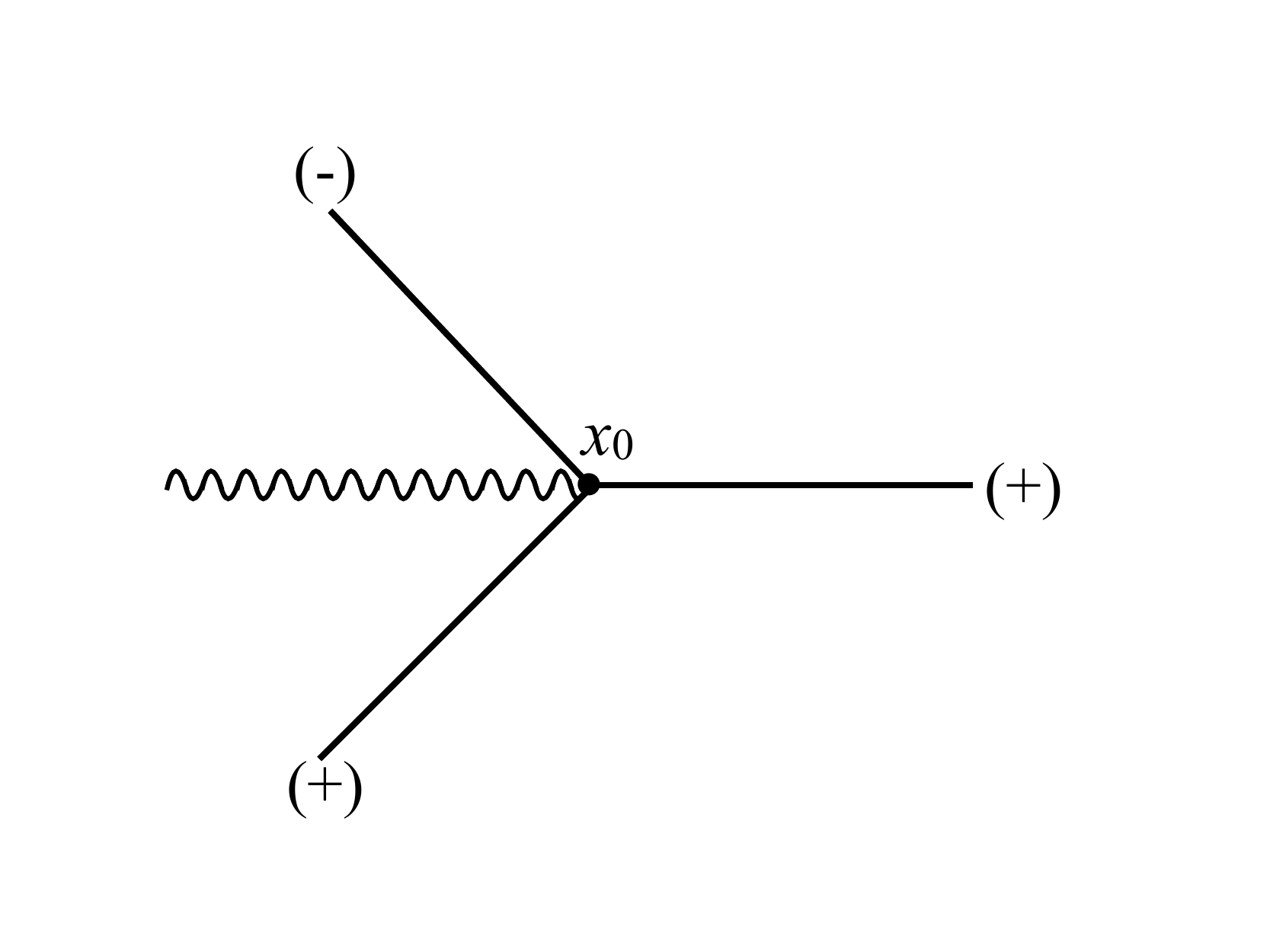}
	\end{subfigure}
	~\hfill
	\begin{subfigure}[h]{0.64\textwidth}
		\caption{\underline{Monodromies for Airy-type EWKB}}	\label{Figure: MonodromyEWKB}
		\vspace{10pt}
		\includegraphics[width=\textwidth]{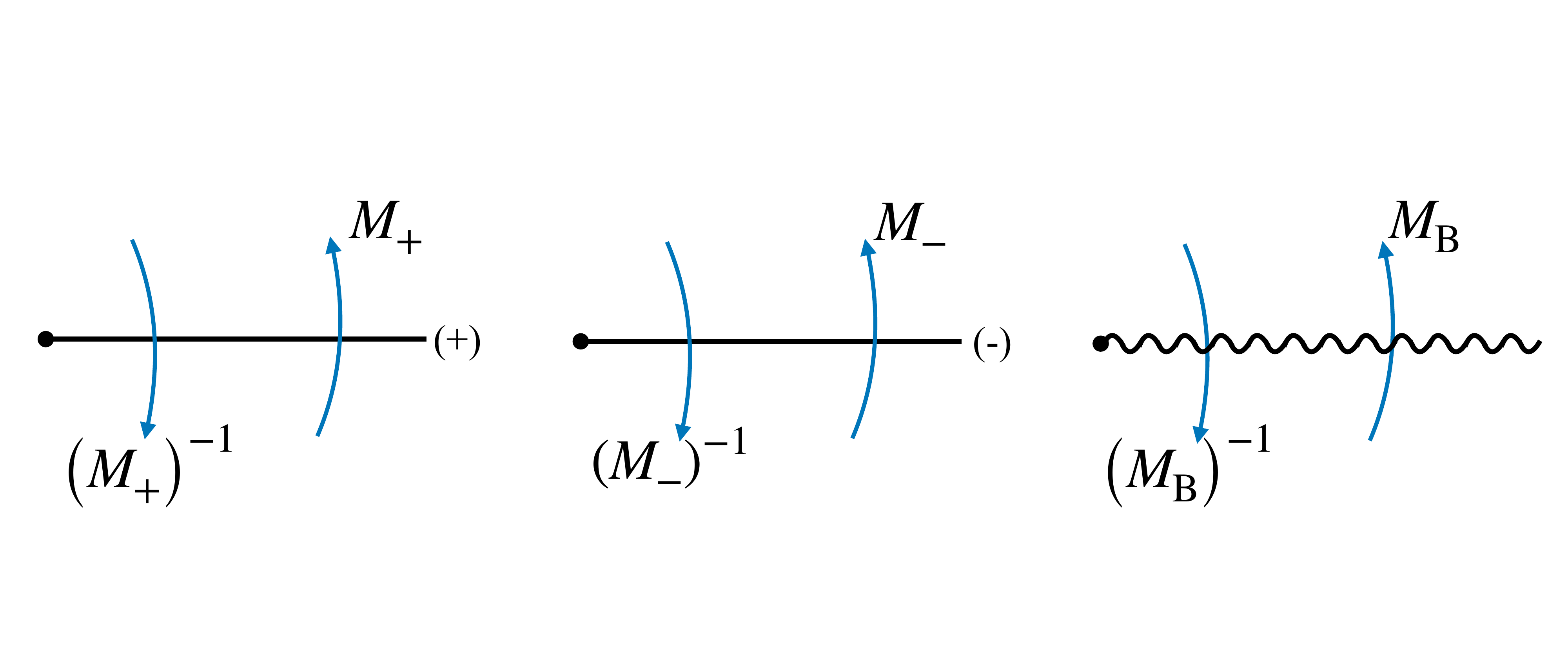}
	\end{subfigure}
	\caption{\textbf{(a)} Stokes diagram for the Airy potential around a simple turning point $x_0$. The labels $(\pm)$ indicate which WKB solution is non-Borel summable across each Stokes line. \textbf{(b)} Airy-type monodromies $M_\pm$ and $M_B$ acting on the Borel-resummed solutions across the Stokes lines (solid lines) and the branch cut (wavy line).} \label{Figure: StokesDiagram_AiryType}
\end{figure}

It is well-known that around simple turning points such that $Q'(x_0)\neq 0$, the Schr\"odinger equation \eqref{Schr\"odingerEquation} is approximated to the Airy equation,
\begin{equation}
	\left[-\frac{\hbar^2}{2}\frac{\mrmd^2}{\mrmd x^2} + x - E \right] \f(x) = 0 \,, \label{AiryEquation}
\end{equation}
for which the EWKB analysis can be performed exactly. In Fig.~\ref{Figure: AiryDiagram}, we illustrated the Stokes diagram for the Airy equation where the labels $(+)$ and $(-)$ for the individual curves indicate that the corresponding solutions, $\psi^{(+)}$ and $\psi^{(-)}$, respectively, are non-Borel summable. The discontinuities across the Stokes curves are represented by the monodromy matrices, which are defined as
\begin{equation}\label{MonodromyMatrices_Airy}
	M_+ = \begin{pmatrix}
		1  & i \\
		0 & 1
	\end{pmatrix} ,  \quad 
	M_- = \begin{pmatrix}
		1  & 0 \\
		i & 1
	\end{pmatrix} , \quad 	
	M_\mrmB = \begin{pmatrix}
		0  & -i \\
		-i & 0
	\end{pmatrix} \, ,
\end{equation}
and their action is described in Fig.~\ref{Figure: MonodromyEWKB}. Note that the last matrix $M_\mrmB$ encodes the discontinuity of the branch cut emanating from $x_0=E$. Other than these discontinuities corresponding to the Stokes curves, both asymptotic series are Borel summable. On the $x$-plane, these regions separated by the Stokes curves are called the \textit{Stokes sectors}.

A very important theorem by Aoki, Kawai, and Takei states that when $V(x)$ is a polynomial, the approximate mapping between~\eqref{Schr\"odingerEquation} and~\eqref{AiryEquation} can be made exact in the EWKB framework~\cite{AKT1}. More specifically, this map holds for the Borel-summable solutions in each Stokes sector. This means that when $x_0$ is chosen as one of the simple turning points of a generic polynomial $V(x)$, the geometry of the Stokes diagram emanating from $x_0$ becomes the same as the one in Fig.~\ref{Figure: AiryDiagram} up to details of the exact form of the curves and the sign choice that determines the discontinuous solution. 

As a consequence of this theorem, the entire Stokes diagram for any polynomial $V(x)$ is given by a collection of Airy-type diagrams emanating from the simple turning points defined by $E=V(x)$. Since these diagrams are of Airy type, the monodromy matrices in~\eqref{MonodromyMatrices_Airy} determine the discontinuities for general $V(x)$. This underlies the power of the EWKB analysis, which allows complicated problems to be treated in terms of simple Airy-type connection formulas.

For generic polynomials $V(x)$, the only additional ingredient is to connect the Stokes sectors associated with different turning points. This requires changing the normalization point, which modifies the Borel summed solutions $\\psi^{(\pm)}$. For changing the normalization point from $x_0$ to $x_1$, this modification is encoded by the \textit{Voros matrix} as
\begin{equation}\label{VorosMatrix}
	N_{x_0,x_1} = \begin{pmatrix}
		e^{\mcalV_{x_0,x_1}}  & 0 \\
		0 & e^{-\mcalV_{x_0,x_1}}
	\end{pmatrix} \, ,
\end{equation}
where $\mcalV$ is given by the integral between the turning points $x_0$ and $x_1$
\begin{equation}
	\mcalV_{x_0,x_1} = \int_{x_0}^{x_1} \mrmd x\, \ts(x,\hbar)\,. \label{Airy_HalfCycle}
\end{equation}

\begin{figure}
	\centering
	\includegraphics[width=0.9\textwidth]{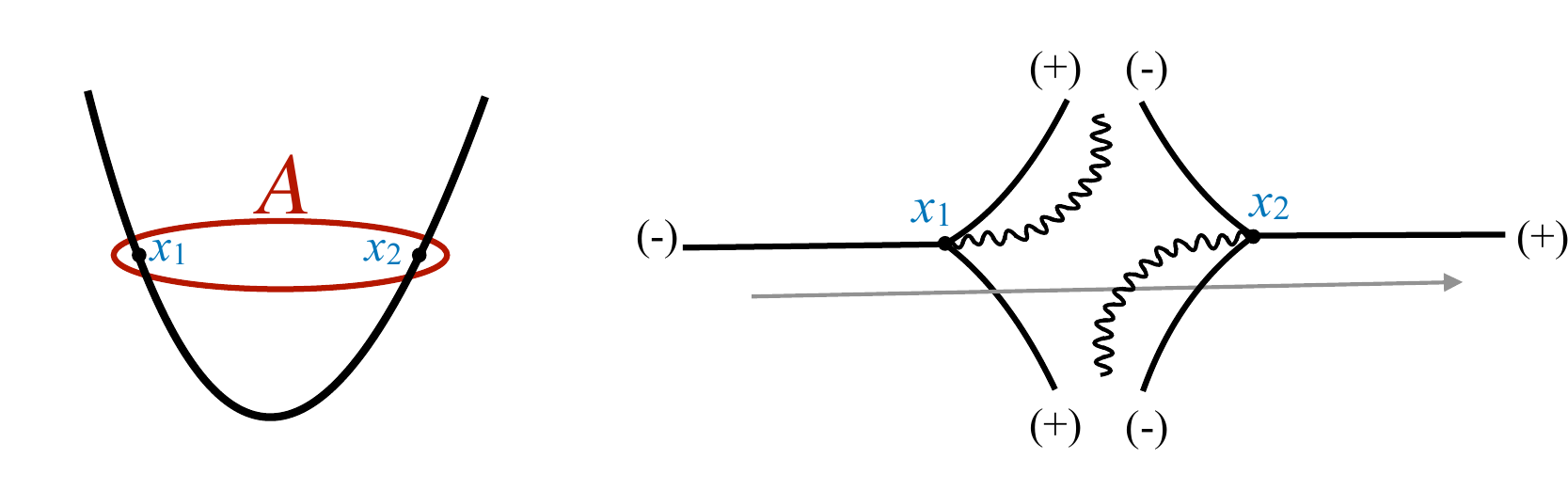}
	\caption{Potential (left) and Stokes diagram (right) for the harmonic oscillator~\eqref{HarmonicOscillator}. The $A$-cycle, enclosing the two turning points $x_1=-\sqrt{2E}$ and $x_2=\sqrt{2E}$, encodes the connection problem between the left and right asymptotic regions.} \label{Figure: HarmonicOscillator}
\end{figure}

Having the monodromy matrices~\eqref{MonodromyMatrices_Airy} and the Voros matrix~\eqref{VorosMatrix}, any connection problem for a polynomial $V(x)$ is encoded in the corresponding Stokes diagram. Let us briefly illustrate the connection problem in the case of a simple harmonic oscillator,
\begin{equation}
	\left[-\frac{\hbar^2}{2}\frac{\mrmd^2}{\mrmd x^2} + \frac{x^2}{2} - E\right]\f(x) = 0 \, . \label{HarmonicOscillator}
\end{equation} 
The corresponding Stokes diagram is depicted in Fig.~\ref{Figure: HarmonicOscillator}. Connecting the left and right asymptotic regions requires incorporating the monodromy matrices. In addition to that, since the natural normalization points for the solutions  in those regions, i.e. $\psi_\mrmL$ and $\psi_\mrmR$ are $x=-\sqrt{2E}$, and $x=+\sqrt{2E}$ respectively, we need to change the normalization points to relate $\psi_\mrmL$ and $\psi_\mrmR$ by using~$\eqref{VorosMatrix}$. 

The connection problem between the right and left regions is formulated as
\begin{equation}
	\Psi_\mrmL = M_+ N_{x_1,x_2} M_\mrmB M_- \Psi_\mrmR \, ,  \label{Connection_HarmonicOscillator}
\end{equation}
where $\Psi_i = \begin{pmatrix}\psi^{(+)}_i \\ \psi^{(-)}_i\end{pmatrix}$ and the integral in the Voros matrix is 
\begin{equation}
	\mcalV_{x_1,x_2} = \frac{1}{2}\oint_A \mrmd x\, \ts(x,\hbar)\, . \label{VorosHarmonic}
\end{equation}
Note that the Voros matrix introduces a very important object to the picture: The WKB cycle associated with the curve $Q(x) = V(x) - E$. In Fig.~\ref{Figure: HarmonicOscillator}, it is labeled as $A$-cycle and in this case, it contains all information about the quantized theory. 

At this point, we stress that the quantization scheme in the Schr\"odinger setting requires additional information, stemming from the boundary condition. For example, in the harmonic oscillator problem, the natural boundary conditions are $\psi^{(+)}(+\infty)=0$ and $\psi^{(-)}(-\infty)=0$, which would keep the system normalizable. Imposing this in~\eqref{Connection_HarmonicOscillator}, we obtain the Bohr-Sommerfeld QC, which we write as
\begin{equation}
	1 + \Pi_A = 0 \, , \label{BohrSommerfeld_QC}
\end{equation}
where 
\begin{equation}
	\Pi_A = \exp\left\{\oint_A \mrmd x\, \ts(x,\hbar) \right\}\, , 
\end{equation}
is the exponentiated action corresponding to the $A$-cycle.

\begin{figure}
	\begin{subfigure}[h]{0.48\textwidth}
		\includegraphics[width=\textwidth]{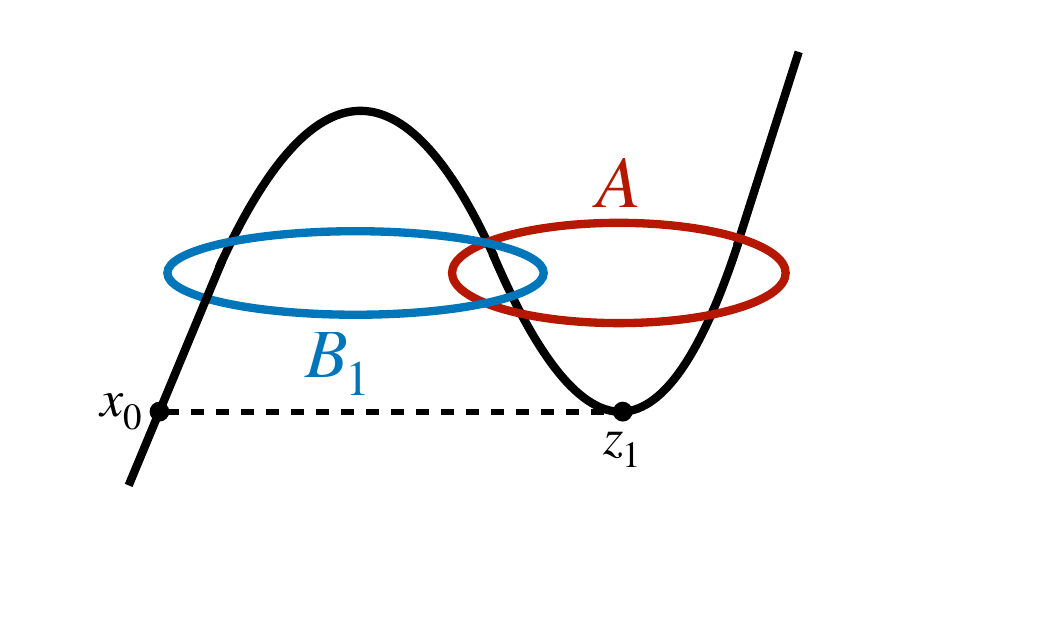}
		\caption{}	
		\label{Figure: Bounce_Generic}
		\vspace{10pt}
	\end{subfigure}
	~\hfill
	\begin{subfigure}[h]{0.48\textwidth}
		\includegraphics[width=\textwidth]{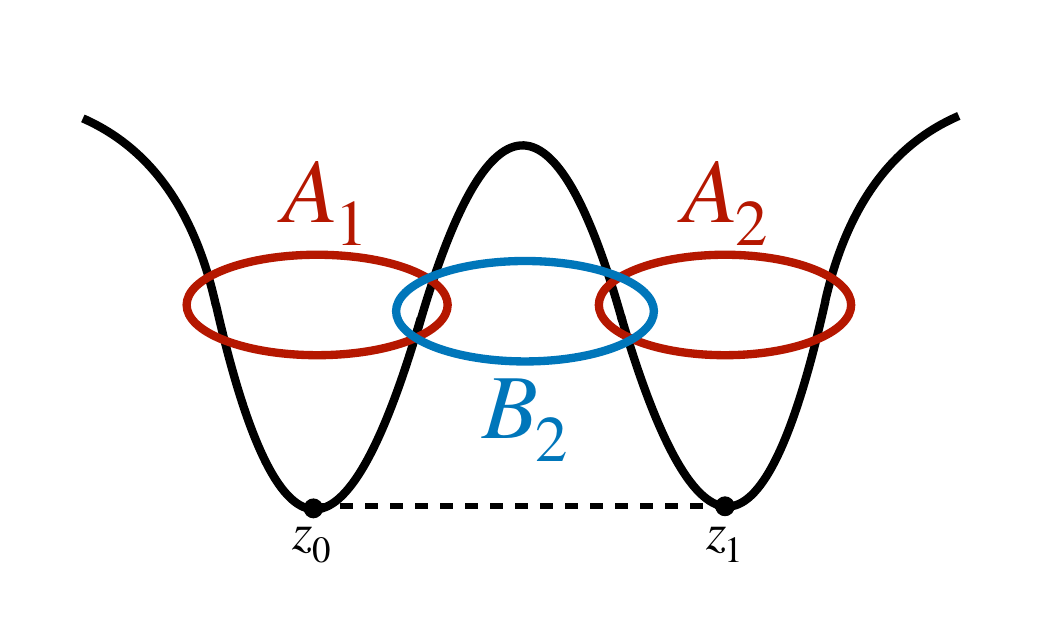}
		\caption{}	
		\label{Figure: Bion_Generic}
		\vspace{10pt}
	\end{subfigure}
	\caption{Two representative WKB-cycle configurations which have different expressions in the Weber-type EWKB approach. \textbf{(a)} A configuration with one saddle point $z_1$ and one simple turning point $x_0$ at $E=E_{\rm cr}$, involving an $A$-cycle and a tunneling $B_1$-cycle. \textbf{(b)} A symmetric double-well configuration with two minima at the same critical level $E=E_{\rm cr}$, involving $A_{1,2}$ and the tunneling cycle $B_2$.} \label{Figure: Generic_Potentials}
\end{figure}

For a higher-order polynomial $V(x)$, even though the connection problem becomes more complicated and the physical information is encoded by more WKB cycles, at its core, it stays the same. The main addition to the harmonic oscillators is the tunneling cycles, which we call $B$-cycles. In Fig.~\ref{Figure: Generic_Potentials}, we illustrate two cases with different types of $B$-cycles, which we consider in this paper. In Fig.~\ref{Figure: Bounce_Generic}, there is one minimum and one simple turning point at $E=E_\mrmcr$, but in Fig.~\ref{Figure: Bion_Generic}, two minima of $V(x)$ are at the same critical level $E=E_\mrmcr$. It turns out that the $B$-cycles have slightly different characters in these two cases. To understand this difference, we turn our focus to a slightly different setup in the EWKB framework, which we call the \textit{Weber-type} EWKB.

\begin{figure}[t]
	\centering
	\includegraphics[width=0.8\textwidth]{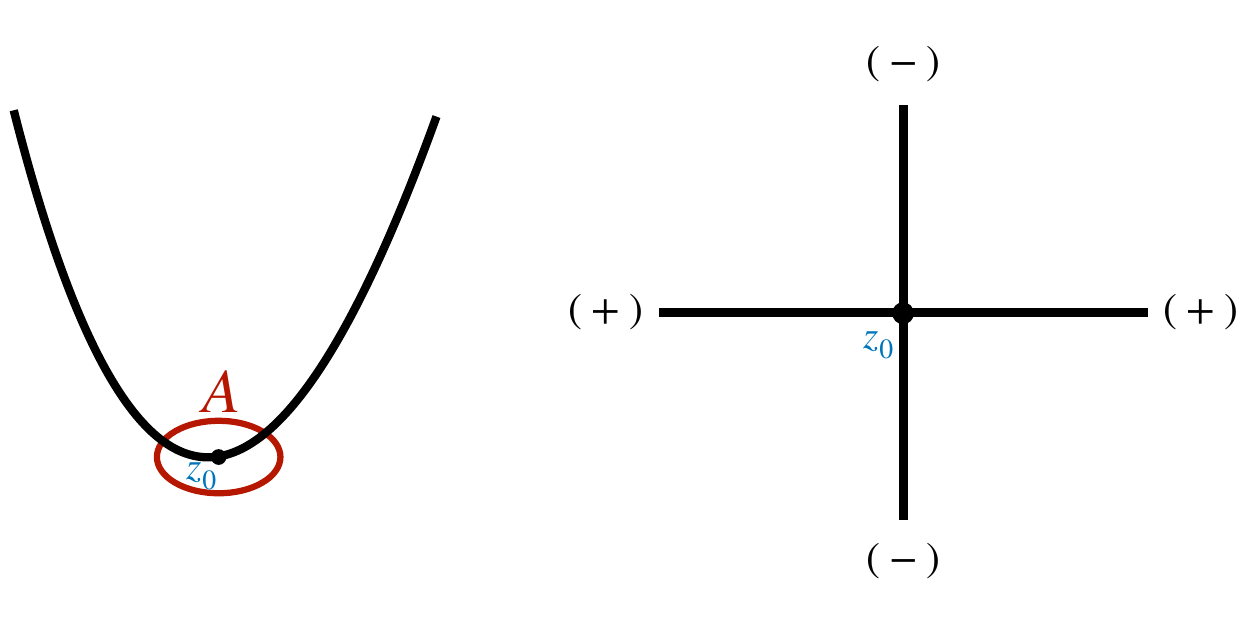}
	\caption{\textbf{(Left)} Harmonic potential with a minimum at $x=z_0$. $A$-cycle represents the limiting case of an Airy-type cycle at classical energy $E=0$. \textbf{(Right)} The Weber-type Stokes diagram emanating from the saddle point at $x=z_0$.}
 \label{Figure: Harmonic_Weber}
\end{figure}

\paragraph{\underline{Weber-type approach}:}As the name suggests, the Weber-type EWKB utilizes the Weber equation, which we write in the following form: 
\begin{equation}
	\left[-\hbar^2 \frac{\mrmd^2}{\mrmd y^2} + \frac{y^2}{4} - \hbar \tilde{E} \right]\tilde{\f}(y) = 0 \,.\label{WeberEquation}
\end{equation}
Note that this can also be considered as a (rescaled) harmonic oscillator but with an important difference that the energy parameter is now written as a quantum term rather than a classical one. This means that in~\eqref{WeberEquation}, the classical energy is set to be zero. Then, the Stokes curves~\eqref{StokesCurve} emanate from the saddle point, i.e.,~double turning point, at the $E=0$ level. For~\eqref{WeberEquation}, we illustrate the potential and the Stokes diagram in Fig.~\ref{Figure: Harmonic_Weber}. 

Note that from a quantum-mechanical perspective, having an energy parameter as $\hbar \tilde E$ is not a problem, as it is well-known that the quantized energy around a locally harmonic well starts with $O(\hbar)$. Then, in some sense, the parameter in~\eqref{WeberEquation} can be considered as a rescaling of the energy in Airy-type approach by setting $E = \hbar \tilde{E}$. This relationship makes these two approaches using~\eqref{HarmonicOscillator} and~\eqref{WeberEquation} become equivalent to each other. 

Similarly to the generalization of the setup for the Airy equation to generic polynomials, the Weber equation~\eqref{WeberEquation} forms a basis for cases when $V(x)$ is locally harmonic, i.e., $V(x) \sim (x - z_0)^2$ around a saddle point $x=z_0$. This is possible due to another theorem by Aoki, Kawai, and Takei~\cite{AKT2} which introduces an exact mapping between~\eqref{WeberEquation} and the Schr\"odinger equation~\eqref{Schr\"odingerEquation} with the rescaling $E=\hbar \tilde{E}$:
\begin{equation}
	\left[-\frac{\hbar}{2}\frac{\mrmd^2}{\mrmd x^2} + V(x) - \hbar \tilde{E} \right]\tilde{\f}(x) = 0 \,.\label{Schr\"odingerEquation_Weber}
\end{equation}
Then, when we focus on the classical energy levels corresponding to the saddle points of $V(x)$, the Weber-type Stokes diagrams in Fig.~\ref{Figure: Harmonic_Weber} become the building blocks for general locally harmonic polynomials $V(x)$.  

\paragraph{\underline{Dictionary}:}Provided by the exact mapping theorems, we infer that for any locally harmonic $V(x)$, we can turn an Airy-type setup to a Weber-type one by simply rescaling the energy parameter as $E=\hbar\tilde{E}$. It turns out that in the Weber-type analysis, the WKB actions $\Pi_A$ and $\Pi_B$ have universal forms~\cite{DP1,Sueishi:2021xti,Misumi:2024gtf,Misumi:2025ijd}, which lead to greater computational power than the direct Airy-type approach, especially for higher-order polynomials. The precise link between the expressions for these two setups is given by a dictionary, which was first\footnote{In fact, before its utilization in the EWKB framework, an equivalent mapping has been used in the uniform WKB formalism, which turns out to be closely related to the Weber-type EWKB approach, but it lacks the geometric information that Stokes diagrams provide. See e.g.~\cite{Dunne:2014bca} for more discussion about the uniform WKB in this direction.} put forward in~\cite{Sueishi:2021xti} and then generalized in~\cite{Misumi:2024gtf,Misumi:2025ijd}.

For the two cases illustrated in Fig.~\ref{Figure: Generic_Potentials}, the dictionary reads
\begin{align}
	\Pi_A &
	= e^{-2\pi i\mcalF_\mrmP(\frac{E}{\hbar},\hbar)} \, ,  \label{Dictionary_Perturbative} 
\end{align}
and
\begin{align} 
	\Pi_{B_1} &
	=  e^{-\mcalG_1(\frac{E}{\hbar},\hbar) }\,\frac{\sqrt{2\pi} \left(\frac{\hbar}{\mcalC_{1}}\right)^{-\mcalF_\mrmP(\frac{E}{\hbar},\hbar)}}{\G\left(\frac{1}{2} + \mcalF_\mrmP\left(\frac{E}{\hbar},\hbar\right) \right)} \, , \label{Dictionary_NonPerturbative}
	\qquad \Pi_{B_2} 
	=  e^{-\mcalG_2(\frac{E}{\hbar},\hbar) }\,\frac{2\pi \left(\frac{\hbar}{\mcalC_{2}}\right)^{-2\mcalF_\mrmP(\frac{E}{\hbar},\hbar)}}{\left[\G\left(\frac{1}{2} + \mcalF_\mrmP\left(\frac{E}{\hbar},\hbar\right) \right)\right]^2}\, .  
\end{align}
The constants $\mcalC_1$ and $\mcalC_2$ are given by the following integrals: 
\begin{align}
	\mcalC_1 &= 2 \, \o_0 \,  \left(z_1 - x_0\right)^2 \exp\left\{2\int_{x_0}^{z_1}\mrmd x\,  \left[\frac{\o_0}{\sqrt{2 V(x)}} - \frac{1}{z_1 - x}\right]\right\}  \, ,  \label{PrefactorLog1} \\ \nonumber \\ 
	\mcalC_2 &= 2 \, \o_0  \left(z_1 - z_0\right)^2\, \exp\left\{\int_{z_0}^{z_1}\mrmd x\, \left[\frac{\o_0}{\sqrt{2 V(x)}} - \frac{1}{x- z_0} - \frac{1}{z_1 - x}\right] \right\} \, , \label{PrefactorLog2}
\end{align}
where $\o_0$ corresponds to the local curvature (harmonic frequency) of the wells located at $x_0$ and $x_1$.
Note that the expression for $\Pi_{B_2}$ in~\eqref{Dictionary_NonPerturbative} and the constant $\mcalC_2$ in~\eqref{PrefactorLog2} assume that the $B_2$-cycle lies between two wells with the same curvature. If the curvatures are different, the expressions differ slightly as discussed in~\cite{Misumi:2025ijd}. For our purposes in this paper, the general formulas are not needed.

The important objects in~\eqref{Dictionary_Perturbative} and \eqref{Dictionary_NonPerturbative} are the exponents $\mcalF_\mrmP$ and $\mcalG$. In general, they are represented by series expansions in $\hbar$ of the forms, 
\begin{align}
	\mcalF_{\mrmP} =  \sum_{m=0}^\infty \mcalF_{\mrmP,k}\,  \hbar^k \, , \qquad  \mcalG_i = \frac{S_i}{\hbar} + \sum_{m=1}^\infty\mcalG_m \hbar^m\, , \label{Action_Expansions}
\end{align}
where $S_i$ corresponds to the classical action of the corresponding $B_i$-cycle. 

In the dictionaries~\eqref{Dictionary_Perturbative} and~\eqref{Dictionary_NonPerturbative}, $\mcalF_\mrmP$ and $\mcalG$ represent the perturbative and non-perturbative information, respectively. Note that $\mcalF_\mrmP$ appears in the expression for $\Pi_{B_i}$. This means that the non-perturbative action contains the information about the perturbative one, indicating resurgence. A greater manifestation of resurgence is established for genus-one curves via an explicit differential equation between $\mcalF_\mrmP$ and $\mcalG$~\cite{Alvarez2,Alvarez3,Dunne:2014bca,Basar:2015xna,Gahramanov:2015yxk,Gorsky:2014lia, Misumi:2024gtf, Misumi:2025ijd, Cavusoglu:2023bai,Cavusoglu:2024usn}. 

Finally, the non-perturbative part of the dictionary in~\eqref{Dictionary_NonPerturbative} links the actions $\Pi_{B_i}$ to the instanton configurations in the path integral picture. We will come back to this relationship in Section~\ref{Section: semi-classical_ITW}, where we discuss the semi-classical-quantization perspective for the inverted triple-well system.

\subsection{Resurgent spectrum in exact WKB} \label{Section: Spectrum_EWKB} As we discussed in Section~\ref{Section: BriefIntro_EWKB}, a Stokes curve normalized at a turning point $x=x_0$ is a result of Borel singularities of the formal WKB series~\eqref{Asymptotic_WaveFunction}, and it represents the discontinuities of the Borel-summed WKB solutions. However, this is different from the discontinuities of physical observables for which we need to address the global structure of a Stokes diagram.

\begin{figure}[t]
	\centering
	\begin{subfigure}[b]{0.45\textwidth}
		\centering
		\caption{\underline{Potential and Stokes diagram at $\t_\hbar = 0$.}}
		\includegraphics[width=\textwidth]{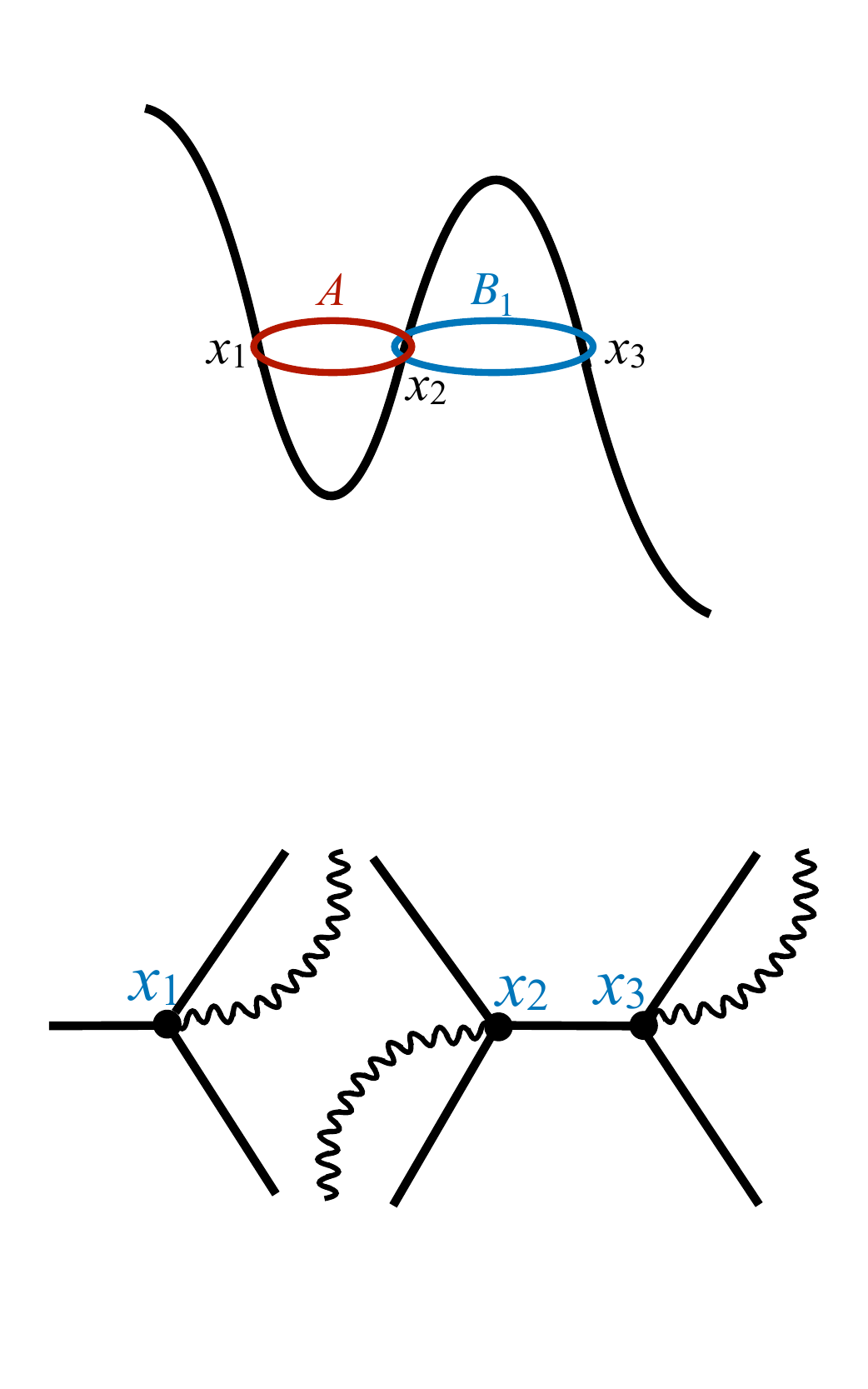}
		\label{Figure: StokesGeneric_Degenerate}
	\end{subfigure}
	\hfill
	\begin{minipage}[b]{0.48\textwidth}
		\centering
		\begin{subfigure}{\textwidth}
			\centering
			\caption{\underline{Stokes diagram for $\t_\hbar > 0$}}
			\includegraphics[width=0.9\textwidth]{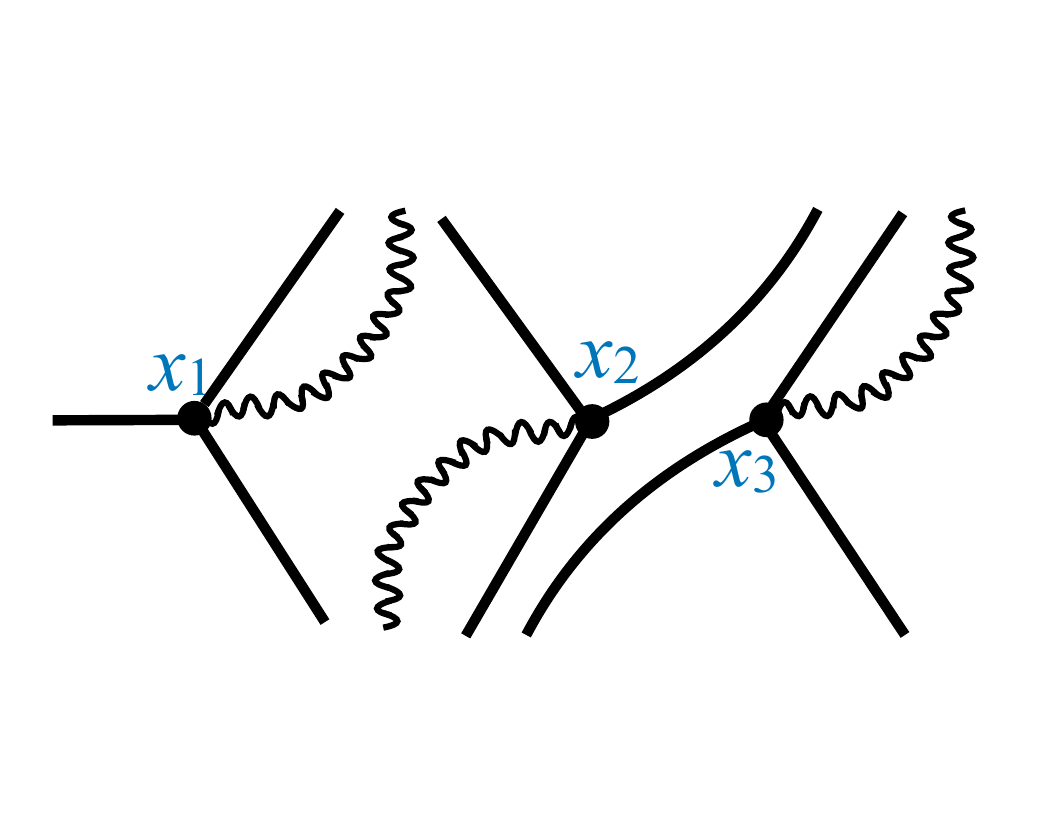}
			\label{Figure: StokesGeneric_Deformed1}
		\end{subfigure}
		\vspace{1em}
		\begin{subfigure}{\textwidth}
			\centering
			\caption{\underline{Stokes diagram for $\t_\hbar < 0$}}
			\includegraphics[width=0.9\textwidth]{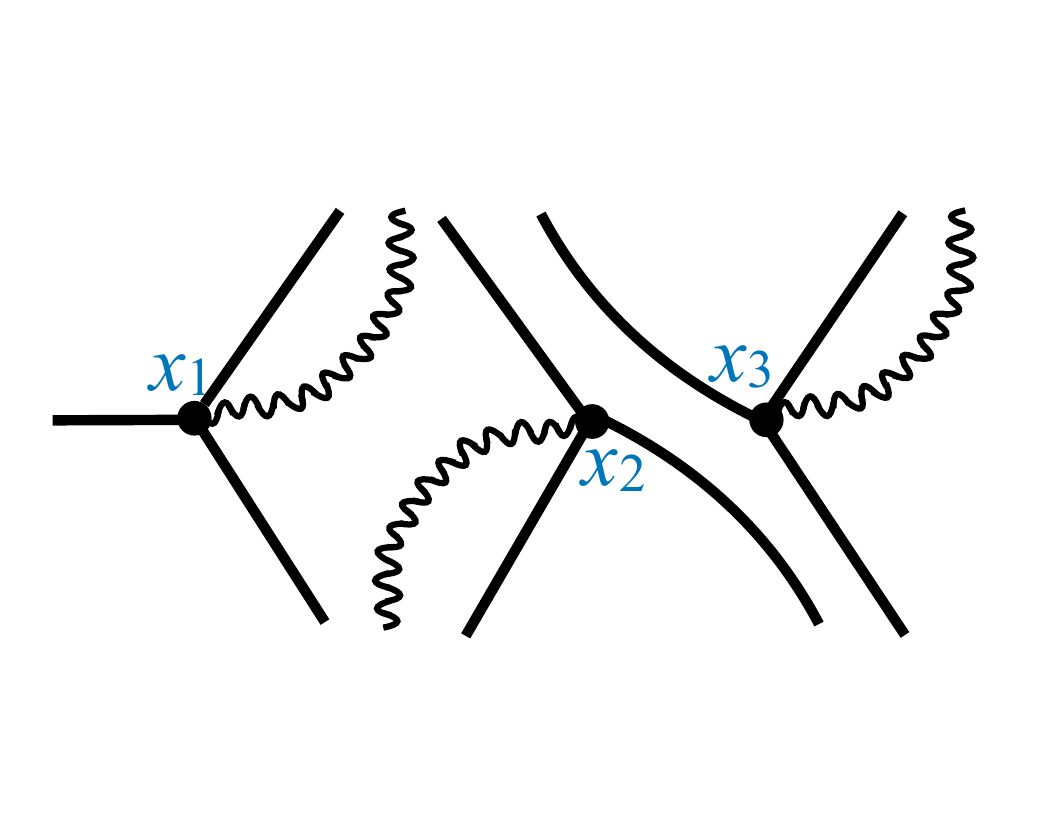}
			\label{Figure: StokesGeneric_Deformed2}
		\end{subfigure}
	\end{minipage}
	\caption{Stokes diagrams for a well-barrier system as the phase of $\hbar=|\hbar|e^{i\theta_\hbar}$ is varied. \textbf{(a)} At $\theta_\hbar=0$ the diagram is degenerate, with a Stokes line connecting the simple turning points $x_2$ and $x_3$. \textbf{(b)-(c)} Turning on $\theta_\hbar\neq 0$ breaks the degeneracy into two analytic continuations related by the Stokes automorphism of the $A$-cycle.}
	\label{Figure: StokesGeneric}
\end{figure}

For a generic polynomial $V(x)$, let us first consider a part of a Stokes diagram corresponding to a well and barrier region. Treating $\hbar$ as a complex parameter and defining $\hbar = |\hbar| e^{i\t_\hbar}$, we can draw the corresponding Stokes diagrams for $\t_\hbar = 0$, $\t_\hbar>0$, and $\t_\hbar<0$ as in Fig.~\ref{Figure: StokesGeneric}. The diagram at $\t_\hbar = 0$ in Fig.~\ref{Figure: StokesGeneric_Degenerate} stands out as a special case, since the Stokes line in the tunneling region connects two simple points $x_2$ and $x_3$. We call such a Stokes diagram as \textit{degenerate}, for which the standard Airy-type monodromies cannot be applied directly. Instead, one needs to turn on the phase $\t_\hbar$, which leads to either of the diagrams in Fig.~\ref{Figure: StokesGeneric_Deformed1} or Fig.~\ref{Figure: StokesGeneric_Deformed2}, depending on the sign of $\t_\hbar$. 

It turns out that the connection problem encoded by these two diagrams is related to a Stokes automorphism. In the particular case of Fig.~\ref{Figure: StokesGeneric}, the discontinuity is due to the non-Borel summability of the $A$-cycle action $\Pi_A$. The associated Stokes discontinuity is~\cite{DP1,DDP2,Iwaki1} 
\begin{equation}
	\mfrS_{\t_\hbar}: \msfS_{\t_\hbar^+}\left[\Pi_A\right] \mapsto \msfS_{\t_\hbar^-}\left[\Pi_A\right] \left(1 + \Pi_B \right)\, , \label{StokesAuto_Acycle}
\end{equation}
where the action $\Pi_{B}$ should be Borel summable. 

In general, for a system with $m$ $A$-cycles and $n$ $B$-cycles, the Stokes automorphism~\eqref{StokesAuto_Acycle} acting on each perturbative action $A_i$ is written as 
\begin{equation}
	\mfrS_{\t_\hbar}^{\nu_i}: \msfS_{\t_\hbar^+}\left[\Pi_{A_i}\right] \mapsto \msfS_{\t_\hbar^-}\left[\Pi_{A_i}\right]  \prod_{j=1}^{n} \left(1 + \Pi_{B_j} \right)^{\nu_i\, \<A_i,B_j\>} \, . \label{StokesAuto_Acycle_general}
\end{equation}
where $i$ and $j$ refer to different $A$- and $B$-cycles, respectively, $ \left\<A_i,B_j\right\>$ is the intersection number between the $A_i$ and $B_j$ cycles, and for all $B$-cycles, we assumed $\msfS_{\t_\hbar^+}\left[\Pi_{B_j}\right]=\msfS_{\t_\hbar^-}\left[\Pi_{B_j}\right]$. 

This is a very useful formula for the EWKB analysis of spectral problems. For example, let us consider a generic polynomial potential $V(x)$ with $n$ $A$-cycles and $m$ $B$-cycles at a given classical energy level $E$ and denote the exact QCs at $\t = \t_\hbar^\pm$ by 
\begin{equation}
	D_{\t_\hbar^\pm}\left(\Pi_{A_1}, \Pi_{A_2}\dots\Pi_{A_m} ,\Pi_{B_1},\Pi_{B_2}\dots,\Pi_{B_n}\right) = 0 \, . \label{ExactQC_general}
\end{equation}
Then, when there are degenerate Stokes lines, the $A$-cycles at $\t = \t_\hbar^\pm$, which have non-zero intersection with the corresponding $B$-cycles, are related via~\eqref{StokesAuto_Acycle_general}. As a result, the Stokes automorphism~\eqref{StokesAuto_Acycle_general} maps the entire QC at $\t= \t_\hbar^+$ to the one at $\t = \t_\hbar^-$.

In terms of the QCs, the Stokes automorphism reads as
\begin{equation}
	\mfrS_{\t_\hbar}: D_{\t_\hbar^+} \mapsto D_{\t_\hbar^-} \, . \label{StokesAuto_QC}
\end{equation}
Then, in a similar manner to the median summations, we can consider half-Stokes automorphisms by setting $\nu_i=\pm \frac{1}{2}$ and define \textit{median quantization conditions} (median QC) as
\begin{equation}
	D_\med = \mfrS_{\t_\hbar}^{1/2} D_{\t_\hbar^+} = \mfrS_{\t_\hbar}^{-1/2} D_{\t_\hbar^-} = 0 \,.\label{medianQC}
\end{equation}
Note that depending on the intersection number $\<A_i,B_j\>\in \mbbZ$, half-Stokes automorphisms might lead to fractional powers such that \[(1+\Pi_{B})^{\pm 1/2} = 1 \pm \frac{1}{2}\Pi_B + O\left(\Pi_B^2\right)\, . \]  Here we pick the branch convention to normalize the half-Stokes automorphism for $\Pi_{B}=0$ to unity, i.e. $\mfrS_{\t_\hbar}^{\pm 1/2}\Big\vert_{\Pi_B = 0} = \mathrm{Id}$. 

Similarly to the median summations, the median QC probes the singular Stokes ray at $\t = \t_\hbar$. Its careful investigation leads to the median summed trans-series in a straightforward manner. This powerful approach prevents us from the hassle of Borel summations of a trans-series representation of the spectrum $E(\hbar)$.

\paragraph{\underline{Solving quantization conditions}:} We note that in relation to the spectrum and $D_\med$ are in fact functional relations because $\Pi_{A_i}$ and $\Pi_{B_j}$ are also functions of $\mcalF_\mrmP(\hbar)$ and $\mcalG(\hbar)$, which induce the $\hbar$ expansions forming the trans-series~\eqref{Trans-seriesGeneral}. Let us briefly review how this trans-series is formed from the exact QCs~\eqref{ExactQC_general}:

If we only want to concentrate on the perturbative spectrum, the problem becomes an equivalent of the simple harmonic oscillator, and exact QC reduces to the Bohr-Sommerfeld equation~\eqref{BohrSommerfeld_QC} for each $A$-cycle. Then, in terms of $\mcalF_\mrmP = \frac{i}{2\pi}\log \Pi_{A}$, we have
\begin{equation}
	\mcalF_\mrmP(E_\mrmP(\hbar,N),\hbar) = N + \frac{1}{2}\, . \label{Perturbative_QC_general}
\end{equation}
Except when $V(x) \propto x^2 $, this is a series expansion in $\hbar$ as in~\eqref{Action_Expansions}, its inversion yields the perturbative spectrum $E_\mrmP(\hbar,N)$.

The trans-series~\eqref{Trans-seriesGeneral} can be considered as (infinitely many) non-perturbative corrections around this series. Then, it is convenient to promote $\mcalF_\mrmP$ to a non-perturbative object and define
\begin{equation}
	\mcalF = \mcalF_\mrmP + \d_\mrmNP \, . \label{NPcorrections_general}
\end{equation}
Then, solving $D_{\t_\hbar^\pm} = 0$ around $\mcalF = \mcalF_\mrmP$ with a $\d_\mrmNP$ expansion leads to a trans-series solution without the leading perturbative order, which is set by the perturbative QC in~\eqref{Perturbative_QC_general}. Finally, to relate this construction with the spectrum~\eqref{Trans-seriesGeneral}, we consider the energy parameter $E(\mcalF,\hbar)$ and expand it around $\mcalF_\mrmP$ for $\d_\mrmNP \ll 1$ as
\begin{equation}
	E \simeq E_\mrmP + \d_\mrmNP \left[\frac{\dee E_\mrmP}{\dee \mcalF}\right]_{\mcalF=\mcalF_\mrmP} + \frac{1}{2}\d^2_\mrmNP \left[\frac{\dee^2 E_P}{\dee \mcalF^2}\right]_{\mcalF = \mcalF_\mrmP} + O\left(\d_\mrmNP^3\right)\, , \label{Spectrum_TransSeries}
\end{equation}
which corresponds to the trans-series expansion of $E(\hbar,N)$. 

Note that \eqref{Spectrum_TransSeries} contains imaginary ambiguities hidden in $\d_\mrmNP$ terms. Moreover, additional ambiguities arise when each series associated with $\frac{\dee^k E_\mrmP}{\dee \mcalF^k}$ is Borel summed. As stated above, the physical spectrum, on the other hand, should be obtained via the median summation of the entire trans-series, which should remove all the ambiguities and lead to a single-valued spectrum. 

In Section~\ref{Section: Trans-seriesDirect}, we will consider the reconstruction of the spectrum for PT-symmetric and (anti-)resonance systems using the solution scheme that we described above. A detailed analysis in this way will also reveal the resurgence structure of the physical spectrum. Note that the median-summed spectrum can also be obtained by solving the median QC~\eqref{medianQC} directly. The associated resurgence structure, on the other hand, requires incorporation of the Alien calculus in a similar way to Section~\ref{Section: Resurgence_review}, which we will discuss next, before finishing this section.

\paragraph{\underline{Alien calculus of quantization conditions}:} The median QC in~\eqref{medianQC} is obtained by the Stokes automorphism $\mfrS_{\t_\hbar}^{\pm 1/2}$ acted on $D_{\t_\hbar^\pm}$, leading to single valued well-defined function. We can act the Stokes automorphism and, more specifically, Alien derivatives on the median QC to understand the associated discontinuity quantitatively. In the following, we are going to discuss this investigation is closely related to the resurgence structure of the spectral function $E(\mcalF,\hbar)$.


Motivated by~\eqref{AlienExpansion_MedianSummed}, we first formally define $D_\med^{(\nu)} = \mfrS_{\t_\hbar}^{\nu} D_\med$ and express the Alien derivative expansion of the exact QC as
\begin{equation}
	D_\med^{(\nu)} = \sum_{k=0}^\infty \frac{\nu^k}{k!}\,  \mathfrak{D}_k \, , \label{AlienDerivativeExpansion_QC}
\end{equation}
where $\mathfrak{D}_k  = \dD^k_{\t_\hbar} \left[D_\med\right]$. Note that, since the dependence on $\hbar$ is via $\Pi_A\left(\mcalF(\hbar)\right)$, to compute the terms for $k \geq 2$ in \eqref{AlienDerivativeExpansion_QC} properly, we need to use the chain rule in~\eqref{AlienChainRule} as
\begin{equation}
	\dD_{\t_\hbar}\left[\mfrD_k\left(\Pi_A(\mcalF)\right)\right] = \dD_A\left[\mfrD_k \right] + \frac{\mrmd \mfrD_k}{\mrmd \mcalF}\dD_{\t_\hbar}\left[\mcalF\right]\, , \label{AlienChainRule_QC}
\end{equation}
where we introduced $\dD_A$ to specify it acts in as a partial Alien derivative on $\mfrD_k$. More specifically, it acts on the actions $\Pi_{A}$ and using \eqref{StokesAuto_Acycle_general}, its specific operation becomes
\begin{equation}
	\dD_A\Pi_A^{\a} = \a \Pi_A^{\a} \sum_{i=1}^n \nu_i \log\left(1 + \Pi_{B_i}\right) = \a \Pi_A^{\a} \log\left[\prod_{i=1}^{n} \left(1+\Pi_{B_i}\right)^{k_i}\right]\, , \label{AlienDerivative_Action}
\end{equation}
where $k_i=\nu\, \<A,B_i\>$. Then, it is also convenient to define a new object for the action of $\dD_A$ on the median QC as
\begin{equation}
	D_k = \dD_A^k \left[D_\med\right]\, . \label{AlienDerivative_QCaction}
\end{equation}
In addition to that $\dD_{\t_\hbar}\left[\mcalF\right]$ in \eqref{AlienChainRule_QC} induces another trans-series. We define it as
\begin{equation}
	\mcalF^{(\nu)} = \sum_{k=0}^\infty \frac{\nu^k}{k!} \mcalF_k \, , \label{AlienDerivative_Perturbative}
\end{equation}
where $\mcalF^{(0)} = \mcalF_0$ and $\mcalF_k = \dD^k_{\t_\hbar}\left[\mcalF_0\right]$. Note that the non-perturbative ansatz~$\eqref{NPcorrections_general}$ is linked to $\mcalF^{(\pm \frac{1}{2})}$, but the organization of \eqref{AlienDerivative_Perturbative} is different from \eqref{NPcorrections_general}. More specifically, $\mcalF_0$ corresponds to the median summation of the expression in \eqref{NPcorrections_general}, i.e.,
\begin{equation}
	\mcalF_0 = \msfS_\pm\left[\mfrS_{\t_\hbar}^{\pm 1/2} \mcalF\right] \,,
\end{equation} 
and $\mcalF_{k \geq 1}$ encodes the resurgence structure when $\nu =\pm \frac{1}{2}$.

This construction is on an equal footing with~\eqref{AlienExpansion_MedianSummed} and reveals the entire resurgence structure of the spectrum in a more abstract way. To understand the precise connection, let us consider a generalized QC:
\begin{equation}
	D^{(\nu)}_\med(\mcalF_0) = \mfrD_0(\mcalF_0) + \nu \mfrD_1(\mcalF_0) + \frac{\nu^2}{2} \mfrD_2(\mcalF_0) + \dots = 0\, , \label{QC_Alien_generalized}
\end{equation}
where $\mfrD_0(\mcalF_0) = D^{(0)}_\med(\mcalF_0)$ by the definition and $\mfrD_{k\geq 1}$ terms encode the entire resurgence structure in an organized manner. 

We note that the QC in \eqref{QC_Alien_generalized} yields $\mfrD_k=0$ at each order. Then, one can obtain the resurgence structure of the spectrum by rewriting $\mfrD_k$ in terms of $D_k$ and $\mcalF_k$,  and solving \eqref{QC_Alien_generalized} order by order for $\mcalF_k$. To the first two orders, we obtain


\begin{equation}
	D_0(\mcalF_0) = 0 \, , \label{Alien_MedianQC}
\end{equation}
and 
\begin{equation}
	D_1(\mcalF_0) + \mcalF_1 \frac{\dee D_0}{\dee \mcalF}\bigg|_{\mcalF=\mcalF_0} = 0 \, , \label{Alien_FirstResurgence}
\end{equation}
The equation in~\eqref{Alien_MedianQC} is the median QC in~\eqref{medianQC}, as expected. The one in~\eqref{Alien_FirstResurgence}, on the other hand, can be re-expressed as
\begin{equation}
	\mcalF_1(\mcalF_0) 
	= - D_1(\mcalF_0) \left(\frac{\dee D_0}{\dee \mcalF}\right)_{\mcalF = \mcalF_0}^{-1} \, .  \label{Alien_EQC_First}
\end{equation}
The higher-order terms in~\eqref{AlienDerivative_Perturbative} can be obtained in the same manner.

\paragraph{\underline{Connection to the spectrum}:} Finally, these order-by-order solutions for $\mcalF_k$ allow us to construct the generalized trans-series of the spectrum, which we express in general as 
\begin{align}
	E(\mcalF^{(\nu)}(\hbar),\hbar) &= E(\mcalF_0(\hbar),\hbar) + \nu \left(\mcalF_1 \frac{\dee E}{\dee \mcalF}\right)_{\mcalF = \mcalF_0(\hbar)} \nonumber \\
	&\quad + \frac{\nu^2}{2}\left( \mcalF_2 \left(\frac{\dee E}{\dee \mcalF}\right)+ \mcalF_1^2 \frac{\dee^2 E}{\dee \mcalF^2} \right)_{\mcalF = \mcalF_0(\hbar)} + O(\nu^3) \, ,  \label{Spectrum_TransSeries_Alien}
\end{align}
where we used \eqref{AlienDerivative_Perturbative} and expanded in $\nu$. Note that in~\eqref{Spectrum_TransSeries_Alien}, each expression depends on the median summed action $\mcalF_0 = \mcalS_{\t_\hbar^\pm}\left[\mcalF\right]$. In this sense, the series is the analytic continuation of the median summed resurgent function $E(\mcalF^{(0)}(\hbar),\hbar)$ and each Stokes jump in the analytic continuations is represented by the value of $\nu$. 

In connection to the spectral problem, \eqref{Spectrum_TransSeries_Alien} corresponds to the Borel summed asymptotic trans-series expansion \eqref{Spectrum_TransSeries} and $E(\mcalF^{(0)}(\hbar),\hbar) = E(\mcalF_0(\hbar),\hbar)$ is the median summed spectrum. The terms $O(\nu^{n\geq 1})$, on the other hand, are linked to the Stokes discontinuity of the resurgent function $E(\mcalF^{(\nu)},\hbar)$ and they correspond to the part which is canceled out during the Borel summation procedure. Therefore, we observe that the application of the Alien calculus on the quantization conditions as discussed above provides both the median summed spectrum and the associated resurgence structure in an organized manner.

To return to the level of the asymptotic trans-series solution, it is necessary to remove the Borel summations $\msfS_{\t_\hbar^\pm}$ and reduce $\mcalF_0$ back to a collection of the asymptotic solutions. Then, for $\nu = \pm \frac{1}{2}$, \eqref{Spectrum_TransSeries} is recovered by setting $\mcalF_0 = \mcalF_\mrmP + \d_\mrmNP$ and expanding around $\mcalF_\mrmP$ for $\d_\mrmNP$. In order to see the link between \eqref{Spectrum_TransSeries_Alien} and the trans-series \eqref{Spectrum_TransSeries}, let us consider the Alien derivative expansion of the energy as
\begin{align}
	\mfrS^\nu_{\t_\hbar}E(\mcalF,\hbar) =  E^{(\nu)}(\mcalF,\hbar) = \sum_{k=0}^{\infty}  \frac{\nu^k}{k!} E_k(\mcalF,\hbar)\, ,\label{AlienExpansion_Spectrum}
\end{align}
where we set $E^{(0)} = E_0$ and $E_k = \dD^k_{\t_\hbar}[E_0]$. Note that in \eqref{AlienExpansion_Spectrum}, $\mcalF$ is considered as a variable rather than a resurgent function itself. 

In this sense, it is an expansion of ``partial'' Alien derivatives acting on the median summed spectrum $E^{(0)}(\mcalF_0(\hbar),\hbar)$. Then, to relate them with the expression in \eqref{Spectrum_TransSeries_Alien}, we can utilize the chain rule \eqref{AlienChainRule} as
\begin{align}
	0=\dD_{\t_\hbar}\left[E_0(\mcalF_0(\hbar),\hbar)\right] = \dD_{\t_\hbar}[E_0(\mcalF,\hbar)]\bigg|_{\mcalF=\mcalF_0(\hbar)} + \frac{\dee E}{\dee \mcalF}\Bigg\vert_{\mcalF=\mcalF_0(\hbar)} \dD_{\t_\hbar}\mcalF_0\, .\label{ChainRule_Spectrum}  
\end{align}
where the overall Alien derivative $\dD_{\t_\hbar}\left[E(\mcalF_0(\hbar),\hbar)\right]$ vanishes due to the inverse functional relationship between $E(\mcalF(\hbar),\hbar)$ and $\mcalF(E(\hbar),\hbar)$ \cite{vanSpaendonck:2023znn}. Then, the $k=1$ term in \eqref{AlienExpansion_Spectrum}, which we evaluate at $\mcalF=\mcalF_0(\hbar)$, becomes
\begin{equation}
	E_1(\mcalF_0,\hbar) = \dD_{\t_\hbar}[E_0(\mcalF,\hbar)]\bigg|_{\mcalF=\mcalF_0(\hbar)} =- \mcalF_1(\mcalF_0) \frac{\dee E_0}{\dee \mcalF}\bigg\vert_{\mcalF=\mcalF_0(\hbar)} \, . \label{AlienDerivative_Energy1}
\end{equation}
Similarly, the vanishing second Alien derivative $\dD^2_{\t_\hbar}\left[E_0(\mcalF_0(\hbar),\hbar)\right]=0$ leads to the following relationship:
\begin{equation}
	\dD_{\t_\hbar}^2\left[E_0(\mcalF,\hbar)\right]\bigg\vert_{\mcalF=\mcalF_0(\hbar)} =- 2 \dD_{\t_\hbar}\left[E_1(\mcalF,\hbar)\right]\bigg\vert_{\mcalF=\mcalF_0(\hbar)} - \left(\mcalF_2\frac{\dee E_0}{\dee \mcalF} + \mcalF_1 \frac{\dee^2 E_0}{\dee \mcalF^2}\right)\bigg\vert_{\mcalF=\mcalF_0(\hbar)} \, ,\label{AlienDerivative_Energy2}
\end{equation}
where $\dD_{\t_\hbar}[E_1(\mcalF,\hbar)]\big\vert_{\mcalF=\mcalF_0(\hbar)} = - \mcalF_1\frac{\dee E_1}{\dee \mcalF}\big\vert_{\mcalF=\mcalF_0(\hbar)}$ in the same way as \eqref{AlienDerivative_Energy1}. 

With the identifications in \eqref{AlienDerivative_Energy1} and \eqref{AlienDerivative_Energy2}, we observe that the expansion \eqref{AlienExpansion_Spectrum} up to $O(\nu^2)$ has the same terms as \eqref{Spectrum_TransSeries_Alien} with opposite signs. This is the sign of resurgence cancellation between the Borel summations, which are induced by the Alien derivative expansion in \eqref{AlienExpansion_Spectrum}, and the trans-series in \eqref{Spectrum_TransSeries_Alien}, which corresponds to the genuine non-perturbative contributions associated with the physical system. We finally note that $-2\dD_{\t_\hbar}[E_1(\mcalF,\hbar)]\big\vert_{\mcalF=\mcalF(\hbar)}$ in \eqref{AlienDerivative_Energy2} is linked to the Borel ambiguity of the $O(\nu)$ term in \eqref{Spectrum_TransSeries_Alien}, which doesn't enter the cancellation of the Borel ambiguity arising from the $E(\mcalF_0(\hbar),\hbar)$ term.

We note that the organization of the trans-series \eqref{Spectrum_TransSeries_Alien} is quite different from the form in \eqref{Spectrum_TransSeries}, where the connection to the divergent perturbation series is clearly stated. In reality both \eqref{Spectrum_TransSeries} and \eqref{Spectrum_TransSeries_Alien} assume the existence of resurgent function $E(\mcalF(\hbar),\hbar)$ and rely on its expansions. In \eqref{Spectrum_TransSeries}, the expansion is performed around the perturbative action $\mcalF_\mrmP$, while in \eqref{Spectrum_TransSeries_Alien}, we expand around the median summed solution directly. 

Despite its abstractness, in comparison to the direct solutions to $D_{\t_\hbar^\pm} =0$, which leads to~\eqref{Spectrum_TransSeries}, the use of the median QC and Alien calculus yields a direct and concrete separation of the median-summed and resurgence-structure parts. Since the latter never appears in the spectrum, this separation is very significant. In Section~\ref{Section: ExactSolutions}, we utilize this approach to construct exact non-perturbative solutions directly. It is also very useful to figure out how the resurgence cancellations take place when we do not have enough quantitative power to reveal them numerically. This last point helps us to understand the semi-classical quantization picture, which will be reviewed in Section~\ref{Section: semi-classical_ITW} and thoroughly discussed in Section~\ref{Section: Trans-seriesDirect} for non-Hermitian inverted triple-well potential.

\paragraph{\underline{Minimal Trans-series}:} Utilizing the Alien calculus, it is also possible to construct the trans-series starting from the asymptotic perturbation series $E_\mrmP$ as well. Recently, a thorough analysis in this direction was pursued in \cite{vanSpaendonck:2023znn}. Although we do not pursue a detailed analysis in this direction, we will briefly elaborate on this and reveal the so-called \textit{minimal trans-series} constituting the entire resurgent trans-series. Let us first express the action of the Stokes automorphism on the perturbative series:
\begin{align}
	\mfrS^\nu_{\t_\hbar}E_\mrmP &= \left(\sum_{k=0}^\infty \frac{\nu^k}{k!} \dD_{\t_\hbar}^k\right)E_\mrmP =  \left(\sum_{k=0}^\infty \frac{\nu^k}{k!} \left(\sum_{m=1}^\infty \dD_{mS}\right)^k\right)E_\mrmP   \nonumber \\
	& = E_\mrmP + \nu \left(\dD_S + \dD_{2S}\right)E_\mrmP + \frac{\nu^2}{2}\dD_S^2 E_{\mrmP}\, + \dots \, . \label{MinimalPerturbative_Alien}
\end{align}
where we assumed the singularities are located at $mS$ for $m\in \mbbN^+$ and ignored the contributions beyond the $2S$ level. Note that in \eqref{MinimalPerturbative_Alien}, the order $\nu^k$ indicates the number of times $\dD_{\t_\hbar}$ acted on $E_\mrmP$. We also observe that the action of $\dD_{mS}^k$ induces an exponential contribution of order $e^{-(m+k)\frac{S}{\hbar}}$; see \eqref{PointedAlien}. The important point of the expansion in \eqref{MinimalPerturbative_Alien} is that all the terms at different orders are connected to the perturbative expansion through a chain of Alien derivatives $\dD_{mS}$. 

The expansion \eqref{MinimalPerturbative_Alien} is an example of the resurgence structure of a minimal trans-series. It is possible to have other minimal trans-series, depending on the physical system. For example, in the symmetric double-well potential \cite{vanSpaendonck:2023znn}, the one-instanton level energy, i.e. $E_\mcalI$, is not connected to the perturbative energies through a resurgence relation. Therefore, applying the Stokes automorphism to $E_{\mcalI}$ induces another minimal resurgence structure of the form
\begin{equation}
	\mfrS_{\t_\hbar}^\nu E_\mcalI =  E_\mcalI + \nu \left(\dD_S + \dD_{2S}\right)E_\mcalI + \frac{\nu^2}{2}\dD_S^2 E_{\mcalI}\, + \dots \, .
\end{equation}

In general, combining all such minimal trans-series leads to the full trans-series of the physical problem under consideration. Let us finally note that even if there are multiple minimal trans-series associated with the spectrum, it is possible that some sectors might vanish. In the PT-symmetric ITW problem, we encounter such a case: At the exceptional point, the non-perturbative corrections which don't belong to the minimal trans-series associated with the perturbative expansion vanish, leaving the entire spectrum to be governed by the resurgence structure encoded by \eqref{MinimalPerturbative_Alien}.

\section{Quantum mechanics of inverted triple-well}
\label{sec:ITW}

\subsection{Setup and numerical solutions}\label{Section: Setup_Numerical_ITW}
\begin{figure}
	\begin{subfigure}[h]{0.48\textwidth}
		\caption{\underline{$x_0 < 1$}}	\label{Figure: ITW1}
		\includegraphics[width=\textwidth]{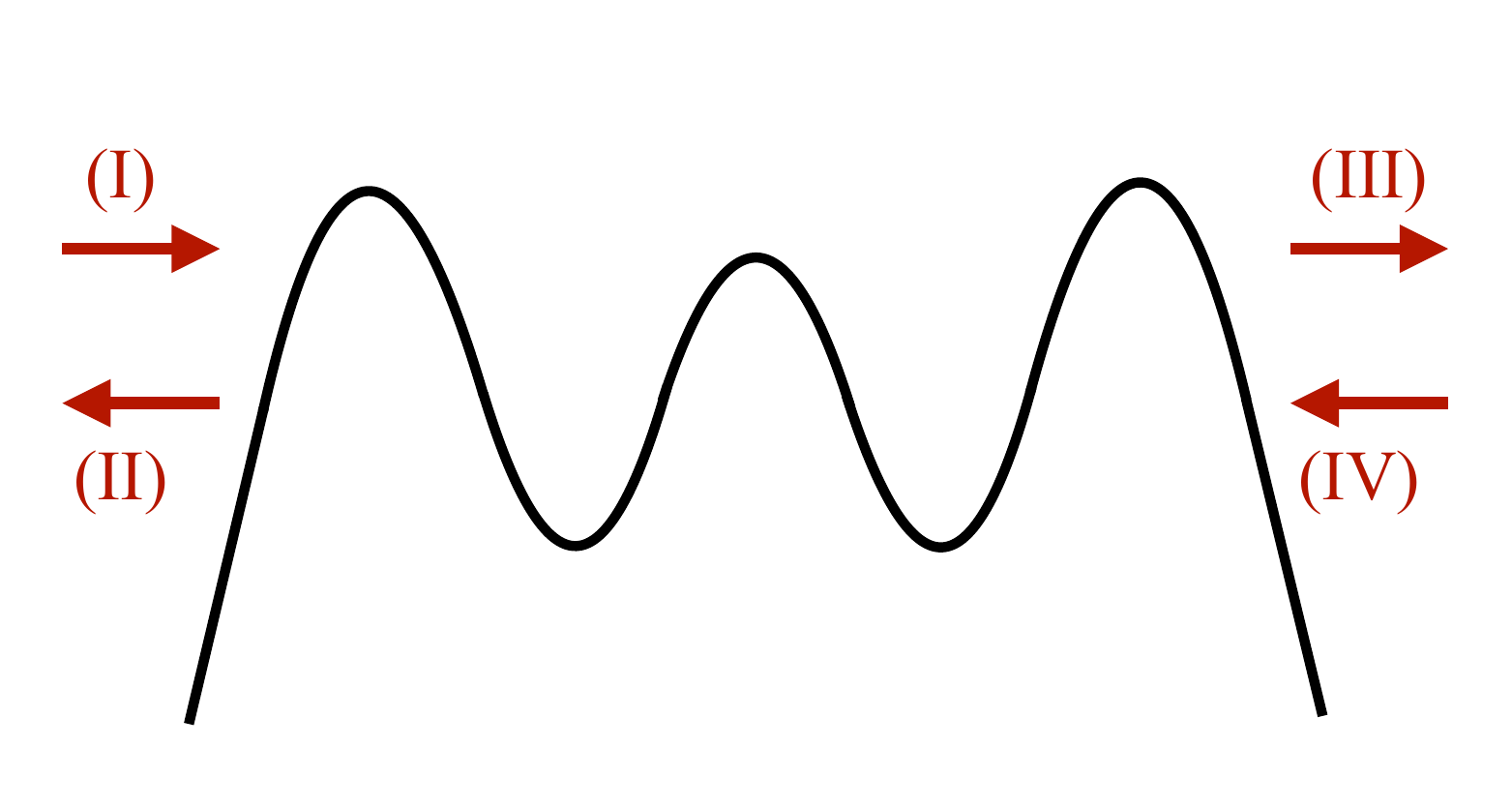}
		\vspace{10pt}
	\end{subfigure}
	~\hfill
	\begin{subfigure}[h]{0.48\textwidth}
		\caption{\underline{$x_0 > 1$}} \label{Figure: ITW2}
		\includegraphics[width=\textwidth]{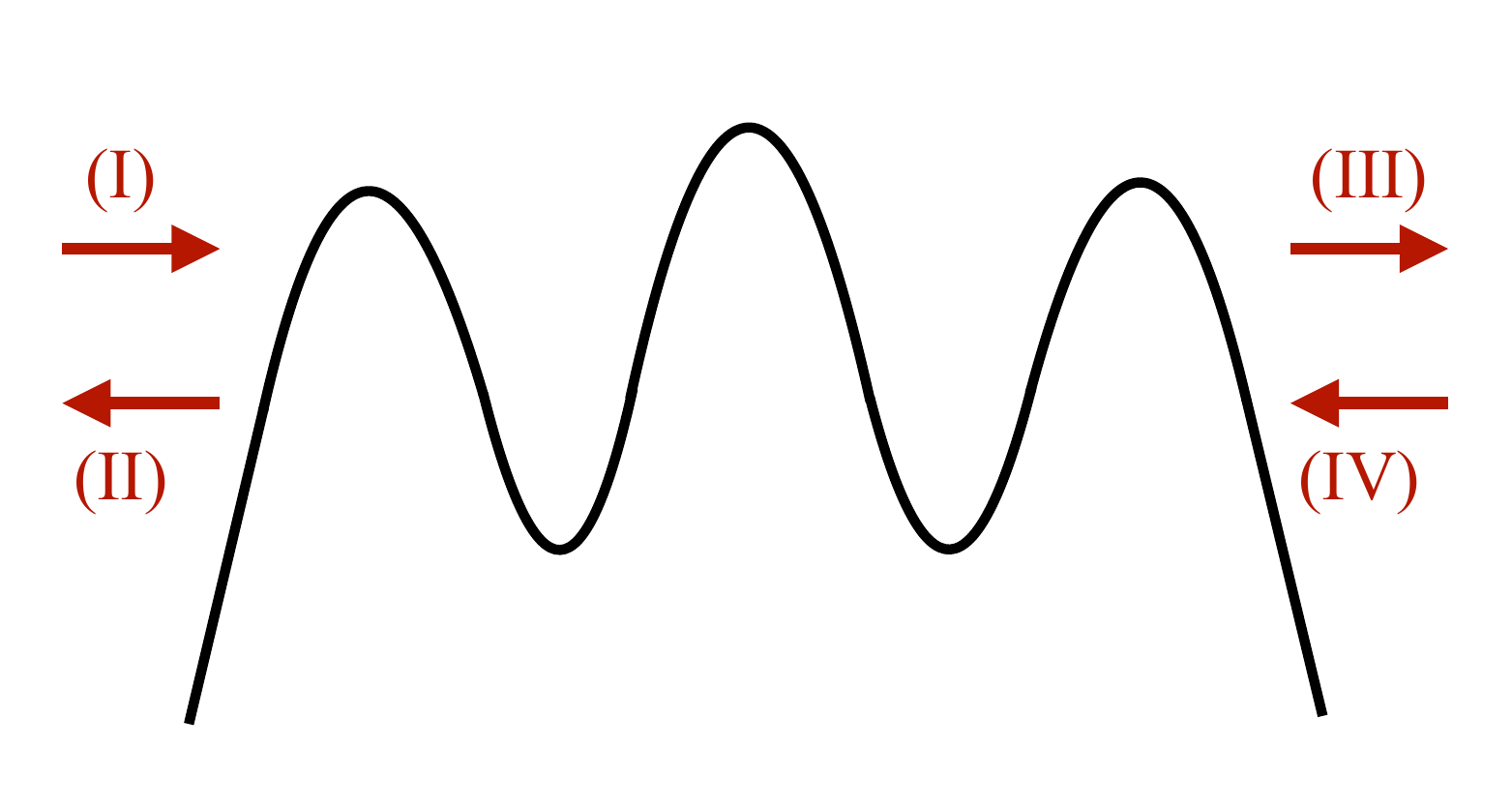}
		\vspace{10pt}
	\end{subfigure}
	\caption{Inverted triple-well potential for two different values of parameter $x_0$. The incoming and outgoing waves are labeled as $(\mathrm{I})$--$(\mathrm{IV})$ and their pairing is used to define the PT-symmetric, resonance, and anti-resonance boundary conditions.} \label{Figure: ITW_Potential}
\end{figure}

Let us introduce the inverted triple-well (ITW) potential of the form
\begin{equation}
	V =- \frac{1}{2}(x^2-2^2) (x^2-x_0^2)^2\, , \label{ITW_potential}
\end{equation}
such that $x_0\in \left(-2,2\right)$. This potential, illustrated in Fig.~\ref{Figure: ITW_Potential}, has a parity symmetry at the classical level and keeps the shape of an inverted triple-well for all $x_0$ in the given interval. 
Since the classical potential is unbounded, ITW is a quintessential example of non-Hermitian quantum mechanics. In some sense, it can be considered as an idealized version of an open quantum system. Then, the system-environment interaction results in outward and inward probability flows, which are enumerated by $(\mrmI)$-$(\mrmIV)$ in Fig.~\ref{Figure: ITW_Potential}, and the relationship between these fluxes determines whether the quantized system has equilibrium or non-equilibrium character. 

When ITW is tackled via the Schr\"odinger equation, a choice of the flow direction is made when the boundary conditions are set by requiring which solutions survive at $x=\pm \infty$. This amounts to imposing boundary conditions that select two of the waves $(\mrmI)$-$(\mrmIV)$ in Fig.~\ref{Figure: ITW_Potential} and set the others to zero. When the waves $(\mrmI)$ and $(\mrmIII)$, or equivalently $(\mrmII)$ and $(\mrmIV)$, are chosen to be non-vanishing and the others are set to zero, the quantized system can be in equilibrium if the inward and outward fluxes balance each other. These choices define a PT-symmetric system.

When the waves $(\mrmII)$ and $(\mrmIII)$ are chosen as the surviving solutions at $x=\pm \infty$, there is an outward flow that is not compensated by an inward one, indicating that any state located originally in a finite region would eventually decay to $x=\pm \infty$. Such systems have resonance characters. On the other hand, if $(\mrmI)$ and $(\mrmIV)$ waves are chosen to non-zero, the flow is inward and the probability of finding a quantum state around the center increases in time. Such systems are called anti-resonance, since they have the opposite characters to the resonance ones. In fact, the resonance and anti-resonance systems are time-reversal to each other, and, in exactly solvable settings, it has been shown that picking forward and backward time-evolution directions favors resonance and anti-resonance systems, respectively~\cite{Ordonez:2016wyh}.

\begin{figure}
	\centering
	\begin{subfigure}[b]{0.6\textwidth}
		\centering
		\caption{\underline{Real eigenvalues for a PT-symmetric system}}
		\includegraphics[width=\textwidth]{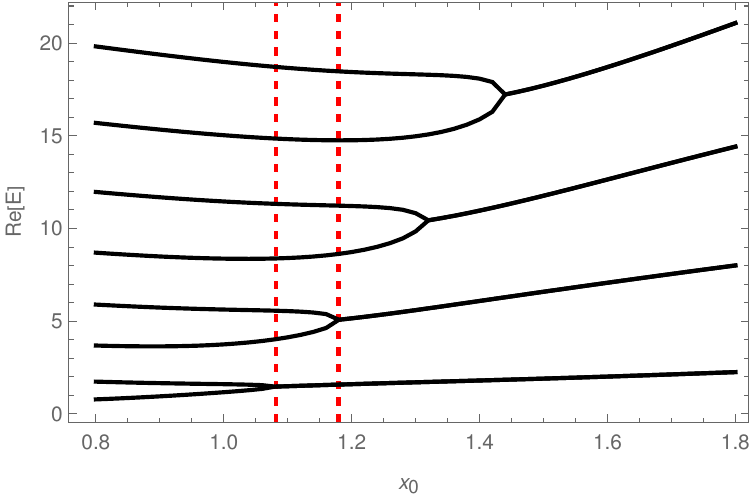}
		\label{Figure: PTsymmetric_Real}
	\end{subfigure}
	
	\vspace{-1em}
	
	\centerline{\underline{Imaginary eigenvalues for a PT-symmetric system}}
	\vspace{0.1em}
	
	\begin{subfigure}[b]{0.42\textwidth}
		\centering
		\caption{\underline{Imaginary energy for $N=0,1$}}
		\includegraphics[width=\textwidth]{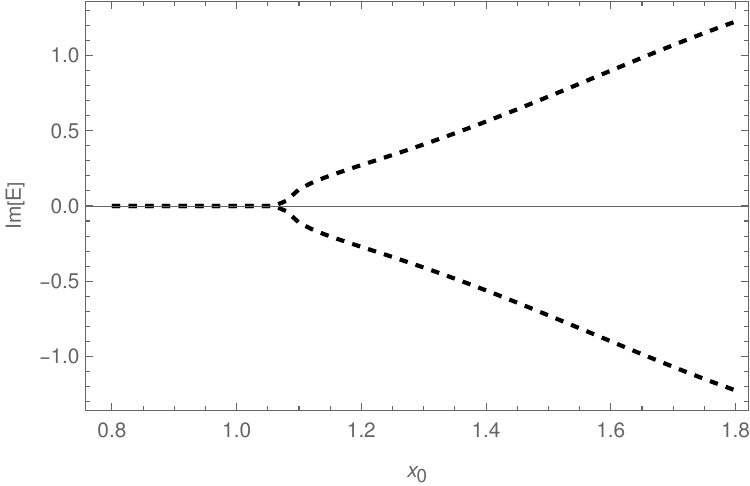}
		\label{Figure: PTsymmetric_Imag0}
	\end{subfigure}
	\hfill
	\begin{subfigure}[b]{0.42\textwidth}
		\centering
		\caption{\underline{Imaginary energy for $N=2,3$}}
		\includegraphics[width=\textwidth]{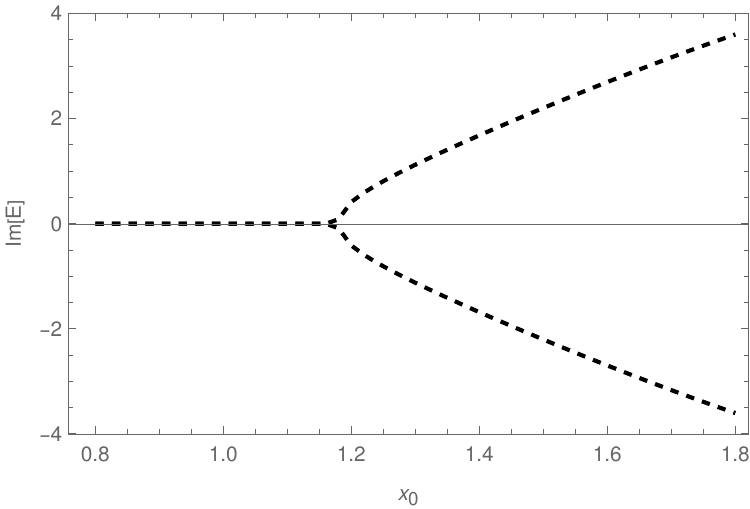}
		\label{Figure: PTsymmetric_Imag1}
	\end{subfigure}
	\vspace{1em}
	\begin{subfigure}[b]{0.42\textwidth}
		\centering
		\caption{\underline{Imaginary energy for $N=4,5$}}
		\includegraphics[width=\textwidth]{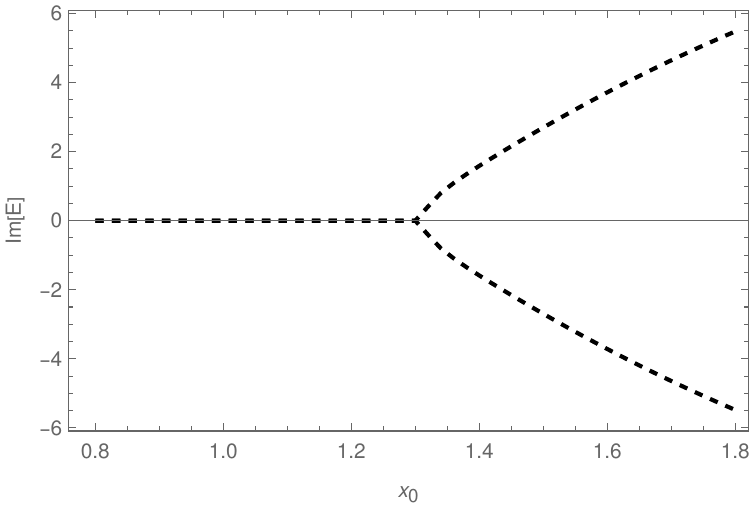}
		\label{Figure: PTsymmetric_Imag2}
	\end{subfigure}
	\hfill
	\begin{subfigure}[b]{0.42\textwidth}
		\centering
		\caption{\underline{Imaginary energy for $N=6,7$}}
		\includegraphics[width=\textwidth]{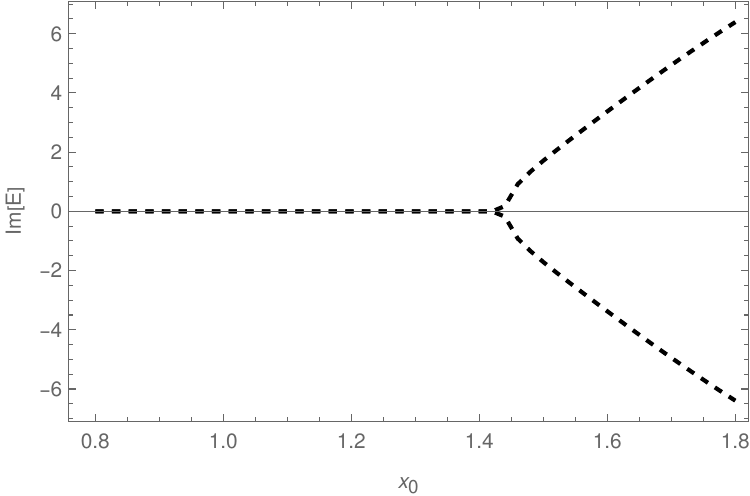}
		\label{Figure: PTsymmetric_Imag3}
	\end{subfigure}
	\caption{Numerical eigenvalues of the PT-symmetric inverted triple-well system as functions of $x_0$ for energy levels $N=0,1, \dots ,7$. \textbf{(a)} Real parts of the low-lying eigenvalues; the dashed vertical lines mark the first two exceptional points. \textbf{(b)--(e)} Imaginary parts for each pair that coalesce at the exceptional points, showing the onset of complex-conjugate branches beyond the respective critical values.}
	\label{Figure: PTsymmetric_Eigenvalues}
\end{figure}

The PT-symmetric case is more subtle. The corresponding boundary conditions reflect the time-symmetric nature of the system, since the two PT-symmetric choices are mapped into each other under time reversal, i.e., the right-moving waves $(\mrmI)$-$(\mrmIII)$ are exchanged with the left-moving waves $(\mrmII)$-$(\mrmIV)$. This suggests the equilibrium character of the PT-symmetric system when the inward and outward probability fluxes balance each other. In such a regime, the spectrum can remain real, which is one of the crucial characteristics of PT-symmetric quantum mechanics. However, there is no guarantee that this balance survives after quantization, and the quantized system may acquire complex eigenvalues in some parameter regions. These regions are called a \textit{broken} PT-symmetric phase, in contrast to the \textit{unbroken} phase with real eigenvalues.

Since the main criterion for $\mathcal{PT}$-symmetry breaking is the emergence of complex eigenvalues, it is natural to examine the spectrum numerically in different parameter regimes. For this purpose, we employ a sinc pseudospectral discretization \cite{gaudreau2015computing} of the Schrödinger operator along the complex contour \cite{Bender:2007nj} \[ z(u)=u-i\alpha\sqrt{1+u^2}\] Here, $\alpha$ controls the asymptotic directions of the contour. Since the potential in \eqref{ITW_potential} is sextic, we choose $\alpha=\tan(\pi/6)$, so that the contour approaches the directions $-\pi/6$ and $-5\pi/6$, along which the wavefunctions are required to vanish in the $\mathcal{PT}$-symmetric case according to the prescription of \cite{Kamata:2024tyb}. The contour is then truncated at $u=\pm L$, where the wavefunctions are numerically negligible, thereby approximating the vanishing boundary conditions at infinity. We use $N=500$ grid points and $L=12$, and diagonalize the resulting non-Hermitian matrix for each value of the physical parameter $x_0$. Numerical accuracy is assessed by checking convergence with respect to $N$ and $L$.

For the ITW potential~\eqref{ITW_potential}, we observe that when $x_0$ exceeds a critical value $x_\mrmcr^{(i)}$ 
complex eigenvalues begin to appear in the spectrum, signaling the PT-symmetry breaking. We illustrate this transition in Fig.~\ref{Figure: PTsymmetric_Eigenvalues}. At $x_0 =x_\mrmcr^{(i)}$, the real parts of a pair of eigenvalues coalesce, and for $x_0 >x_\mrmcr^{(i)}$ they develop complex-conjugate imaginary parts. This pattern, namely the coalescence of the real parts and the emergence of the imaginary parts in eigenvalues, is also familiar from other quantum-mechanical examples of PT-symmetry breaking; see, e.g.,~\cite{Bender:2012ts,Bender:2023cem}. 

Further remarks on the plots in Fig~\ref{Figure: PTsymmetric_Eigenvalues} are as follows: First, the critical points $x_\mrmcr^{(i)}$ for each pair have different values with a hierarchy $x_\mrmcr^{(1)}<x_\mrmcr^{(2)}<x_\mrmcr^{(3)}<\dots$. This means a PT-symmetry breaking condition, which determines $x_\mrmcr^{(i)}$, should depend on the energy level parameter. The plots in Fig.\ref{Figure: PTsymmetric_Eigenvalues} also suggest a square root branch cut behavior around the critical points. This is an indication of the critical points also being \textit{exceptional points}, rather than diabolic ones. At an exceptional point, along with the pair of eigenvalues, their eigenvectors coalesce, and in contrast to a diabolic point, where two distinct states have the same energy eigenvalues and are degenerate, the system ceases to be diagonalizable. In one-dimensional quantum mechanics with decaying boundary conditions, however, the latter is not possible, as there is only one independent eigenfunction for each eigenvalue \cite{hille1968lectures}. Therefore, the points where two eigenvalues coincide in Fig.~\ref{Figure: PTsymmetric_Eigenvalues} are exceptional points. Later in Section~\ref{Section: PTsymmetric_median}, we will verify the double-valued nature and the branching behavior around the critical points $x_\mrmcr^{(i)}$ from an analytical perspective.

\subsection{Semi-classical picture}\label{Section: semi-classical_ITW}
Having the numerical solutions for the eigenvalues, we turn to the main purpose of this paper, i.e., the analytic quantization of non-Hermitian quantum mechanics of ITW. As we stated before, our main approach is the EWKB framework, which is particularly helpful in determining the PT-symmetry breaking analytically. In doing so, we provide a concrete link to the semi-classical quantization in the path-integral formalism and describe physical properties in this language. For this reason, before analyzing the ITW potential using the EWKB formalism, we first introduce the basic building blocks for the semi-classical analysis of the ITW potential.

We first observe that, since~\eqref{ITW_potential} has two symmetric wells for any $x_0\in \left(-2,2\right)$, the perturbative spectra localized in these wells are degenerated to all orders. Note also that the perturbative series is real, meaning that when $x_0>x_\mrmcr$ the PT-symmetry breaking should be induced by non-perturbative effects. Another important feature of the perturbative sector of ITW is that the series diverges factorially at large orders, which is non-Borel summable. This is a general characteristic of perturbative sectors when the corresponding potentials have minima at the same level. In our particular case, we infer the perturbation expansion has the following form:
\begin{equation}
	E_\mrmP(\hbar) = \sum_{k=0}^{\infty} \a^{(1)}_k \hbar^k + \sum_{k=0}^\infty \a^{(2)}_k \hbar^k \, , \qquad \a^{(i)}_k \sim S_i^{-k-\g_i}\G\left(k+ \eta_i+1\right) \, , \label{PerturbationSeries_general}
\end{equation}
where $\g_i$ and $\eta_i$ will be determined later. Note that the perturbation series can be organized in a single sum, but separating it into two parts as in~\eqref{PerturbationSeries_general} will help us for our purpose.

We also note that the perturbation series in~\eqref{PerturbationSeries_general} leads to non-perturbative ambiguities upon the Borel-summation procedure. These ambiguities must participate in the so-called large- and low-orders resurgence cancellation, since the physical observables should be single-valued. Therefore, regardless of other physical properties, there should be compensating (ambiguous) imaginary non-perturbative contributions, to which we now turn our focus.

\begin{figure}
	\centering
	\includegraphics[width=0.8\textwidth]{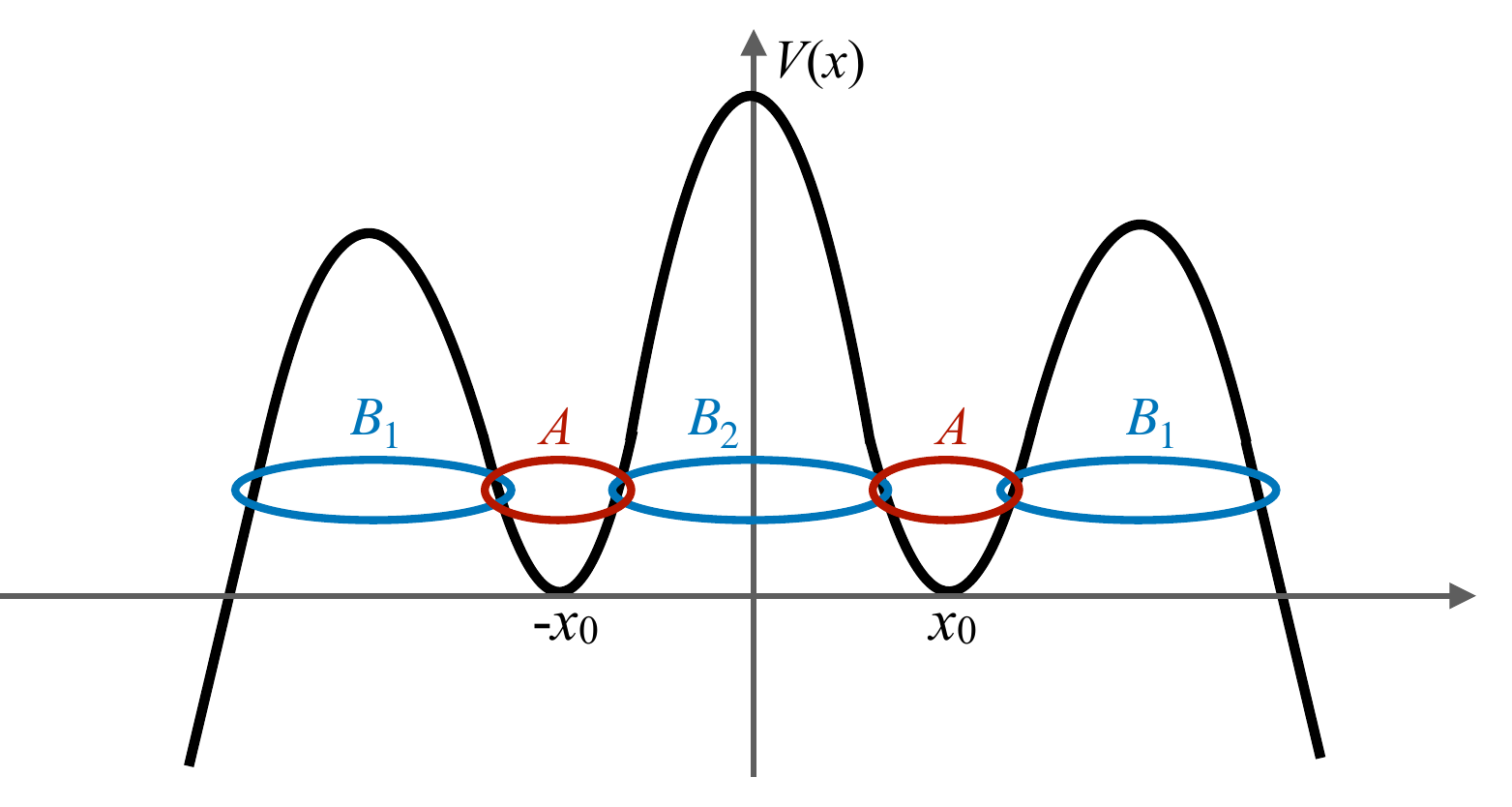}
	\caption{The WKB-cycle structure of the inverted triple-well potential. The outer $B_1$ cycles are associated with bounce configurations, while the central $B_2$ cycle between the two inner wells is related to the bion configuration. The red $A$-cycles are the perturbative cycles around the wells.}
 \label{Figure: InverseTW_cycles}
\end{figure}
The non-perturbative sector of ITW is governed by two types of configurations. One of them is linked to the \textit{bounce} solution, which, in our setting, appears in the $E=0$ limit of the $B_{1}$-cycle depicted in Fig.~\ref{Figure: InverseTW_cycles}. We label a bounce event contributing to the trans-series as $[\mcalB]$. For the ITW in~\eqref{ITW_potential} and at the leading order in the non-perturbative sector, it reads
\begin{equation}
	[\mcalB]_\pm \simeq  \mp i \frac{e^{-\frac{S_1}{\hbar}}}{2\sqrt{2\pi}N!}\left(\frac{\mcalC}{\hbar}\right)^{N+\frac{1}{2}}\, , \label{Bounce_LeadingOrder}
\end{equation}
where $\pm$ subscript in $\left[\dots\right]_\pm$ refers to the configurations for $\sgn\left(\Im\hbar\right) =\pm 1$. We note that in \eqref{Bounce_LeadingOrder} they arise with opposite imaginary signs, indicating the so-called non-perturbative ambiguity. We also note that  in \eqref{Bounce_LeadingOrder}, the factor $N$ corresponds to the energy levels,
\begin{align}
	S_1 &= 2 \int_{x_0}^2 \mrmd x\, \sqrt{E -V(x)} =  \frac{1}{2} x_0\left(x_0^2 + 2\right)\sqrt{4 - x_0^2} + 8 \left(1- x_0^2\right) \arccot\left[\sqrt{\frac{2-x_0}{2+x_0}}\right] \, , \label{BounceAction_ITW}
\end{align}
is the classical action for a bounce and 
\begin{equation}
	\mcalC = x_0^3 \left(4 - x_0^2\right)^{5/2}\, , \label{One_LoopDeterminant}
\end{equation}
is the one-loop determinant term. The expression in~\eqref{Bounce_LeadingOrder} is sometimes called fugacity~\cite{Callan:1977gz} in multi-instanton expansions (or the multi-bounce expansion in the present case) in the dilute-gas treatment~\cite{Callan:1977pt,Bogomolny80,Zinn-Justin:1981qzi,Zinn-Justin:1983yiy,Behtash:2018voa}. Although we do not delve into the dilute-gas picture, we will keep the language in this paper.

We note that bounce solutions in quantum mechanics and quantum field theory were originally introduced long ago in the context of the vacuum decay~\cite{Coleman:1977py, Callan:1977pt}. 
In the exact quantization, however, a bounce does not necessarily signal a complex spectrum nor an instability. These are intrinsic characteristics of a quantum system and are mainly determined by the boundary conditions in our setting. Then, to uncover the precise role of the bounce solution in the non-perturbative spectrum, one should address the exact quantization techniques, which we discuss in Section~\ref{Section: Trans-seriesDirect} (see also~\cite{Misumi:2025ijd} for a recent discussion). 

The other types of non-perturbative configurations in the ITW system are linked to the \textit{bion} solution. In our setting, a bion is associated with the $E=0$ limit of the $B_2$-cycle in Fig.~\ref{Figure: InverseTW_cycles}. We find its classical action as
\begin{align} 
	S_2 &= 2\int_{-x_0}^{x_0}\mrmd x\, \sqrt{E - V(x)} \nonumber \\
	&= 4\pi \left(1-x_0^2\right) + x_0 \left(x_0^2 + 2\right)\sqrt{4 - x_0^2} + 16 \left(1-x_0^2\right)\arccot\left[\sqrt{\frac{2+x_0}{2-x_0}}\right] \, . \label{BionAction_ITW}
\end{align}
Unlike the bounce, a bion has a topological character, since it consists of a combination of instanton and anti-instanton events. Therefore, in this paper, we represent one-bion event as $[\mcalI \mcalIbar]$, emphasizing the non-trivial inner structure that is in contrast to the one-bounce event $[\mcalB]$. 

At the leading order, a single instanton (or anti-instanton) event contribution is 
\begin{align}
	[\mcalI] = [\mcalIbar] \simeq \frac{e^{-\frac{S_2}{2\hbar}}}{\sqrt{2\pi}N!}\left(\frac{\mcalC}{\hbar}\right)^{N+\frac{1}{2}} \, , \label{Instanton_LeadingOrder}
\end{align}
which can be considered as the fugacity of the instanton molecules in the dilute-gas system. Note that $\mcalC$ is the same as in~\eqref{One_LoopDeterminant}. It is well-known that in Hermitian quantum mechanics, $[\mcalI]$ (and $[\mcalIbar]$) is responsible for the real splitting of the perturbative degeneracy in a double-well system. As we will discuss later, for ITW, the role of $[\mcalI]$ stays the same if they appear in the trans-series solutions. Contrary to the Hermitian double-well case, this is not guaranteed, and we will show in Section~\ref{Section: PT-symmetric_Direct} that in the broken PT-symmetric phase, no $[\mcalI]$ (nor $[\mcalIbar]$) event contributes to the spectrum.

Another known fact in Hermitian systems is that when instanton and anti-instanton events are combined to form bions, it leads to an imaginary ambiguity. At the leading order, this ambiguous contribution becomes
\begin{equation}
	[\mcalI \mcalIbar]_\pm \simeq \mp i\frac{e^{-\frac{S_2}{\hbar}}}{2 N!}\left(\frac{\mcalC}{\hbar}\right)^{2N + 1} + \Re\left([\mcalI\mcalIbar]_\pm\right)\, , \label{Bion_LeadingOrder}
\end{equation}
where the subscripts of $\left[\dots\right]_\pm$ again refers to the configurations for $\sgn\left(\Im\hbar\right) =\pm 1$. We also note that the real part in~\eqref{Bion_LeadingOrder} is single-valued. Finally, we would like to emphasize that the expressions~\eqref{Bounce_LeadingOrder}, \eqref{Instanton_LeadingOrder}, and~\eqref{Bion_LeadingOrder} are on an equal footing with their counterparts in the semi-classical analysis of Hermitian settings (see, e.g., \cite{Coleman:1977py,Callan:1977pt,Bogomolny80,Zinn-Justin:1983yiy} for classical papers and~\cite{Jentschura:2011zza,Dunne:2014bca,Misumi:2014raa, Misumi:2014jua,Misumi:2014rsa,Misumi:2014bsa,
 Misumi:2016fno,Fujimori:2018kqp,Basar:2015xna, Misumi:2015dua,Dunne:2020gtk, Behtash:2015loa,Behtash:2015zha,Behtash:2017rqj,Fujimori:2017oab, Fujimori:2017osz,Behtash:2018voa, Sueishi:2019xcj, Fujimori:2022lng} more modern examples). 

Having introduced the basic building blocks, let us return to the semi-classical quantization of the ITW system. Since the perturbation series is divergent and non-Borel summable and there are various non-perturbative configurations, it is natural to turn to the resurgence theory and examine the intricate relationships between different parts contributing to the spectrum. Before discussing the trans-series structures of PT-symmetric and (anti-)resonance systems in the next section, we now elaborate the resurgence cancellations between the Borel-summed perturbation series and ambiguous non-perturbative configurations.

As we mentioned in Section~\ref{Section: Resurgence_review}, this cancellation procedure is called the Bogomolny-Zinn--Justin (BZJ) mechanism~\cite{Bogomolny77,Bogomolny80,Zinn-Justin:1981qzi,Zinn-Justin:1983yiy} and has been discussed inteisively in Hermitian setups~\cite{Zinn-Justin:2004vcw,Zinn-Justin:2004qzw, Basar:2013eka, Dunne:2013ada, Dunne:2014bca,Misumi:2014raa, Misumi:2014jua,Misumi:2014rsa,Misumi:2014bsa,
 Misumi:2016fno,Fujimori:2018kqp,Basar:2015xna, Misumi:2015dua,Dunne:2016qix,Dunne:2020gtk, Behtash:2015loa,Behtash:2015zha,Behtash:2017rqj, Kozcaz:2016wvy, Dunne:2016jsr, Fujimori:2016ljw, Basar:2017hpr, Fujimori:2017oab, Fujimori:2017osz,Behtash:2018voa, Sueishi:2019xcj, Fujimori:2022lng, Cavusoglu:2023bai,Misumi:2025ijd} (see also~\cite{Kamata:2023opn,Misumi:2025ijd} for related discussions in PT-symmetric systems without referencing to the PT-symmetry breaking). The BZJ mechanism is needed for the spectrum to be unambiguous. As we mentioned before, this should also be the case in non-Hermitian systems, where the final spectrum must be single-valued. This means that the Borel ambiguity of the perturbation series in~\eqref{PerturbationSeries_general} or the other non-perturbative ambiguities arising from configurations such as $[\mcalB]_\pm$ in \eqref{Bounce_LeadingOrder} or $[\mcalI\mcalIbar]_\pm$ in \eqref{Bion_LeadingOrder} should vanish in the final spectrum.


When the spectrum is complex, some part of the imaginary contributions should remain. In this paper, we will encounter two possible ways leading to a complex spectrum: 
\begin{enumerate}
	\item First possibility is that the imaginary contributions from the Borel summation and non-perturbative configurations do not agree. Then, the cancellation takes place partially and results in an imaginary contribution in the Borel-summed trans-series. In our discussion, we observe such an emergence of complex spectra in the resonance and anti-resonance cases, as well as in the broken PT-symmetric phase. 
	
	A crucial point in such cases is that different analytic continuations, which originally lead to ambiguities, must lead to a single unambiguous result. Therefore, even if they do not coincide exactly, the imaginary parts of the Borel-summed perturbation series and non-perturbative configurations should still act in harmony. 
	
    \item Another possibility is that the imaginary non-perturbative contributions are not related to any perturbative series in the pre-Borel-summed trans-series. In this case, as there is no counterpart to this imaginary contribution, it directly leads to a complex spectrum. We encounter such non-perturbative configurations in the case of the broken PT-symmetry.
\end{enumerate}

We emphasize that in all cases, the resurgence cancellations take place to some extent. In fact, as we will discuss in Section~\ref{Section: Trans-seriesDirect} and Section~\ref{Section: ExactSolutions}, there is a subpart in the full trans-series which is the same for all of the cases regardless of the other physical properties. It can be called a \textit{minimal} trans-series, in comparison to the full trans-series~\cite{vanSpaendonck:2023znn}. The leading-order non-perturbative configurations of this subpart are $[\mcalB]_\pm$ and $\Im\left([\mcalI\mcalIbar]_\pm\right)$, along with the perturbative expansion~\eqref{PerturbationSeries_general}. For all choices of the boundary conditions, they participate in the resurgence cancellations in the same way. In fact, as we discussed above, this is necessary since the perturbation series is non-Borel summable, which leads to a non-perturbative ambiguity and needs non-perturbative counterparts to fix it. Moreover, since it is in the same form for all the cases, this non-perturbative counterpart should be the same. 

Let us finally numerically verify $[\mcalB]_\pm$ in~\eqref{Bounce_LeadingOrder} and $\Im \left([\mcalI\mcalIbar]_\pm\right)$ in~\eqref{Bion_LeadingOrder} are indeed the events that take part in this cancellation. Being imaginary exponentially-suppressed contributions, they are intimately related to a divergent series via the dispersion relation~\cite{Bender:1973rz,Lipatov:1977cd}. For this reason, we express the divergent series associated with the bounce $(i=1)$ and bion $(i=2)$ configurations as
\begin{equation}
	\mcalE^{(i)} = \sum_{k=0}^\infty \b_k^{(i)} \hbar^k \, , \label{PerturbativeExpansions_divided}
\end{equation} 
For $[\mcalB]_\pm$ in~\eqref{Bounce_LeadingOrder}, we obtain the associated large-order behavior of the coefficients as
\begin{equation}
	\b_k^{(1)} \sim \frac{\mcalC^{N+\frac{1}{2}}}{2\pi\sqrt{2\pi} N!} \frac{\G\left(k + N + \frac{1}{2}\right)}{S_1^{k+N+\frac{1}{2}}}\, ,\label{LargeOrder_Bounce}
\end{equation}
and for $\Im\left([\mcalI\mcalIbar]_\pm\right)$, we get
\begin{equation}
	\b_k^{(2)} \sim \frac{\mcalC^{2N+1}}{2\pi N!} \frac{\G(k+2N+1)}{S_2^{k+2N+1}} \, . \label{LargeOrder_Bion}
\end{equation}

\begin{figure}
\centering
\begin{subfigure}[h]{0.48\textwidth}
	\caption{\underline{$x_0 = \frac{1}{2}$}}	\label{Figure: DivergenceBion}
	\includegraphics[width=\textwidth]{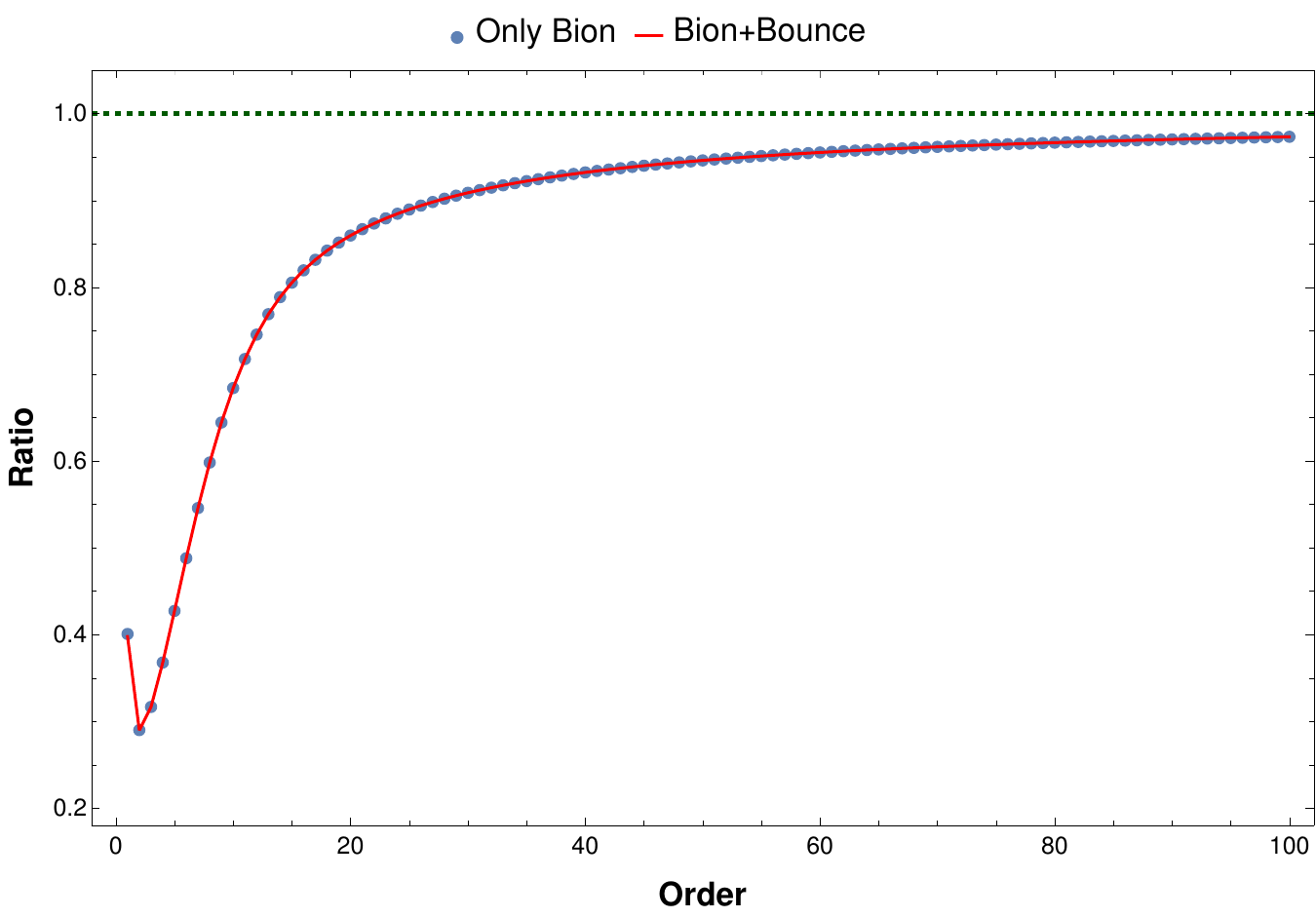}
\end{subfigure}
~\hfill
\begin{subfigure}[h]{0.48\textwidth}
	\caption{\underline{$x_0 = 1.15$}}	\label{Figure: DivergenceBounce}
	\includegraphics[width=\textwidth]{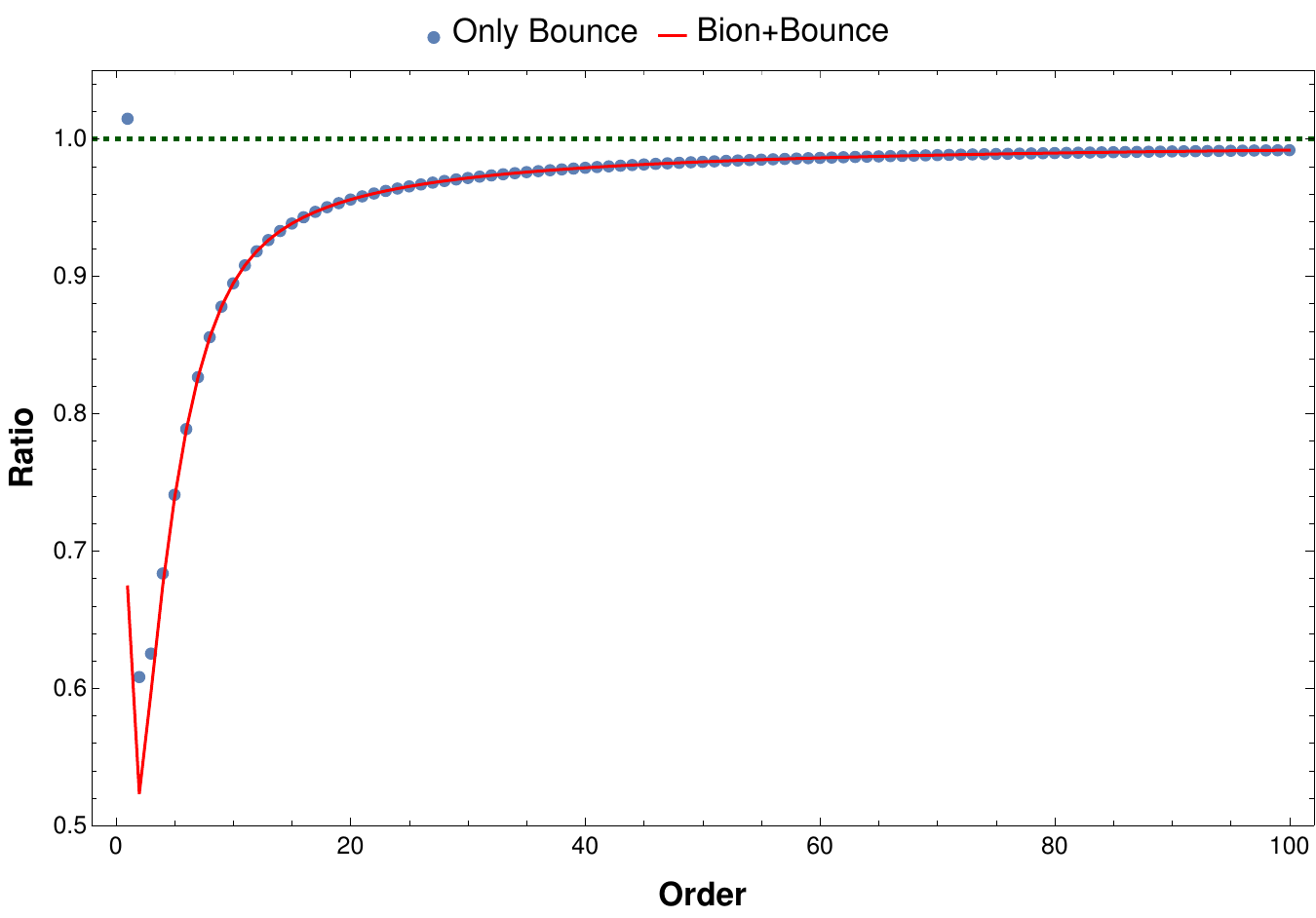}
\end{subfigure}

\vspace{0.0em}
\centering
\begin{subfigure}[h]{0.6\textwidth}
	\caption{\underline{$x_0 = 0.86$}}	\label{Figure: DivergenceBoth}
	\includegraphics[width=\textwidth]{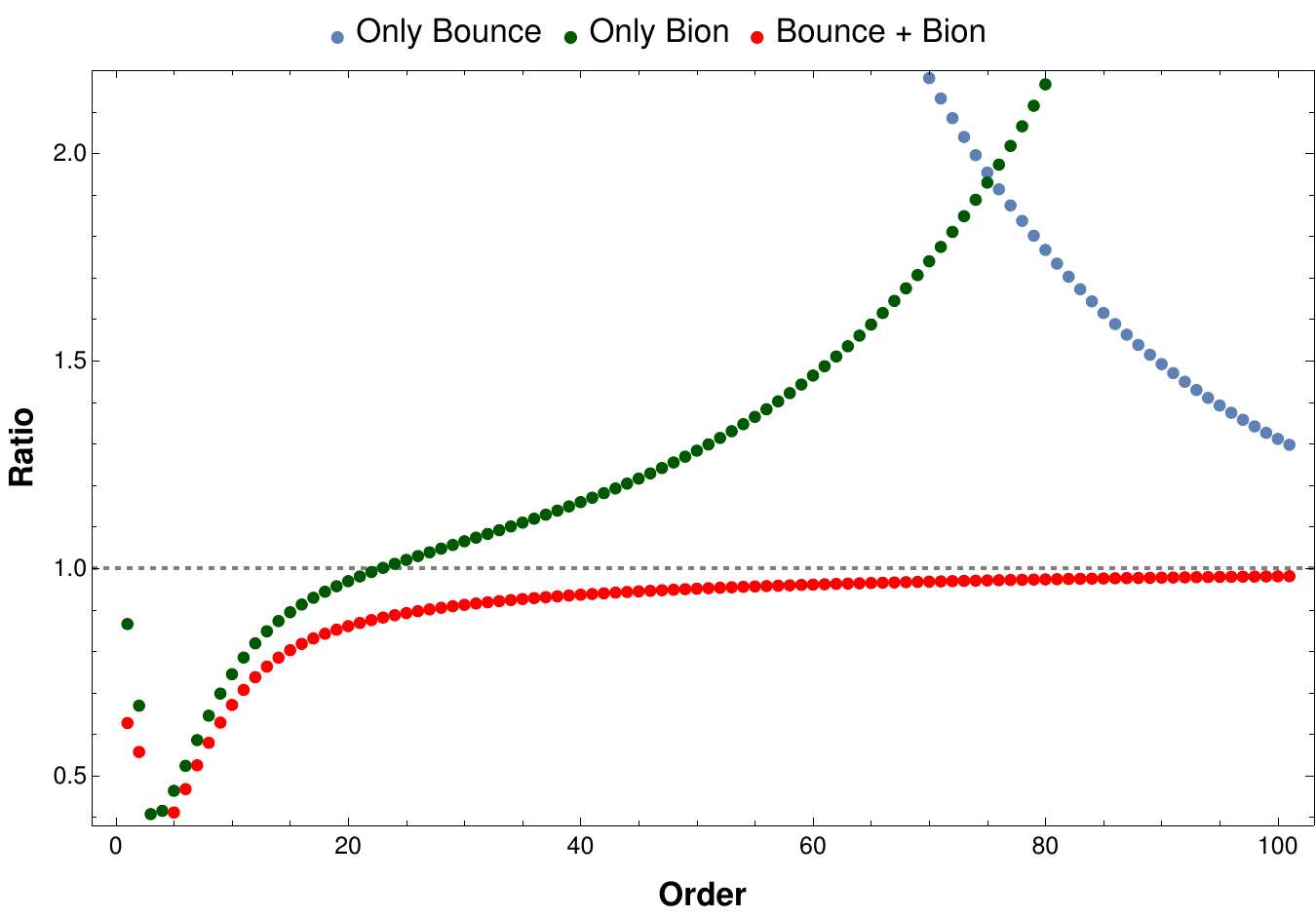}
\end{subfigure}	
\caption{Large-order test of the perturbative coefficients against the semi-classical estimates from the bounce and bion contributions. The ratio plotted compares the exact Bender-Wu coefficients with the asymptotic formulas in three parametric regions. \textbf{(a)} For $x_0=\frac{1}{2}$, the bion configuration dominates. \textbf{(b)} For $x_0=1.15$, the bounce configuration dominates. \textbf{(c)} For $x_0 = 0.86$, no configuration dominates. In \textbf{(a)} and \textbf{(b)}, the dominant bion or bounce contribution is shown by {\color{darkblue} blue dots}, while the combination of both bounce and bion is represented by {\color{red} red line}. In \textbf{(c)}, the ratios involving only bounce ({\color{darkblue} blue dots}) and bion ({\color{ForestGreen}green dots}) do not reproduce the exact result. Instead, the combination of bounce and bion ({\color{red}red dots}) captures the correct large-order behavior.}
\label{Figure: DivergenceCompare}	
\end{figure}

In Fig.~\ref{Figure: DivergenceCompare}, we compare~\eqref{LargeOrder_Bounce} and \eqref{LargeOrder_Bion} with the exact perturbation series, which is obtained using the Bender-Wu package \cite{Sulejmanpasic:2016fwr}, for several values of the parameter $x_0$. We note that in the parameter regions where $[\mcalI\mcalIbar]$ and $[\mcalB]$ events dominate, the behaviors of the exact perturbation series agree with \eqref{LargeOrder_Bion} and \eqref{LargeOrder_Bounce}, respectively. This can be seen in Fig.~\ref{Figure: DivergenceBion} for $x_0=\frac{1}{2}$ where the bion configuration dominates, and Fig.~\ref{Figure: DivergenceBounce} for $x_0 = 1.15$ where the bounce does. When $[\mcalB] \simeq [\mcalI\mcalIbar]$, we observe that both contributions should be taken into account to recover the behavior of the exact perturbation series. This is depicted in Fig.~\ref{Figure: DivergenceBoth}, where we demonstrate that the individual contribution from $[\mcalB]$ or $[\mcalI\mcalIbar]$ does not reproduce the behavior of the exact perturbation series.

Note that despite these apparent differences in its non-perturbative counterpart for different values of $x_0$, the perturbation series is the same in any parameter region. This is simply because the properties of the well of $V_\mrmITW(x)$ in~\eqref{ITW_potential} do not change their characters for $x_0\in (-2,2)$. Then, the differences in Fig.~\ref{Figure: DivergenceCompare} are only quantitative, and the large-order behavior of the exact perturbation series should be controlled by both bion and bounce configurations. In fact, a closer look at the large-order estimations from bounce and bion in $x_0=\frac{1}{2}$ and $x_0 = 1.15$, we observe that the subdominant contribution is numerically negligible, which we show in Fig.~\ref{Figure: DivergenceBion} and Fig.~\ref{Figure: DivergenceBounce} by comparing the dominant contribution with the combined contributions of bion and bounce. 

As a result, we conclude that, in general, the large-order perturbation theory is controlled by both $[\mcalB]$ and $[\mcalI\mcalIbar]$, which also validates our separation of the perturbation series in~\eqref{PerturbativeExpansions_divided} into two parts. Finally, for completeness, let us fix the constants $\d_i$ and $\eta_i$ in~\eqref{PerturbationSeries_general}: By comparing~\eqref{LargeOrder_Bounce} and \eqref{LargeOrder_Bion}, we find $\d_1 = \eta_1 = N -\frac{1}{2}$ and $\d_2 = \eta_2 = 2N$.

Despite their fundamental roles in the semi-classical dynamics, as we mentioned above, the bion and bounce configurations need to be supported by more comprehensive quantization schemes to obtain accurate information about the non-perturbative spectrum. At this point, the EWKB framework comes to our aid. In Section~\ref{Section: Trans-seriesDirect}, we will solve the exact QCs to obtain the trans-series structure (up to a finite order) for PT-symmetric and (anti-)resonance cases independently. In addition to that, with the help of the Alien calculus, the EWKB formalism leads to truly \textit{exact} non-perturbative corrections, which we will discuss in Section~\ref{Section: ExactSolutions}. In the PT-symmetric case, this allows us to find the exact analytical semi-classical condition for the PT-symmetry breaking and to determine the exceptional points. Before addressing these solutions, we now return to the connection problem and describe PT-symmetric and (anti-)resonance problems in view of the EWKB framework.


\subsection{Exact WKB formulation of inverted triple-well}\label{Section: EWKB_Setup_ITW}

\begin{figure}
	\begin{subfigure}[h]{0.48\textwidth}
		\caption{\underline{$\Im \, \hbar > 0$}}	\label{Figure: ITW_Stokes1}
		\includegraphics[width=\textwidth]{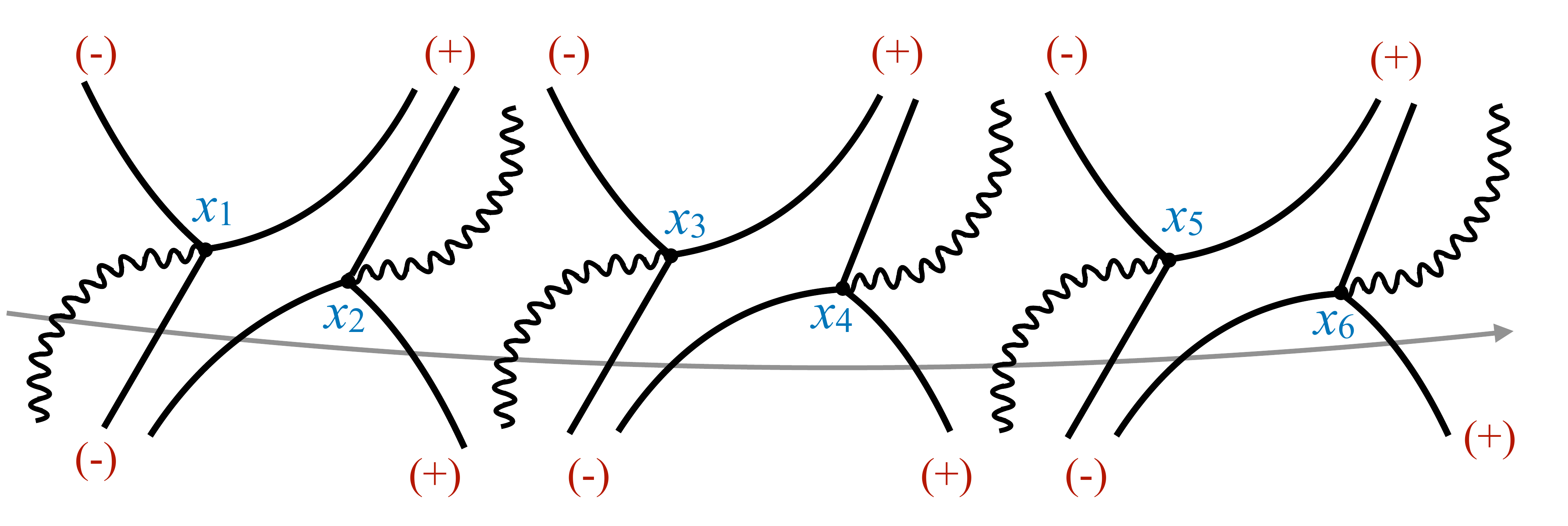}
	\end{subfigure}
	~\hfill
	\begin{subfigure}[h]{0.48\textwidth}
		\caption{\underline{$\Im \, \hbar < 0$}} \label{Figure: ITW_Stokes2}
		\includegraphics[width=\textwidth]{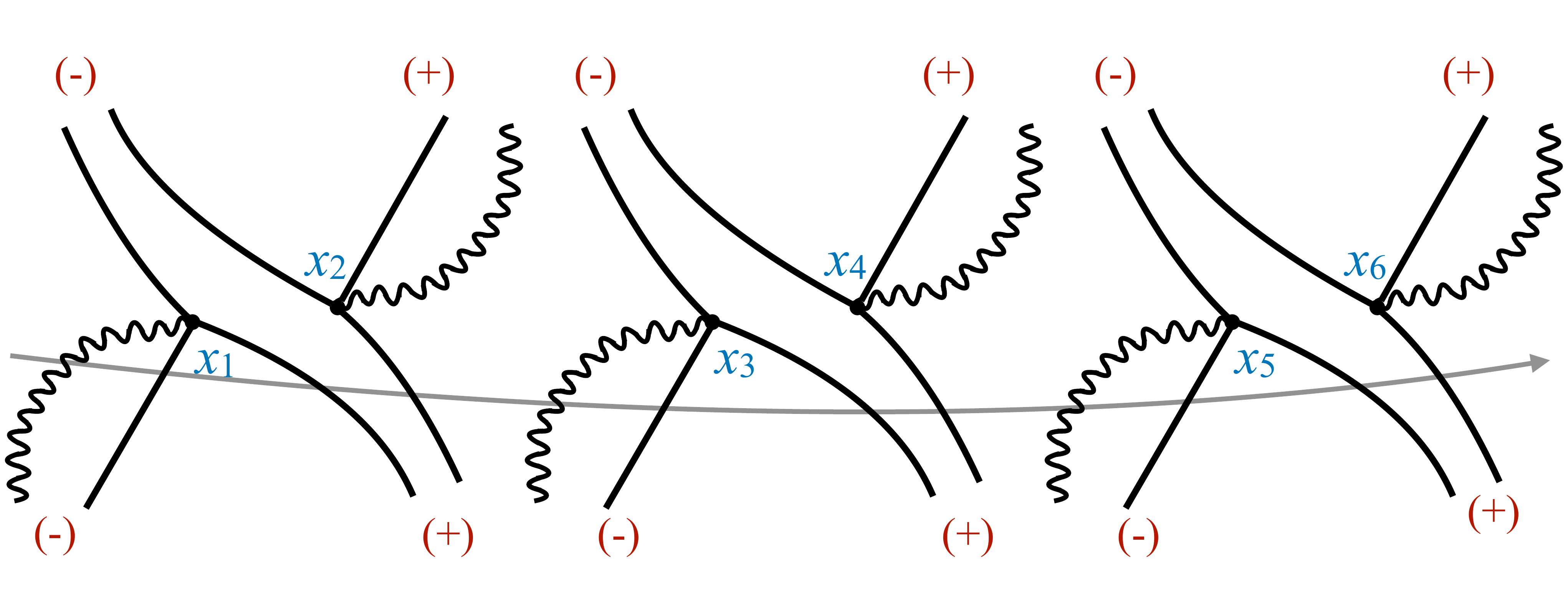}
	\end{subfigure}
	\caption{Analytic continuations of the inverted triple-well Stokes diagram for $\Im\,\hbar=0^\pm$. The two diagrams for $\Im\,\hbar>0$ and $\Im\,\hbar<0$ define the transition matrices $T^{(+)}$ and $T^{(-)}$ connecting the asymptotic solutions between $x=-\infty$ and $+\infty$.}
 \label{Figure: Stokes_ITW}
\end{figure}

We start with $\Im \,\hbar = 0^\pm$ analytic continuations of the Stokes diagrams for the ITW potential, which are depicted in Fig.~\ref{Figure: Stokes_ITW}. For $\Im\, \hbar>0$ and $\Im \,\hbar<0$, we define the transition matrices connecting $x=\pm \infty$, respectively, as
\begin{equation}
	T^{(+)} = \begin{pmatrix}
		T^{(+)}_{1,1}  & T^{(+)}_{1,2} \\
		T^{(+)}_{2,1}  & T^{(+)}_{2,2}
	\end{pmatrix} \, , 
	\quad T^{(-)} = \begin{pmatrix}
		T^{(-)}_{1,1}  & T^{(-)}_{1,2} \\
		T^{(-)}_{2,1}  & T^{(-)}_{2,2} 
	\end{pmatrix} \, . 
\end{equation}
In terms of $T^{(\pm)}$, the connection problem is formulated as
\begin{equation}
	\Psi_\infty^{(\pm)} = T^{(\pm)} \Psi_{-\infty}^{(\pm)}  \, , \label{connectionGeneral}
\end{equation}
where $\Psi^{(\pm)}_{\pm\infty}$ is a vector of right- and left-moving waves at $x=\pm \infty$. Note that even though we take into account the Borel-summed solutions, they are not needed to determine the direction of the corresponding waves in the asymptotic regions. The reason is that the leading order exponential parts, i.e.,
\[\f^{(\pm)}\sim e^{\pm \s(x)}\, , \quad \s(x) = \frac{1}{\hbar}\int_{x_0}^x\mrmd z\, \sqrt{V(x)- E}\, \] are not involved in the divergent expansions in \eqref{Asymptotic_WaveFunction} and they appear as the only exponential prefactors after the Borel summations at $x=\pm\infty$, which leads to an analytic function without any discontinuity\footnote{Even when one of the asymptotic WKB expansion in \eqref{Asymptotic_WaveFunction} is not Borel-summable, the pre-factor $e^{\pm \s(x)}$ would appear in the summed expression. This can be seen from the Laplace integral in \eqref{Borelsummed_solution}, which induces a singularity at $t=\pm\s(x)$.}. Therefore, the direction of the waves is determined by the imaginary sign of the leading order solution as in the conventional WKB method. 

In our setting, on the other hand, the necessary imaginary sign is linked to the branch-cut convention, which we pick as $\sqrt{V(x) - E} = + i \sqrt{E-V(x)}$. Then, we express the solution vectors in \eqref{connectionGeneral} as
\begin{align} 
	\Psi_{\pm \infty} &=\begin{pmatrix}
		\psi^{(\rightarrow )}_{\pm\infty} \\ \psi^{(\leftarrow)}_{\pm \infty} 
	\end{pmatrix} \, , 
\end{align}
where $\left(\substack{\rightarrow \\ \leftarrow}\right)$ represent the direction of the associated waves with the leading order behavior as
\begin{equation}
	\psi^{\left(\substack{\rightarrow \\ \leftarrow}\right)} \sim \exp\left\{\pm \frac{i}{\hbar} \int \mrmd x\, \sqrt{E-V(x)}\right\}\, .
\end{equation}
We also removed the superscripts $(\pm)$ in \eqref{connectionGeneral} to simplify the notation. The following discussion is not affected by analytic continuation depending on the sign of $\Im \hbar$.

Using these identifications, we finally can pick the quantization conditions for the PT-symmetric or (anti-)resonance cases in accordance with the discussion in Section~\ref{Section: Setup_Numerical_ITW}. First, let us express the connection formula \eqref{connectionGeneral} explicitly\footnote{For the sake of simplicity of the notation, we removed the $\pm$ superscripts in the wave functions which are associated to $\Im\hbar>0$ and $\Im\hbar<0$ analytic continuations and does not play a role in the upcoming discussion. We reintroduce the superscripts below, once we set the quantization conditions explicitly.} as
\begin{align}
	\psi_\infty^{(\rightarrow)} &= T_{1,1} \psi_{-\infty}^{(\rightarrow)} + T_{1,2} \psi_{-\infty}^{(\leftarrow)} \, , \label{mixing1}\\ 
	\psi_\infty^{(\leftarrow)} &= T_{2,1} \psi_{-\infty}^{(\rightarrow)} + T_{2,2} \psi_{-\infty}^{(\leftarrow)}\, . \label{mixing2}
\end{align}
Of course, a complete solution is a linear combination of the right- and left-moving ones. For example, at $x=+\infty$, we express it as
\begin{equation}
	\psi_{\infty} = \mathcal{R}_{\infty} \psi^{\left(\rightarrow\right)}_{\infty} + \mcalL_{\infty} \psi^{\left(\leftarrow\right)}_{\infty} \, . \label{linearCombination}
\end{equation}
Then, if we want to look at a right-moving solution, flowing toward $+\infty$, we set $\mcalL_{\infty} =0$. In contrast, setting $\mathcal{R}_{\infty}=0$, we can pick left-moving solutions coming from $+\infty$. 

When we consider the connection problem, $\psi^{(\rightarrow)}_{+\infty}$ and $\psi^{(\leftarrow)}_{+\infty}$ are written in terms of the solutions at $x=-\infty$ through \eqref{mixing1} and \eqref{mixing2}. Then, imposing one of the right-moving or left-moving solutions at $x=-\infty$ requires eliminating the other one from the connection problem. 

For example, if we require right-moving solutions at both $x=\pm\infty$, we can first set $\mcalL_\infty=0$ in \eqref{linearCombination}, which eliminates $\psi^{(\leftarrow)}_\infty$. However, in order to prevent $\psi_{-\infty}^{(\leftarrow)}$ from entering the solution through \eqref{mixing1} as 
\begin{equation}
	\psi_{\infty} = \mathcal{R}_\infty \psi_{\infty}^{(\rightarrow)} = \mathcal{R}_\infty \left[T_{1,1} \psi_{-\infty}^{(\rightarrow)} + T_{1,2} \psi_{-\infty}^{(\leftarrow)}\right]\, ,
\end{equation}
we also need to set $T_{1,2}=0$. This is nothing but the quantization condition, determining the spectrum of the associated quantum system. We note that this procedure is equivalent to the one applied in Hermitian problems and non-Hermitian ones with complex boundary conditions. In those cases, the guiding principle is the exponentially decaying/diverging behaviors of the solutions. In our setting, however, since we impose the boundary conditions on the real line, all the solutions have oscillatory parts, and the directions of waves impose the character of the physical system as discussed in Section~\ref{Section: Setup_Numerical_ITW}.

Finally, following these principles, we set the quantization conditions for PT-symmetric, resonance and anti-resonance systems:
\begin{itemize}
	\item For the PT-symmetric case, the waves are either right-moving $(\rightarrow)$ or left-moving $(\leftarrow)$. Then, to pick $(\rightarrow,\rightarrow)$, we impose $T^{(\pm)}_{1,2} =0$, while for $(\leftarrow,\leftarrow)$ we impose $T^{(\pm)}_{2,1} = 0$.
	\item The resonance wave flows outwards toward real infinities. Then, to pick $(\leftarrow, \rightarrow)$ solutions, we set $T^{(\pm)}_{2,2} = 0$.
	\item Finally, for the anti-resonance case, we set $T^{(\pm)}_{1,1}=0$ which chooses inward wave flow, i.e. $(\rightarrow, \leftarrow)$
\end{itemize}
The exact QCs for all possible cases are summarized in Table~\ref{Table: ExactQuantizationCondition_ITW}. 
 
\begin{table}
	\caption{Exact quantization conditions for the inverted triple-well. } \label{Table: ExactQuantizationCondition_ITW}
	\vspace{-10pt}
	\begin{center} 
	\begin{tblr}{ |c | c | }
		\hline 
		$\bm{\Im \,\hbar >0}$ &   \\
		\hline  
		PT-symmetric 1 & $T^{(+)}_{1,2} = i \O \left\{ (1+A)^2 + B_1 (1+A) + B_2(1+B_1) \right\} = 0$ \\
		\hline  
		PT-symmetric 2 & $T^{(+)}_{2,1} = -i \O \left\{ (1+A)^2 + B_1 (1+A) + B_2(1+B_1) \right\} = 0$ \\
		\hline
		Anti-resonance & $T^{(+)}_{1,1} = \O \left\{(1+A)^2 + B_2  \right\} = 0 $ \\
		\hline
		Resonance & $T^{(+)}_{2,2} = \O \left\{(1+A+B_1)^2 + B_2(1+B_1)^2 \right\}=0$\\
		\hline 
		\hline 
		$\bm{\Im \,\hbar <0}$ &   \\
		\hline  
		PT-symmetric 1 & $T^{(-)}_{1,2} = i \O \left\{(1+A)^2 + A B_1 (1+A) + A^2 B_2 (1+B_1) \right\} =0$ \\
		\hline  
		PT-symmetric 2 & $T^{(-)}_{2,1} = -i \O \left\{(1+A)^2 + A B_1 (1+A) + A^2 B_2 (1+B_1) \right\} =0$ \\
		\hline
		Anti-resonance & $T^{(-)}_{1,1} = \O \left\{(1 + A + A B_1)^2 + A^2 B_2 (1+B_1)^2\right\} = 0$ \\
		\hline
		Resonance & $T^{(-)}_{2,2} = \O \left\{ (1+A)^2 + A^2 B_2 \right\} = 0$ \\
		\hline 
	\end{tblr}
\end{center}
\end{table}

In all equations, we defined 
\begin{align}
    \O = \left(A^{-2}B_1^{-2} B_2^{-1}\right)^{1/2} 
\end{align}
for convenience. In this way, until the QCs are solved to get the spectrum, all the computations are carried out exactly, and the relationship among $T^{(\pm)}_{i,j}$ becomes manifest.

Given all these terms, we first observe that
\begin{equation}
	  \mathsf{C} T^{(\pm)}_{1,2} = T^{(\mp)}_{2,1}\, , \qquad  \mathsf{C} T^{(\pm)}_{1,1} = T^{(\mp)}_{2,2} \, ,\label{ITW_complexConjugate}
\end{equation}
where $\mathsf{C}$ is the complex-conjugation operator. As expected, $\mathsf{C}$ acts as a time-reversal operator, mapping resonance to anti-resonance and also PT-symmetric to PT-symmetric with waves moving in the opposite directions. Note that one may be confused with the roles of the time-reversal operator $\mathsf{C}$ and the Stokes automorphism, as both map the QCs for $\Im \,\hbar>0$ to the ones for $\Im \,\hbar <0$ and vice versa. We will revisit the issue of time-reversal and elaborate its role at the end of this subsection.


Next, we show how the entire connection problem transforms under the Stokes automorphism. We know that the physics for $\Im \,\hbar >0$ and $\Im \,\hbar <0$ are mapped to each other via a Stokes automorphism, which should reveal itself in the connection problem in \eqref{connectionGeneral}. To observe the one-to-one map precisely, in addition to the transition matrices $T^{(\pm)}_{i,j}$, we apply the Stokes automorphism to the solutions $\Psi^{(\pm)}_{\mp\infty}$. Applying the Stokes automorphism onto the transition matrices 
is the standard treatment, which we discussed in Section~\ref{Section: Spectrum_EWKB}. The transformation of $\Psi^{(\pm)}_{\mp\infty}$, on the other hand, is not widely discussed in the literature. In our discussion, we follow the construction in~\cite{Fujimori:2025kkc}. 

The Stokes automorphism for ITW~\eqref{ITW_potential} is given by 
\begin{equation}
	\mfrS^\n_A: A\mapsto A(1+B_1)^{\n}(1+B_2)^\n\, . \label{StokesAuto_ITW}
\end{equation}
For the PT-symmetric parts, the action is straightforward:
\begin{equation}
	\mfrS_A T_{1,2}^{(+)} = T_{1,2}^{(-)} \, , \quad \quad \mfrS_A T^{(+)}_{2,1} = T^{(-)}_{2,1}\, . \label{StokesAuto_ITW_PT}
\end{equation}
For resonance and anti-resonance cases, on the other hand, we have
\begin{align}
	\mfrS_A T^{(+)}_{1,1} &= \O (1+ B_1)^{-1} \left\{(1+A)^2 + A^2 B_1^2(1+B_2) + 2 A B_1 (1+ A + A B_2) \right\} \nonumber \\
	&= (1+ B_1)^{-1} T^{(-)}_{1,1}\, , \label{StokesAuto_ITW_Res}
\end{align}
and 
\begin{align}
	\mfrS_A T^{(+)}_{2,2} &= \O (1 + B_1) \left\{(1+A)^2 + A^2 B_2\right\} \nonumber \\
	& = (1+B_1) T^{(-)}_{2,2} \, . \label{StokesAuto_ITW_AntiRes}
\end{align}
The remaining factors are cancelled, once the Stokes automorphism is applied to the vector of the WKB solutions (wave-functions), which are given as
\begin{equation}
	\Psi^{(+)}_\infty = \S\Psi^{(-)}_\infty \, , \qquad \Psi^{(+)}_{-\infty} = \S^{-1}\Psi^{(-)}_{-\infty} \, ,
\end{equation}
where 
\begin{equation}\label{StokesAutomorphism_WaveFunction}
	\S = \begin{pmatrix}
		(1+B_1)^{-1/2} & 0\\
		0  & (1+ B_1)^{1/2}
	\end{pmatrix} \, .
\end{equation}
Then, in view of~\eqref{StokesAuto_ITW_PT}-\eqref{StokesAuto_ITW_AntiRes} and \eqref{StokesAutomorphism_WaveFunction}, the entire connection problem transforms under the Stokes automorphism as 
\begin{equation}\label{StokesAuto_Combined}
	\Psi^{(+)}_\infty = T_+ \Psi_{-\infty}^{(+)} \; \xlongrightarrow{\tilde{\mfrS}} \;  \S\Psi^{(-)}_\infty = \begin{pmatrix}
		(1+B_1)^{-1}T^{(-)}_{1,1}  & T^{(-)}_{1,2}\\
		T^{(-)}_{2,1}  & (1+B_1) T^{(-)}_{2,2}
	\end{pmatrix} \S^{-1} \Psi_{-\infty}^{(-)}\, ,
\end{equation}
where the right-hand side can be arranged as
\begin{equation}
	\Psi^{(-)}_\infty = T_- \Psi^{(-)}_{-\infty}\, .
\end{equation}
Having the entire system being mapped between $\Im \,\hbar = 0^\pm $ cases, we summarize our findings and make the following observations: 
\begin{itemize}[wide]
	
	\item The two analytic continuations of the connection problem are related to each other exactly via Stokes automorphism, which we schematically represent as
	\begin{equation}
		\tilde{\mfrS}: 	\Psi^{(+)}_\infty = T_+ \Psi_{-\infty}^{(+)}\mapsto 	\Psi^{(-)}_\infty = T_- \Psi_{-\infty}^{(-)}\,
	\end{equation}
	where $\tilde{\mfrS}$ is a combination of both Stokes automorphisms in \eqref{StokesAuto_ITW} and \eqref{StokesAutomorphism_WaveFunction} and $\pm$ superscripts refer to $\sgn\left(\Im\hbar\right) = \pm 1$. 
	
	The individual elements of the connection matrices, on the other hand, are related via time-inversion, which is represented by the complex conjugation operator in \eqref{ITW_complexConjugate}. If we consider the wave directions at $x=\pm \infty$ for each case, we observe that the complex conjugation reverses their directions. This makes the time reversal property explicit. 	Schematically, the reversal of the wave directions in \eqref{ITW_complexConjugate} can be summarized as follows
	\begin{align}
		\mathrm{PT\text{-}symmetric}:\qquad  &\mathsf{C}T_{1,2}^{(\pm)} = T_{2,1}^{(\pm)} \quad  \Longrightarrow  \mcalT: \left(\rightarrow,\rightarrow\right) \mapsto \left(\leftarrow, \leftarrow\right) \\
		\mathrm{(Anti\text{-})Resonance}:\qquad  &\mathsf{C}T_{1,1}^{(\pm)} = T_{2,2}^{(\pm)} \quad  \Longrightarrow  \mcalT: \left(\rightarrow,\leftarrow\right) \mapsto \left(\leftarrow, \rightarrow\right) \, , 
	\end{align}
	where $\mcalT$ is the time-reversal operator. 
	\item A complex conjugation and a Stokes automorphism are equivalent only if there is time-reversal symmetry at the quantum level. Note that this is always the case for Hermitian systems. However, the equivalence is not generalized to the resonance and anti-resonance cases, where the time-reversal symmetry is broken. We expect that this observation also holds for more general potentials as well. 
    
    We also note that recently, in the case of IDW~\cite{Kamata:2023opn}, the PT-symmetric case was linked to the resonance system by a one-parameter Stokes automorphism (more precisely, the half-Stokes automorphism $\mfrS^{-1/2}$). It was shown that this link can be understood as a new formulation of the ABS conjecture~\cite{Ai:2022csx}. Our analysis above, however, shows that this relationship does not generalize to the ITW system. 
    
	\item Recalling the relations in~\eqref{ITW_complexConjugate}, the forms of the exact QCs for the resonance and anti-resonance cases are exchanged by the Stokes automorphism up to a complex conjugation. Note that the complex conjugation has a crucial role here, as it compensates the sign difference which is introduced by $\Im \hbar >0$ and $\Im \hbar <0$ analytic continuations, and ensures the signs of the imaginary parts of the resonance spectrum staying intact, i.e., $\Im \, E <0$.
	Same arguments apply to the anti-resonance solutions with $\Im\, E >0$ for both $\Im \,\hbar = 0^\pm$. Note that this is in complete agreement with the previous point that the Stokes automorphism and complex conjugation (time reversal) are different operations. 
		
	\item Fixing boundary conditions is basically equivalent to choosing the direction of the time evolution. This becomes evident in the spectra of resonance and anti-resonance cases via the imaginary signs of their respective eigenvalues. In the PT-symmetric case, however, the direction of time is lost due to the time-reversal symmetry; but as we showed in~\eqref{ITW_complexConjugate}, the PT-symmetric choices are time-reversed partners of each other. 
	
	Then, the only role of the Stokes automorphism becomes mapping the analytic continuation around $\Im\, \hbar =0$, which is its original purpose. This is, in fact, crucial for the completeness of our discussion, since the physical quantities lie on the Stokes ray corresponding to $\Im\, \hbar = 0$, which is reached by the action of half-Stokes automorphism, i.e., $\mfrS^{\pm 1/2}_A$, and since the time direction should not be affected by the direction of the $\Im\, \hbar \rightarrow 0^\pm$ limits, the Stokes automorphism must not be related to the time-reversal.
\end{itemize}

\section{Trans-series from exact WKB}\label{Section: Trans-seriesDirect}

In this section, we will solve the exact quantization conditions in Table~\ref{Table: ExactQuantizationCondition_ITW} and obtain the trans-series structures of the spectra for PT-symmetric, resonance, and anti-resonance cases. For each case, we will demonstrate how the resurgence cancellations occur and lead to the spectra that are consistent with the physical properties of the corresponding system. Although our analysis in this section is based on the EWKB formalism, we will link it to the semi-classical quantization which we introduced in Section~\ref{Section: semi-classical_ITW}. In this way, we will pursue our discussion in close connection with the path integral formalism, revealing the similarities and differences in the semi-classical trans-series and the roles of bions and bounces, in each system for different analytic continuations of the $\hbar$ parameter.

\subsection{PT-symmetric case}\label{Section: PT-symmetric_Direct}

In the PT-symmetric case, the QCs $T_{1,2}^{\pm}=-T_{2,1}^{(\pm)} = 0$ in Table~\ref{Table: ExactQuantizationCondition_ITW} can be expressed on an equal footing as  quadratic algebraic equations in $1+\Pi_A^{\pm 1}$:
\begin{equation}
	\left(1+ \Pi_A^{\pm 1}\right)^2 + \Pi_{B_1}\left(1+\Pi_A^{\pm 1}\right) + \Pi_{B_2}\left(1+ \Pi_{B_1}\right) = 0\, , \qquad \Im \hbar =0^\pm \, .
\end{equation}
Solving this equation for $1+\Pi_A^{\pm 1}$, we isolate the perturbative part and obtain
\begin{equation}
	1+\Pi_{A}^{\pm 1} = -\frac{1}{2}\left[\Pi_{B_1} + \ve \sqrt{\Pi_{B_1}^2 - 4 \Pi_{B_2}\left(1+\Pi_{B_1}\right)} \right]\, ,\label{EQC_1}
\end{equation} 
and later using the Weber-type expression $\Pi_A^{\pm 1} = e^{\mp 2\pi i\mcalF}$ via the dictionary we introduced in Section~\ref{Section: BriefIntro_EWKB}, we re-write \eqref{EQC_1} as
\begin{align}
	2 e^{\mp i \pi \mcalF} \cos \pi \mcalF = -\frac{\Pi_{B_1}}{2} + \ve \Big[\Disc_\mrmPT\left(\Pi_{B_1}\Pi_{B_2}\right)\Big]^{1/2}\, . \label{EQC_PT_both}
\end{align}
In \eqref{EQC_PT_both}, we define $\ve = \pm 1$, which signals a non-perturbative splitting in the spectrum. Moreover, we re-scaled the discriminant of the quadratic equations and defined 
\begin{equation}
	\Disc_\mrmPT = \frac{\Pi_{B_1}^2}{4} - \Pi_{B_2}\left(1+\Pi_{B_1}\right)\, .  \label{discriminantPT}
\end{equation} 
The reason behind this rescaling will be clear when we specify the terms associated with the bounce and bion contributions to the quantized system.

In Section~\ref{Section: Spectrum_EWKB}, we discussed the construction of the trans-series solution around a perturbative saddle. Accordingly, we define
\begin{equation}
	\mcalF = N+\frac{1}{2} + \d_\mrmPT \, . \label{NPcorrections_definition_PT}
\end{equation}
Then, we can solve \eqref{EQC_PT_both} for $\d_\mrmPT$ after expanding it for $\d_\mrmPT \ll 1$. In our case, this approach is applicable in the below-barrier-top region where both $\Pi_{B_1}$ and $\Pi_{B_2}$ are exponentially suppressed\footnote{More precisely, $\Pi_{B_1}(\mcalF,\hbar)$ and $\Pi_{B_2}(\mcalF,\hbar)$ are considered as series expansions in $\hbar \ll 1$ and their exponentially suppressed characters gradually disappear while approaching to barrier top region as they begin to be order $O(\hbar^0)$ rather than $O(e^{-1/\hbar})$.~This means that the (multi-)bounce or (multi-)bion configurations are no more suppressed, indicating more and more terms should be taken into account for good approximations as one approaches to the barrier top region. This continuous quantitative change, however, does not play an important role in our general discussion except when we discuss the barrier-top region in the PT-symmetric case in Section~\ref{Section: PTsymmetric_median}. Therefore, we keep treating $\Pi_{B_{1,2}}$ as exponentially suppressed, unless it is stated otherwise.}. 

Note that the quantitative relationship between $\Pi_{B_1}$ and $\Pi_{B_2}$ depends on the size of the corresponding barriers and their quantitative dominance changes as the parameter $x_0$ is altered. 
In the following, we will examine~\eqref{EQC_PT_both} in two asymptotic regimes corresponding to $\Pi_{B_1}^2 \ll \Pi_{B_2}$ and $\Pi_{B_2}\ll \Pi_{B_1}^2$. By uncovering the corresponding trans-series, we will show that these limits correspond to the unbroken and broken phases of the PT-symmetric theory, which was shown numerically in Fig.~\ref{Figure: PTsymmetric_Eigenvalues}.  

The relationship between $\Pi_{B_1}^2$ and $\Pi_{B_2}$ becomes important when one deals with the square-root term in~\eqref{EQC_PT_both}.~It is straightforward to see that assuming $\Pi_{B_1}^2 \ll \Pi_{B_2}$ or $\Pi_{B_2} \ll \Pi_{B_1}^2$ leads to two distinct expansions, indicating different physical features in these asymptotic limits.~Expanding directly in terms of $\Pi_{B_1}$ and $\Pi_{B_2}$, however, could be tricky and possibly misleading, particularly when the higher order non-perturbative corrections to the trans-series is considered. This is because the actual small parameter $\d_\mrmPT$ appears also in $\Pi_{B_1}$ and $\Pi_{B_2}$, and a second expansion in $\d_\mrmPT$ would be needed if the square-root term is expanded directly for $\Pi_{B_1}^2 \ll \Pi_{B_2}$ or $\Pi_{B_2} \ll \Pi_{B_1}^2$.

Instead, we utilize the EWKB dictionary in~\eqref{Dictionary_Perturbative} and \eqref{Dictionary_NonPerturbative} directly and re-arrange~\eqref{EQC_PT_both} as
\begin{align}
	\frac{1}{\pi}\cos\left(\pi \mcalF\right) = -\tPi_{B_1} - \ve\left[\tPi_{B_1}^2 - \tPi_{B_2}^2 \left(1+\Pi_{B_1}\right)\right]^{1/2} \, , \label{EQC_PT_rearranged}
\end{align}
where 
we re-defined the actions on the right-hand side as
\begin{align}
	\tPi_{B_1} &=  \frac{e^{\pm i\pi \mcalF}}{2\cdot 2\pi}\Pi_{B_1} = \, \frac{e^{-\mcalG^{(1)}}}{2\sqrt{2\pi}}\, \frac{e^{\pm i \pi \mcalF}}{\G\left(\frac{1}{2} + \mcalF\right)}\, \left(\frac{\mcalC}{\hbar}\right)^\mcalF \, , \label{B1_redefined} \\
	\tPi_{B_2} & = \frac{e^{\pm i \pi \mcalF}}{2\pi} \Pi^{1/2}_{B_2}= \, \frac{e^{-\frac{1}{2 }\mcalG^{(2)}}}{\sqrt{2\pi}} \,\frac{ e^{\pm i \pi \mcalF}}{\G\left(\frac{1}{2} + \mcalF \right)}\,  \left(\frac{\mcalC}{\hbar}\right)^\mcalF\, . \label{B2_redefined}
\end{align}
The main reason behind this re-definition is being able to treat the contributions from the $B_1$- and $B_2$-cycles on an equal footing in~\eqref{EQC_PT_rearranged}. In order to see its benefit clearly, let us express $\mcalF = N+\frac{1}{2} + \d_\mrmPT$ and expand $\tPi_{B_{1,2}}$ and $\Pi_{B_1}$ in $\d_\mrmPT \ll 1$. The corresponding series-expansions are
\begin{align}
	\tPi_{B_1} &= \pm i (-1)^N \mcalK_1 \sum_{k=0}^\infty H_k^{(1)} \d_\mrmPT^k \, , \label{b1_fugacity} \\ 
	\tPi_{B_2} &= \pm i (-1)^N \mcalK_2 \sum_{k=0}^\infty H_k^{(2)} \d_\mrmPT^k \, , \label{b2_fugacity}\\
	\Pi_{B_1} & = 4 \pi \mcalK_1 \sum_{k=0}^\infty P_k^{(1)} \d_\mrmPT^k\, ,\label{b1_fugacity2}
\end{align}
where
\begin{equation}
	\mcalK_1 = \frac{e^{-\frac{S_1}{\hbar}}}{2\sqrt{2\pi}\, N!} \left(\frac{\mcalC}{\hbar}\right)^{N+\frac{1}{2}}\, , \qquad \mcalK_2= \frac{e^{-\frac{S_2}{2\hbar}}}{\sqrt{2\pi}\, N!} \left(\frac{\mcalC}{\hbar}\right)^{N+\frac{1}{2}}\, , \label{ITW_fugacities}
\end{equation}
and the expressions $S_{1,2}$ and $\mcalC$ are written in~\eqref{BounceAction_ITW}, \eqref{BionAction_ITW} and \eqref{One_LoopDeterminant}. Note that we also recover the leading order approximation of bounce and half-bion (i.e., instanton or anti-instanton) events in \eqref{Bounce_LeadingOrder} and \eqref{Instanton_LeadingOrder} with $\mcalK_{1,2}$ being fugacities in respective dilute gas systems.

Another important addition to the original definitions of $\Pi_{B_{1,2}}$ in \eqref{Dictionary_NonPerturbative} is the phase $e^{\pm i \pi \mcalF} = \pm i (-1)^N e^{\pm i \pi \d_\mrmPT}$. The expansion of the non-perturbative phase $e^{\pm i \pi \d_{\mrmPT}}$ in $\d_\mrmPT \ll 1$ produces imaginary contributions at each term $O\left(\d_\mrmPT^k\right)$ for $k\geq 1$, indicating that $H_{k}^{(1,2)}$ have imaginary parts in addition to the real ones, i.e., $P_k^{(1,2)}$, except for $H_0^{(1,2)}$. To see the explicit connection to the original real expansion of $\Pi_{B_{1,2}}$, let us consider the series expansion \[e^{\pm i \pi \d_\mrmPT} = \sum_{n=0}^\infty \frac{(\pm i \pi)^n}{n!}\d_\mrmPT^n\] separately and express each term in \eqref{b1_fugacity} and \eqref{b2_fugacity} as
\begin{equation}
	H_k^{(1,2)} = \sum_{n=0}^k \frac{(\pm i \pi)^n}{n!} P_{k-n}^{(1,2)} \, . \label{separate_expansion}
\end{equation}
Later, this relation helps us to identify the ambiguous and non-ambiguous non-perturbative corrections to the spectrum.

With the rearrangements that we discussed above, the link to the semi-classical quantization in the path-integral formalism becomes manifest. 
More importantly, using~\eqref{b1_fugacity}-\eqref{b1_fugacity2}, we can now understand the competition between the $B_1$- and $B_2$-cycles in terms of $\mcalK_1$ and $\mcalK_2$, rather than the formal and intricate quantities $\Pi_{B_1}$ and $\Pi_{B_2}$. This is possible, since the fugacities $\mcalK_1$ and $\mcalK_2$ are same order to each other and also to the leading-order non-perturbative correction $\d_\mrmPT$. Then, the series in~\eqref{b1_fugacity}-\eqref{b1_fugacity2} are suppressed with respect to $\mcalK_1$ and $\mcalK_2$, and the competition between bion and bounce is now reduced to a simple quantitative relationship between $\mcalK_1$ and $\mcalK_2$.

We can now expand~\eqref{EQC_PT_rearranged} in $\d_\mrmPT \ll 1$. 
For example, the second-order expansion of the right-hand side reads
\begin{align}
	&-\tPi_{B_1} - \ve\left[\tPi_{B_1}^2 - \tPi_{B_2}^2 \left(1+\Pi_{B_1}\right)\right]^{1/2}  \nonumber\\
    & = -\mcalK_1 H^{(1)}_0 -  \ve \sqrt{\mcalK_1^2 \left(H^{(1)}_0\right)^2-\mcalK_2^2 \left(H^{(2)}_0\right)^2- 4\pi \mcalK_1 \mcalK_2^2 P^{(1)}_0 \left(H^{(2)}_0\right)^2 } \nonumber \\
	& -  \d_\mrmPT \left(\mcalK_1 H^{(1)}_1 + \ve \frac{ \mcalK_1^2 H^{(1)}_0 H^{(1)}_1-  \mcalK_2^2 H^{(2)}_0 H^{(2)}_1 -  4\pi \mcalK_1 \mcalK_2^2\left( H^{(2)}_0 H^{(2)}_1	P^{(1)}_0+  \frac{1}{2} P^{(1)}_1 \left(H^{(2)}_0\right)^2  \right)}{\sqrt{\mcalK_1^2 \left(H^{(1)}_0\right)^2-\mcalK_2^2
			\left(H^{(2)}_0\right)^2- 4\pi \mcalK_1 \mcalK_2^2 P^{(1)}_0\left(H^{(2)}_0\right)^2 }}\right) \, . \label{expansionRHS}
\end{align}
Note that the square-root factors are independent of $\d_\mrmPT$. Therefore, it is finally safe to choose a dominant non-perturbative saddle and expand the square-root terms for $\mcalK_1 \ll \mcalK_2$ or $\mcalK_2 \ll \mcalK_1$. 

On the left-hand side of~\eqref{EQC_PT_rearranged}, the expansion of the $\G$ functions leads to another expansion in $\d_\mrmPT$:
\begin{equation}
	-\frac{1}{\pi}\cos\left(\pi \mcalF\right) = \frac{(-1)^N}{\pi} \sin\left(\pi \d_\mrmPT \right)  = (-1)^N\, \sum_{k=1}^{\infty} b_k \d_\mrmPT^k\, , 
\end{equation}
where $b_1=1$, $b_2=0$, and so on. Then, finally, solving the QC~\eqref{EQC_PT_rearranged} order by order in $\d_\mrmPT$, we can obtain the trans-series solutions to the non-perturbative corrections in the asymptotic limits 
up to $O\left(\mcalK_1^2\right)$ and $O(\mcalK_2^2)$:


\begin{itemize}[wide] 
\item \underline{$\mcalK_1 \ll \mcalK_2$}:  We first consider the case when the half-bion is the dominant configuration. In this limit, we found the solution to~\eqref{EQC_PT_rearranged} as
\begin{align}
	\d_\mrmPT^\mrmUBr & \simeq  -\ve \mcalK_2 H^{(2)}_0 + \ve \frac{\mcalK_1^2 \left(H^{(1)}_0\right)^2}{2 \mcalK_2 H^{(2)}_0}  \mp  i \mcalK_1 H^{(1)}_0 + \mcalK_2^2  H^{(2)}_0 H^{(2)}_1 \nonumber \\ 
	&  - 2 \mcalK_1^2   H^{(1)}_0 H^{(1)}_1  +\ve  \mcalK_1 \mcalK_2 \left(  \pm i H^{(1)}_1 H^{(2)}_0  \pm i H^{(1)}_0 H^{(2)}_1  - 2\pi  P^{(1)}_0 H^{(2)}_0\right) \, , \label{PT_NP_Trans-series_Unbroken}
\end{align}
where we kept $H_k^{(1,2)}$ to show they are originating from $\tPi_{B_{1,2}}$ in~\eqref{EQC_PT_rearranged}. In this sense, we observe that the action $\Pi_{B_1}$ in~\eqref{EQC_PT_rearranged} contributes only to the last term of $O(\mcalK_1\mcalK_2)$ in~\eqref{PT_NP_Trans-series_Unbroken}. We will elaborate its role shortly.

If we use~\eqref{separate_expansion} to reveal the ambiguous parts in~\eqref{PT_NP_Trans-series_Unbroken}, it becomes
\begin{align}
	\d_\mrmPT^\mrmUBr &\simeq -\ve \mcalK_2 P_0^{(2)} + \ve\frac{\mcalK_1^2}{\mcalK_2}\,\frac{\left(P_0^{(1)}\right)^2}{2 P_0^{(2)}} \mp i \mcalK_1 P_0^{(1)} + \mcalK_2^2  \left(P_0^{(2)} P_1^{(2)} \mp i\pi \left(P_0^{(2)}\right)^2 \right) \nonumber \\ 
	&  - 2 \mcalK_1^2 \left(P_0^{(1)} P_1^{(1)} \mp i \pi \left(P_0^{(1)}\right)^2 \right) \pm  \ve i \mcalK_1\mcalK_2 \left(P_0^{(1)} P_1^{(2)} + P_1^{(1)} P_0^{(2)} \right) \, . \label{PT_NP_Trans-series_Unbroken2} 
\end{align}
Then, we observe that the leading order contribution is a single instanton effect, i.e., $\left[\mcalI\right]$ or $\left[\mcalIbar\right]$. As expected, it leads to the level splitting of the perturbatively degenerate states and is in accordance with the standard picture in Hermitian quantum theories.

The one-instanton effect is accompanied by $O(\mcalK_1^2 \mcalK_2^{-1})$ term, which also comes with $\ve = \pm 1$ factor but with the opposite sign. Then, we express the first non-perturbative corrections in the trans-series as
\begin{equation}
	\mathrm{LO:} \quad [\mcalI] + \left[\mcalB^{2} \mcalI^{-1}\right] = [\mcalIbar] + [\mcalB^{2} \mcalIbar^{-1}] =   \ve \left(-\mcalK_2 H_0^{(2)} + \frac{\mcalK_1^2 \left(H_0^{(1)}\right)^2}{2 \mcalK_2 H_0^{(2)}}\right) \, ,  \label{LeadingOrder_PT_unbroken}
\end{equation}
Note that since we assumed $\mcalK_1 \ll \mcalK_2$, the effect of $\left[\mcalI\right]$ (or $\left[\mcalIbar\right]$) is never canceled by $\left[\mcalB^2 \mcalI^{-1}\right]$ (or $[\mcalB^2 \mcalIbar^{-1}]$). In fact, the dominance between these two terms interchanges when $\Pi_{B_1}^2 > \Pi_{B_2}$, which, as we will see shortly, leads to a different expansion due to the square-root term in~\eqref{EQC_PT_both}.

\paragraph{\underline{Remark}:} We also note that we defined $\left[\mcalB^2 \mcalI^{-1}\right]\sim \frac{\mcalK_1^2}{\mcalK_2}$ by extending the notation introduced in Section~\ref{Section: semi-classical_ITW} beyond the single bounce and single bion regime. From the semi-classical point of view, $\frac{\mcalK_1^2}{\mcalK_2}$ is a new type of combination between what we call bounce and half-bion fugacities. 
As far as we know, there is no previous discussion on the origin of such contributions in the literature, where the dilute instanton gas connection was mostly discussed in Hermitian settings. In our analysis, it simply stems from the square-root of the discriminant in~\eqref{EQC_PT_both}, which itself is linked to the quadratic nature of the QCs $T_{1,2}^{(\pm)} = 0$ in Table~\ref{Table: ExactQuantizationCondition_ITW}. Despite the lack of a precise connection to the dilute instanton gas picture in the path-integral formalism, we continue using the extended notation of bion and bounce configurations for the brevity of the discussion.


The first ambiguous imaginary contribution in~\eqref{PT_NP_Trans-series_Unbroken2} arises at orders $O(\mcalK^2_2)$ and $O(\mcalK_1)$ as
\begin{align}
	 \Im[\mcalB]_\pm + \Im [\mcalI\mcalIbar]_\pm &= \mp \left( \mcalK_1 P_0^{(1)} + \pi \mcalK_2^2 \left(P_0^{(2)}\right)^2 \right) \label{ambiguity_PT_unbroken}\\ 
	& \simeq  \mp \frac{e^{-\frac{S_1}{\hbar}}}{2\sqrt{2\pi}\, N!} \left(\frac{\mcalC}{\hbar}\right)^{N+\frac{1}{2}} \mp   \frac{e^{-\frac{S_2}{\hbar}}}{2 \left(N!\right)^2} \left(\frac{\mcalC}{\hbar}\right)^{2N+1} + O(\hbar) \, , \label{leadingAmbiguity_PT_unbroken}
\end{align}
where we used the fact that $P_0^{(l=1,2)} = 1+ O(\hbar)$ to write the final expression. Note that~\eqref{ambiguity_PT_unbroken} is the sum of the exact contributions from one-bounce and one-bion events. Hence, we used the shorthand representation with the equality sign. The expression in~\eqref{leadingAmbiguity_PT_unbroken}, on the other hand, recovers the leading-order imaginary contributions from one-bounce and one-bion events, which we discussed in Section~\ref{Section: semi-classical_ITW} [see~\eqref{Bounce_LeadingOrder} and \eqref{Bion_LeadingOrder}]. As we verified numerically, $[\mcalB]_\pm + \Im[\mcalI\mcalIbar]_\pm$ is canceled against the Borel ambiguity of the perturbative sector. This shows the PT-symmetry is unbroken in $\mcalK_1 \ll \mcalK_2$ region at least at the leading order.

The remaining terms in~\eqref{PT_NP_Trans-series_Unbroken2} are higher-order non-perturbative corrections. Unfortunately, the quantitative analysis showing the reality of the spectrum is quite cumbersome; therefore, it will not be pursued here. Of course, even with the best possible numerical approach never proves the reality of the spectrum rigorously. This exactness problem will be addressed in Section~\ref{Section: ExactSolutions}, when we consider the median QC in view of the Alien calculus, which will also help us to uncover the resurgence structure more precisely.

Despite the lack of quantitative information, let us finally comment on the roles of remaining terms in~\eqref{PT_NP_Trans-series_Unbroken2}.~The real parts are simply the higher-order corrections contributing to the spectrum directly. They are all products of two successive events; namely, they are two-bounce and one-bion (i.e., one instanton plus one anti-instanton) events and appear in the next-to-leading (NLO) order non-perturbative corrections to the spectrum. Schematically, their contribution can be represented as
\begin{align}
	\mathrm{NLO}&: \quad \Re \left[\mcalB^2\right] + \Re\left[\mcalI\mcalIbar \right] \, . \label{TwoEvents_real_PT} 
\end{align}

The imaginary ambiguous terms, on the other hand, are associated with a two-bounce event and the successive events of one-bounce and a half bion, i.e.,
\begin{equation}
	\Im \left[\mcalB^2\right]_\pm + \left[\mcalB \mcalI\right]_\pm \qquad \mathrm{or} \qquad \Im \left[\mcalB^2\right]_\pm + \left[\mcalB \mcalIbar\right]_\pm\, .
\end{equation}
The contribution from $\left[\mcalB\mcalI\right]_\pm \sim \left[\mcalB\mcalIbar\right]_\pm \sim \mcalK_1 \mcalK_2$ should be linked to a Borel ambiguity of a divergent series. We anticipate this series to be the fluctuations around $\left[\mcalI\right]\sim \left[\mcalIbar\right]$, which govern the leading-order non-perturbative splitting. However, it is possible that there exist higher-order terms participating in the resurgence cancellations in a similar way that we observed for the cancellation between the perturbative sector and $[\mcalB]_\pm$ and $[\mcalI\mcalIbar]_\pm$ events. We do not examine this cancellation qualitatively here but, instead, verify the removal of $\left[\mcalB\mcalI\right]$ (and $\left[\mcalB\mcalIbar\right]$) from the median-summed series in Section~\ref{Section: ExactSolutions}.

Finally, let us comment on the other ambiguous term, which is a two-bounce effect, i.e., $\left[\mcalB^2\right]$. This contribution is not directly related to the cancellation of a real divergent series at a lesser non-perturbative order. Instead, it turns out that $\left[\mcalB^2\right]$ is related to a part of the entire trans-series, which is linked to the original perturbative expansion. This part is also called as \textit{minimal} trans-series in the literature~\cite{vanSpaendonck:2023znn}. In~\eqref{PT_NP_Trans-series_Unbroken2}, it reads as
\begin{equation}
	\d_\mrmNP^\mathrm{minimal} \simeq \mp i\left( \mcalK_1 P_0^{(1)} + \pi\mcalK_2^2 \left(P_0^{(2)}\right)^2 - 2\pi \mcalK_1^2 \left(P_0^{(1)}\right)^2 \right) \label{PT_minimalTransseries}
\end{equation} 
For our discussion, it is important that the minimal series appears across all parameter regimes of the PT-symmetric and (anti-)resonance systems in \textit{exactly} the same form. We will prove its link to the original perturbative sector in Section~\ref{Section: MedianResurgenceStructure_PT}.


After all these cancellations, the median-summed non-perturbative contributions to the spectrum become
\begin{align}
	\mcalS_0 \left[\d_\mrmPT^\mrmUBr\right] \simeq \ve \left(-\mcalK_2 \mcalS_{0}\left[P_0^{(1)}\right] +\frac{\mcalK_1^2}{\mcalK_2} \mcalS_{0}\left[\mcalP_{\mcalB^2\mcalI}\right]   \right) + \mcalK_2^2 \mcalS_{0}\left[P_0^{(2)} P_1^{(2)}\right] - 2\mcalK_1^2 \mcalS_0\left[P_0^{(1)} P_1^{(1)}\right] \, , \label{PT_NP_BorelSummed_Unbroken}
\end{align}
where we introduced $\mcalP_{\mcalB^2\mcalI^{-1}} = \frac{\left(P_0^{(1)}\right)^2}{2P_0^{(2)}}$ as a shorthand notation and $\mcalS_0$ refers to the median summation at $\t_\hbar=0$. This is a real and unambiguous expression and confirms that the $\mcalK_1\ll \mcalK_2$ limit falls into the unbroken PT-symmetric phase. In Section~\ref{Section: ExactSolutions}, we will verify this result using Alien calculus as a limit of the exact solution, which is valid across all parameter regimes.

\paragraph{\underline{Remark}:} Recalling the discussion in Section~\ref{Section: Resurgence_review} and the definition in~\eqref{BorelSummation_OnStokesRay}, we remind ourselves that the median summation corresponds $\mcalS_0 = \msfS_{0^\pm} \circ \mfrS_\hbar^{\mp 1/2}$. However, during the resurgence cancellation procedure, as we explained above, we did not consider the Stokes automorphism in a direct way. Then, in this sense, this procedure is an equivalent but more practically accessible way to consider the removal of the Stokes automorphism and subsequent Borel summations, leading to the median summation $\mcalS_0\left[\d_\mrmPT^\mrmUBr\right]$.

\item \underline{$\mcalK_2 \ll \mcalK_1$}: Now, let us consider the limit when the bounce configuration is dominant. Again, first keeping the imaginary contributions of $e^{\pm i\pi\mcalF}$ hidden, we find the non-perturbative corrections as
\begin{align}
	\d_\mrmPT^\mrmBr & \simeq \mp i (1 + \ve) \mcalK_1 H^{(1)}_0  \pm  i \ve \frac{\mcalK_2^2 \left(H^{(2)}_0\right)^2}{2 \mcalK_1 H^{(1)}_0} - 2(1+\ve) \mcalK_1^2 H^{(1)}_0	H^{(1)}_1    \nonumber \\ 
	&  \quad + \mcalK_2^2\left(  (1+\ve)   H^{(2)}_0 H^{(2)}_1 \pm \frac{2 i \pi \ve\left( H_0^{(2)}\right)^2 P_0^{(1)}}{H_0^{(1)}}\right)  \, .  
	\label{PT_NP_Trans-series_Broken}
\end{align}
Then, separating the imaginary factors using \eqref{separate_expansion}, we got
\begin{align}
	\d_\mrmPT^\mrmBr & \simeq \mp i (1 + \ve) \mcalK_1 P^{(1)}_0  \pm  i \ve \frac{\mcalK_2^2 \left(P^{(2)}_0\right)^2}{2 \mcalK_1 P^{(1)}_0} - 2(1+\ve) \mcalK_1^2 \left(P^{(1)}_0	P^{(1)}_1  \pm i\pi P_0^{(1)} P_0^{(1)} \right) \nonumber \\ 
	&  \quad + \mcalK_2^2\left( (1+\ve) P^{(2)}_0 P^{(2)}_1 \mp i \pi \left(1- \ve\right) \left(P_0^{(2)}\right)^2 \right) \label{PT_NP_Trans-series_Broken2}
\end{align}

Contrary to the previous case, there is no real term to split the perturbative degeneracy. In fact, the events $\left[\mcalI\right]$ and $\left[\mcalB^2 \mcalI^{-1}\right]$ disappear from the trans-series completely. Instead, we observe that there is a new type of event, i.e.~$\left[\mcalI \mcalIbar \mcalB^{-1}\right]_\pm \sim \pm i\frac{\mcalK_2^2}{\mcalK_1}$. Again, we note that to our knowledge, such combinations between non-perturbative saddles have not been observed in Hermitian settings and have not been discussed from a direct dilute instanton gas perspective.

The remaining parts are similar to the bion and bounce combinations from the previous case, but they arise with different pre-factors, which have crucial impacts on the physical properties of the quantized system. Let us investigate their structure in more details:

We first observe that the imaginary contributions of one-bounce and one-bion configurations appear as:
\begin{align}
	&\Im \left((1+\ve)\left[\mcalB\right]_\pm + (1-\ve)[\mcalI \mcalIbar]_\pm\right) \nonumber\\
    &\simeq 	\mp \left( (1 + \ve) \mcalK_1 P^{(1)}_0 + \pi(1-\ve) \mcalK_2^2 \left(P_0^{(2)}\right)^2  \right) \,\nonumber \\ 
	& = \mp \left( \mcalK_1 P_0^{(1)} + \pi \mcalK_2^2 \left(P_0^{(2)}\right)^2\right)  \mp  \ve \left( \mcalK_1 P_0^{(1)} -  \pi \mcalK_2^2\left(P_0^{(2)}\right)^2 \right) \, . \label{LeadingOrder_PT_broken}
\end{align}
The first part of~\eqref{LeadingOrder_PT_broken} is in the same form as~\eqref{ambiguity_PT_unbroken}. Since the perturbative expansion does not change its behavior in the parameter region corresponding to $\mcalK_2 \ll \mcalK_1$, this part links to the Borel ambiguity that participates in the resurgence cancellation, which we verified numerically in Fig.~\ref{Figure: DivergenceCompare}.


The existence of the latter part in~\eqref{LeadingOrder_PT_broken}, on the other hand, has a very significant physical consequence. Since there is no other imaginary ambiguous contribution to the trans-series at the orders $\mcalK_1$ and $\mcalK^2_2$, it remains in the Borel-summed series and contributes to the spectrum. As a result, the eigenvalues of the system become complex. This means that the PT-symmetry is \textit{broken} non-perturbatively. 

A very important feature of the surviving part in \eqref{LeadingOrder_PT_broken} is that it comes with the factor $\ve = \pm 1$, which induces the splitting of the perturbatively degenerate states. Contrary to the unbroken phase, the splitting is an imaginary one, as it was predicted by the numerical solutions in Fig.~\ref{Figure: PTsymmetric_Real}. This is in accordance with the PT-symmetry breaking examples in the literature, which we also find via numerical methods and showed in Fig.~\ref{Figure: PTsymmetric_Eigenvalues}. As we explained above, the splitting in~\eqref{LeadingOrder_PT_unbroken} has a clear link to the bion and bounce configurations through the exact QC~\eqref{EQC_PT_both}, making our approach a first successful example for an analytic explanation.

Finally, we also emphasize that the imaginary parts of both bion and bounce are originally ambiguous terms. However, since it is combined with the splitting factor $\ve$, the total contribution has no ambiguity. More specifically, we remind ourselves that in~\eqref{LeadingOrder_PT_unbroken}, assigning $ \ve = + 1 $ or $\ve = -1$ to the states originally localized around one of the wells is an arbitrary decision. This random choice has no effect on the physical properties of the quantum system. Then, the apparent ambiguity for $\pm \ve$ terms in \eqref{LeadingOrder_PT_broken} is compensated by this randomness, leading to an unambiguous result after the Borel summations.

The numerical solutions in Fig.~\ref{Figure: PTsymmetric_Eigenvalues} indicate that while imaginary contributions to the Borel-summed spectrum should be accompanied with the $\ve = \pm 1$ factor, the real contributions should have the same sign for both perturbative states. Therefore, the terms in~\eqref{PT_NP_Trans-series_Broken2} which do not fall into these categories should not contribute to the physical spectrum. In our discussion, the only way for this is the resurgence cancellations.

In light of these facts, the real parts of two-bounce and one-bion contributions should play a role in the resurgence cancellations. More specifically, we split their contributions as
\begin{align}
	\Re \left((1+\ve)\left[\mcalI\mcalIbar\right]_\pm + (1+\ve)\left[\mcalB^2\right]_\pm  \right) &= - 2 \mcalK_1^2 P_0^{(1)} P_1^{(1)} + \mcalK_2^2 P_0^{(2)} P_1^{(2)}\nonumber \\
	&\quad + \ve \left(-2\mcalK_1^2 P_0^{(1)} P_1^{(1)} + \mcalK_2^2 P_0^{(2)} P_1^{(2)} \right)  \, , \label{PTbroken_real}
\end{align}
where the expression in the second line should cancel out. Comparing with the lower-order terms, we infer that the perturbative side of these cancellations is the fluctuations of $\ve [\mcalB]_\pm$ and $\ve [\mcalI\mcalIbar\mcalB^{-1}]_\pm$ events:
\begin{align}
	\Re\left(\mp i \ve \mcalK_1 \msfS_\pm\left[P_0^{(1)}\right]\right) &\longleftrightarrow \Re\left(\ve [\mcalB^2]_\pm \right) \, , \label{ResurgenceCancellations_broken1} \\
	\Re\left(\pm i\ve \frac{\mcalK_2^2}{2\mcalK_1}\msfS_\pm\left[\mcalP_{\mcalI\mcalIbar\mcalB^{-1}}\right] \right) &\longleftrightarrow \Re \left(\ve [\mcalI\mcalIbar]_\pm\right) \,,\label{ResurgenceCancellations_broken2}
\end{align}
where we introduced $\mcalP_{\mcalI\mcalIbar\mcalB^{-1}} = \frac{\left(P_0^{(2)}\right)^2}{P_0^{(1)}}$ to simplify the notation.  However, as before, it is possible that more than one non-perturbative sector is involved in the cancellations, but numerical computations are too cumbersome to pursue here. 

Note that although there is a lack of numerical capability to demonstrate, the roles of $\Re\left(\ve[\mcalB^2]_\pm\right)$ and $\Re\left(\ve[\mcalI\mcalIbar]\right)$ are quite unconventional and deserve further discussion. We first remark that~\eqref{PTbroken_real} is free from ambiguities and such terms normally do not play a role in the cancellations of ambiguous Borel summations, when the re-summed spectrum is real. However, we recall that the series expansions in~\eqref{ResurgenceCancellations_broken1} and \eqref{ResurgenceCancellations_broken2} have a double sign in front of the $\ve$ factor. Originally, this double-valuedness stems from the analytic continuations $\Im\, \hbar>0$ and $\Im\, \hbar <0$. Note that the $\Im\, \hbar=0^\pm$ limits [upper/lower signs in \eqref{ResurgenceCancellations_broken1} and \eqref{ResurgenceCancellations_broken2}] are linked to $\msfS_\pm$ Borel summations, respectively. Then, combining these fixed signs for each analytic continuation and respective Borel summations leads to a real and single-valued expression as 
\begin{align}
	\mp i \msfS_\pm\left[P_0^{(1)}\right] &\sim \left(\mp i\right) \left(\pm i\right)\left(1+ O(\hbar)\right) = (+1)\left(1 + O(\hbar)\right)\, , \label{Cancellation_singleSign} \\
	\pm i \msfS_\pm[\mcalP_{\mcalI\mcalIbar\mcalB^{-1}}] &\sim (\pm i) (\pm i) \left(1 + \cdots \right) = (-1)(1 +\cdots ) \, , \label{Cancellation_singleSign2}
\end{align} 
where we utilized $P_0^{(l=1,2)} = 1 + O(\hbar)$. Then, their non-perturbative counterparts are expected to be real and single-valued. 

The remaining term in~\eqref{PT_NP_Trans-series_Broken2} is the ambiguous imaginary contribution at $O\left(\mcalK_1^2\right)$. As in the unbroken phase, it is a part of the minimal trans-series, which remains in the form of~\eqref{PT_minimalTransseries} as we expected, and plays a role in the cancellation of the Borel ambiguity emanating from the perturbation series $\msfS_{\pm}\left[E_\mrmP\right]$. 

As a conclusion of this discussion, let us present the Borel-summed non-perturbative corrections in the broken phase together:
\begin{align}
	\mcalS_0 \left[\d_\mrmPT^\mrmBr\right] &\simeq  -i \ve  \left(\mcalK_1 \mcalS_0\left[P_0^{(1)}\right] - \frac{\mcalK_2^2}{\mcalK_1}\mcalS_0\left[\mcalP_{\mcalI\mcalIbar\mcalB^{(-1)}}\right] + 2 \pi \mcalK_1^2 \mcalS_{0}\left[\left(P_0^{(1)}\right)^2\right] + \pi \mcalK_2^2 \left(P_0^{(2)}\right)^2   \right) \nonumber \\ 
	& \quad -2 \mcalK_1^2 \mcalS_0\left[ P_0^{(1)} P_1^{(1)}\right] + \mcalK_2^2 \mcalS_0\left[P_0^{(2)} P_1^{(2)}\right] \, . \label{PT_NP_BorelSummed_Broken}
\end{align}
Although we cannot verify the resurgence cancellations numerically beyond the leading order, in Section~\ref{Section: ExactSolutions}, we will show \eqref{PT_NP_Trans-series_Unbroken2} is indeed the non-perturbative contribution to the spectrum in $\mcalK_2 \ll \mcalK_1$ limit, which falls into the broken PT-symmetric phase.


\end{itemize}


\subsection{Resonance and anti-resonance cases}\label{Section: R_AR_Direct}

We now turn our attention to the resonance and anti-resonance cases. As we discussed above in Section~\ref{Section: EWKB_Setup_ITW}, their exact QCs, and hence their dynamics, are related to each other by time reversal. Both cases induce complex spectra, implying their non-equilibrium nature under the interaction with the surrounding environment. In the following, we will uncover the corresponding trans-series structures on an equal footing with the PT-symmetric one in Section~\ref{Section: PT-symmetric_Direct}.



\subsubsection{Resonance solutions} Let us start with the resonance case. Its exact QCs for $\Im \,\hbar = 0^\pm$ are given in terms of $T^{(\pm)}_{2,2}$ elements of the transition matrices, and they are presented in Table~\ref{Table: ExactQuantizationCondition_ITW}.  

For the $\Im \,\hbar <0$ deformation, it is straightforward to bring $T_{2,2}^{(-)} = 0$ in the following form:
\begin{equation}
	e^{+i\pi\mcalF} 2 \cos\left(\pi \mcalF\right) = i\ve \sqrt{\Pi_{B_2}}\, , \label{EQC_Res1}
\end{equation}
where $\ve=\pm 1$. Equivalently, setting $\mcalF = \frac{1}{2} + N + \d_\mrmR^{(-)}$ 
\begin{equation}
	\frac{1}{\pi} \sin\left(\pi \d_\mrmR^{(-)} \right) = \sum_{k=0}^\infty b_k \left(\d_\mrmR^{(-)}\right)^k =  \ve \mcalK_2 \sum_{k=0} H_k^{(2)} \left(\d_\mrmR^{(-)}\right)^k\, , \label{EQC_R_1}
\end{equation}
where we used the definition of $\mcalK_2$ in~\eqref{ITW_fugacities}. For $\Im \, \hbar >0$, $T_{2,2}^{(+)} = 0$ is also a quadratic equation, and we find its formal solution for $\left(1+\Pi_A\right)$ as
\begin{equation}
	e^{-i \pi \mcalF} 2\cos\left(\pi\mcalF\right) = -\Pi_{B_1} - i\ve  \sqrt{\Pi_{B_2}} \left(1+\Pi_{B_1}\right)\, . \label{EQC_Res2}
\end{equation}
Then, setting $\mcalF = \frac{1}{2} + N + \d_\mrmR^{(+)}$ and using the definitions in~\eqref{ITW_fugacities} and \eqref{b1_fugacity}-\eqref{b1_fugacity2}, we obtain
\begin{align}
	\sum_{k=0}^\infty b_k \d_\mrmR^{(+)} =  - 2 i \mcalK_1 \sum_{k=0}^\infty H_k^{(1)} \d_\mrmR^{(+)} -  \ve\mcalK_2 \left(\sum_{k=0}^\infty H_k^{(2)} \d_\mrmR^{(+)}\right) \left[1+ 2\pi \mcalK_1 \sum_{k=0}^\infty P_k^{(1)} \d_\mrmR^{(+)}\right] \, . \label{EQC_R_2}
\end{align}
Note that, as in the PT-symmetric case [see e.g., \eqref{EQC_PT_rearranged}], the last $\Pi_{B_1}$ on the right-hand side of~\eqref{EQC_Res2} is not affected by the $\frac{1}{2\pi}$ re-scaling and $e^{-i\pi\mcalF}$. Thus, its fluctuation terms are real, and we introduced $2\pi$ factor by hand in the last term in~\eqref{EQC_R_2}.

Both~\eqref{EQC_R_1} and \eqref{EQC_R_2} can be solved order by order in a similar manner as the PT-symmetric system. In the current case, since the square-root terms in~\eqref{EQC_Res1} and \eqref{EQC_Res2} contain only the $\Pi_{B_2}$ action, the order of expansion does not matter. Then, we continued in a straightforward manner and up to orders $O(\mcalK_1^2)$ and $O(\mcalK_2^2)$, we find the non-perturbative corrections of the resonance case for $\Im \,\hbar >0$ and $\Im \,\hbar <0$ as
\begin{align}
	\d_\mrmR^{(+)} &=  -\ve \mcalK_2  H^{(2)}_0 +\mcalK_2^2 H^{(2)}_0 H^{(2)}_1 -2 i \mcalK_1 H^{(1)}_0 -4 \mcalK_1^2 H^{(1)}_0 H^{(1)}_1\nonumber \\ 
	& \quad + 2\ve \mcalK_1 \mcalK_2 \left( i  H^{(1)}_1 H^{(2)}_0+ i H^{(1)}_0 H^{(2)}_{1} - 2 \pi    H^{(2)}_0 P^{(1)}_0\right) \, ,	\label{Resonance_NP_Trans-series1_hide}
\end{align}
and 
\begin{align}
	\d_\mrmR^{(-)} &\simeq 	-\ve \mcalK_2 H^{(2)}_0 + \mcalK_2^2 H^{(2)}_0 H^{(2)}_1\, ,  \label{Resonance_NP_Trans-series2_hide}
\end{align}
respectively. Using~\eqref{separate_expansion}, the complex parts can be uncovered explicitly as
\begin{align}
	\d_\mrmR^{(+)}  & \simeq -\ve  \mcalK_2 P_0^{(2)} {\color{red}-} 2 i \mcalK_1 P_0^{(1)} + \mcalK_2^2 \left(P_0^{(2)} P_1^{(2)} - i \pi \left(P_0^{(2)}\right)^2 \right)    \nonumber \\
	& \quad - 4\mcalK_1^2 \left(P_0^{(1)} P_1^{(1)} {\color{red}-} i\pi \left(P_0^{(1)}\right)^2 \right) {\color{red}+} 2i\ve \mcalK_1 \mcalK_2 \left(P_0^{(1)} P_1^{(2)} + P_1^{(1)} P_0^{(2)} \right) \, , \label{Resonance_NP_Trans-series1}
\end{align}
and
\begin{align}
	\d_\mrmR^{(-)} & \simeq  -\ve  \mcalK_2 P_0^{(2)} + \mcalK_2^2 \left(P_0^{(2)} P_1^{(2)} + i \pi \left(P_0^{(2)}\right)^2 \right) \, . \label{Resonance_NP_Trans-series2}
\end{align}
Note that we highlighted some of the signs with red for future comparison with the anti-resonance trans-series in \eqref{AntiResonance_NP_Trans-series1}.

Interestingly, $\d^{(+)}_\mrmR$ and $\d^{(-)}_\mrmR$ lead to quite different trans-series solutions. Let us now discuss how they lead to a consistent picture when the (median) Borel summations are taken into account:  

\begin{itemize}[wide]
\item We first observe that in \eqref{Resonance_NP_Trans-series1} and \eqref{Resonance_NP_Trans-series2}, $\d_\mrmR^{(\pm)}$ share the same contributions at $O(\mcalK_2)$ and $O(\mcalK_2^2)$ up to complex conjugation. In the semi-classical dilute gas approximation, these terms correspond to the one-instanton and one-bion configurations, i.e.,
\begin{equation}
	\left[\mcalI\right] = \mcalK_2 P_0^{(2)}\, , \quad \left[\mcalI\mcalIbar\right]_\pm = \mcalK_2^2 \left(P_0^{(2)} P_1^{(2)} \mp  i\pi \left(P_0^{(2)}\right)^2\right)\, ,
\end{equation}
where the sign ambiguity is the bion ambiguity for analytic continuations $\arg \hbar = 0^\pm$. 

\item Note that both terms are on an equal footing with the unbroken PT-symmetric system in Section~\ref{Section: PT-symmetric_Direct}. Therefore, their roles in the associated trans-series are the same as well. The real part contributes to the leading-order non-perturbative spectra as
\begin{equation}
	\mathrm{LO:} \qquad \ve[\mcalI] + \Re\left([\mcalI\mcalIbar]_\pm\right) = \ve \mcalK_2 P_0^{(2)} + \mcalK_2^2 P_0^{(2)} P_1^{(2)}  \, , \label{resonance_LeadingOrderSplitting}
\end{equation}
where $\ve=\pm 1$ indicates the real splitting of the perturbatively degenerate energies.

The ambiguous part arising from the one-bion event, i.e., $\Im[\mcalI\mcalIbar]_\pm$, participates in the resurgence cancellations. Recall that it is a term in the minimal trans-series emanating from the perturbative sector; see~\eqref{PT_minimalTransseries}. Then, $\Im [\mcalI\mcalIbar]_\pm$ disappears after the Borel summations. 

\item The case of bounce contributions is quite different: First, we observe that in the trans-series expansion of $\d^{(-)}_\mrmR$ in \eqref{Resonance_NP_Trans-series2}, no term associated with the bounce event $\left[\mcalB\right]_- = +i \mcalK_1 P_0^{(1)}$, appears. This means there is no term that cancels the Borel ambiguity associated with the large order behavior in \eqref{LargeOrder_Bounce} in the perturbative sector, which leads to $\Im \left(\msfS_-\left[E_\mrmP\right]\right) = -i \mcalK_1 P_0^{(1)}$ for $\Im\hbar = 0^-$ limit.

In the expression for $\d^{(+)}_\mrmR$, on the other hand, the term of $O(\mcalK_1)$ resembles the bounce configurations with the identification $2[\mcalB]_+ = - 2i\mcalK_1 P_0^{(1)}$. The overall factor of 2, however, does not match the term in~\eqref{ambiguity_PT_unbroken}. Therefore, the $[\mcalB]_+$ contribution in~\eqref{Resonance_NP_Trans-series2} fixes the ambiguity issue of the Borel summation, which reads as $\Im \left(\msfS_+\left[E_\mrmP\right]\right) = +i \mcalK_1 P_0^{(1)} $ for $\Im\hbar= 0^+$ limit. As a result, an imaginary part remains and the spectrum becomes complex.

Combining both observations, the leading order resurgence cancellations take place at $\Im \,\hbar = 0^\pm$ limits as follows:
\begin{align}
	\Im \left(\msfS_+[E_\mrmP]\right) + \Im[\mcalI\mcalIbar]_{+} + 2 [\mcalB]_+ &\simeq  -i \mcalK_1 P_0^{(1)}\, ,\quad \left(\Im\hbar=0^+\right) \label{resonance_LeadingOrder_Imaginary1} \\ 
	 \Im\left(\msfS_-\left[E_\mrmP\right]\right) + \Im[\mcalI\mcalIbar]_{-}   &\simeq  -i \mcalK_1 P_0^{(1)} \,,\quad \left(\Im\hbar=0^-\right)\label{resonance_LeadingOrder_Imaginary2}
\end{align}
In both cases, the remaining parts represent the leading-order imaginary parts of the spectra. A crucial feature of these results is that there is no ambiguity left, as both results are exactly the same, including their overall imaginary sign. Note that the minus imaginary sign is also fully consistent with the decaying feature of the resonant system.

\paragraph{\underline{Remark}:} In the literature, the results in~\eqref{resonance_LeadingOrder_Imaginary1} and~\eqref{resonance_LeadingOrder_Imaginary2} are known as a bounce contribution that leads to the decay of a metastable state. However, as we discussed thoroughly in the PT-symmetric case, its existence does not always lead to an instability. Similarly, for the $\Im \,\hbar =0^{-}$ limit, we observe in~\eqref{resonance_LeadingOrder_Imaginary2} that the bounce configuration does not contribute at all but that the spectrum is complex. This is solely due to the Borel summation of the perturbative sector, whose imaginary part is canceled only partially. For $\Im \,\hbar = 0^{+}$, on the other hand, as we stated above, the $[\mcalB]_{+}$ event cancels $\Im\left(\msfS_+\left[E_\mrmP\right]\right)$, while leaving an imaginary contribution behind, which is exactly the same as $\Im \left(\msfS_-\left[E_\mrmP\right]\right)$.

This analysis shows that, although the resulting contributions in~\eqref{resonance_LeadingOrder_Imaginary1} and \eqref{resonance_LeadingOrder_Imaginary2} are equivalent to the usual decay analysis based solely on the bounce contribution, the true picture is quite different. We also emphasize that a bounce configuration is naturally multi-valued. In the literature, the imaginary sign is chosen so that the result is consistent with the decaying behavior. In our analysis, the resulting sign is fixed for any choice of analytic continuation, showing full consistency with the physical properties of the quantum system without any ad-hoc manipulation. A detailed comparison between the exact quantization procedure that we discussed in this paper and the standard decay analysis revolving around the bounce event would be very interesting for mathematical completeness. It might also be enlightening for the decay process beyond the standard semi-classical analysis, which includes non-linearities dominating early- and late-time behaviors. 

\item Let us also discuss the higher-order non-perturbative contributions. In the unbroken PT-symmetric case, we indicated a resurgence relationship between $[\mcalB \mcalI]_\pm \sim \pm i\ve \mcalK_1 \mcalK_2$ contributions and the Borel summation of the 1-instanton fluctuations, which reads as $ \msfS_\pm[\mcalI]= \msfS_\pm\left[-\ve \mcalK_1 P_0^{(2)}\right] \sim \mp i\ve \mcalK_1 \mcalK_2 + \dots $, where $\dots$ indicates the possible higher order terms emerging from the Borel summation procedure which we ignore. 

Note that in the expansion of $\d_\mrmR^{(+)}$, the coefficients in front of $O(\mcalK_1)$ and $O(\mcalK_1\mcalK_2)$ do not match. This indicates a partial cancellation between $[\mcalB\mcalI]_+$ and $\msfS_+[-\ve \mcalK_1 P_0^{(2)}]$; and as in the leading-order case, an imaginary contribution remains after the resurgence cancellations. For $\d_\mrmR^{(-)}$ case, on the other hand, since a contribution associated with $[\mcalB \mcalI]_-\sim -i\ve \mcalK_1\mcalK_2$ is lacking, the only imaginary contributions arise via the Borel summation $\msfS_-[-\ve\mcalK_1 P_0^{(2)}]$. 

In both cases, the resulting imaginary part reads as
\begin{align}
	  2[\mcalB\mcalI]_+ + \msfS_{+}[\mcalI] &\simeq i\ve \mcalK_1\mcalK_2  \left(P_0^{(1)} P_1^{(2)} + P_1^{(1)} P_0^{(2)} \right)\, , \quad (\Im\hbar =0^+)\, , \label{resonance_Imaginary_NLO1} \\
	 \msfS_-[\mcalI] &\simeq i\ve \mcalK_1\mcalK_2  \left(P_0^{(1)} P_1^{(2)} + P_1^{(1)} P_0^{(2)} \right)\, , \quad (\Im\hbar =0^-)\, , \label{resonance_Imaginary_NLO2}
\end{align}
where there is no sign ambiguity as it should be. We note that for the $\Im \,\hbar=0^+$ case, the exact resurgence cancellation for $\msfS_+\left[P_0^{(2)}\right]$ may need extra higher-order non-perturbative counterparts. However, this does not change the picture we presented here, as the $i\ve \mcalK_1\mcalK_2$ contribution appears in the Borel-summed spectrum in any case.

\item Finally, let's consider the resurgence relations associated with $O(\mcalK_1^2)$ contributions. 
We first concentrate on the $\d^{(+)}_\mrmR$ expansion and recall the remaining term after the leading order cancellation in \eqref{resonance_LeadingOrder_Imaginary1}. In the broken PT-symmetric case, we discussed the resurgence relationship between $\Re\left(\msfS_+[\mcalB]_+\right)= \Re\left(-i\mcalK_1 \msfS_+[P_0^{(1)}]\right)$ and $\Re[\mcalB^2]_+= -2\mcalK_1^2 P_0^{(1)} P_1^{(1)}$; see~\eqref{ResurgenceCancellations_broken1} and \eqref{Cancellation_singleSign}. Then, we infer their overall contribution to \eqref{Resonance_NP_Trans-series1} becomes
\begin{align}
	\Re\left(\msfS_+[\mcalB]_+\right) + 2\Re[\mcalB^2]_+ &= \Re\left(-i\mcalK_1 \msfS_+[P_0^{(1)}]\right) - 4\mcalK_1^2 P_0^{(1)} P_1^{(1)} \nonumber \\
	&\simeq -2\mcalK_1^2 P_0^{(1)} P_1^{(1)}\, , \label{R_NP_NLO}
\end{align}
which contributes to the real spectrum at the order of 2-bounce events. Again, the complete resurgence relation for $\Re(\msfS_+[\mcalB]_+)$ may need higher-order non-perturbative terms, which we do not consider here.
Note that the term $\Re\left(\msfS_+[\mcalB]_+\right)$ is the real part of the remainder of the resurgence cancellations in \eqref{resonance_LeadingOrder_Imaginary1}. While it contributes to the cancellations in \eqref{R_NP_NLO}, the associated imaginary part contributes to the spectrum directly.

The remaining imaginary part $2\Im [\mcalB^2]_+ = +4 \pi \mcalK_1^2 \left(P_0^{(1)}\right)^2$, on the other hand, is linked to the second singularity of $\msfS_+[E_\mrmP]$, which contributes as $-2\pi i \mcalK_1^2 \left(P_0^{(1)}\right)^2$. However, we observe that the prefactors are not in complete agreement and that there is a remainder
\begin{equation}
	2\Im[\mcalB^2]_+ - 2\pi i\mcalK_1^2 \left(P_0^{(1)}\right)^2 = + 2\pi i \mcalK_1^2 \left(P_0^{(1)}\right)^2\, , \label{minimalR_NLO}
\end{equation} 
which contributes the physical spectrum. 

\item For the trans-series of $\d_\mrmR^{(-)}$, there is no term of $O(\mcalK_1^2)$ in \eqref{Resonance_NP_Trans-series2}. Therefore, both real and imaginary contributions of $O(\mcalK_1^2)$ arise from the Borel summations. One of them simply arises from the second singularity of $\msfS_-[E_\mrmP]$, which induces $-2\pi i\mcalK_1^2$ $\left(P_0^{(1)}\right)^2$. The other contribution arises from the remainder of the leading order cancellation \eqref{resonance_LeadingOrder_Imaginary2}, which leads to $-i \mcalK_1 \msfS_-[P_0^{(1)}] = - \mcalK_1^2 P_0^{(1)} P_1^{(1)}$. Both contributions match with \eqref{R_NP_NLO} and \eqref{minimalR_NLO}, showing the equivalence of physical contributions at $O(\mcalK_1^2)$ for both $\Im\hbar = 0^\pm$ limits. 
\end{itemize}

As a result, we obtain the resulting median-summed non-perturbative contributions, 
\begin{align}
	\mcalS_0\left[\d_\mrmR^{(\pm)}\right] &\simeq -\ve \mcalK_2 \mcalS_0\left[P_0^{(2)}\right] + \mcalK_2^2 \mcalS_0\left[P_0^{(2)} P_1^{(2)}\right] - i\mcalK_1 \mcalS_0\left[P_0^{(1)}\right]\nonumber \\
	&\quad + 2 \mcalK_1^2 \mcalS_0\left[-P_0^{(1)} P_1^{(1)} +\pi i\left(P_0^{(1)}\right)^2\right]  - i \ve\mcalK_1\mcalK_2 \mcalS_0\left[P_0^{(1)} P_1^{(2)} + P_1^{(1)} P_0^{(2)} \right] \, . \label{R_NP_BorelSummed}
\end{align}
Note that the last term introduces an imaginary splitting in addition to the real one, which is induced by the $\ve [\mcalI]$ contribution. This means that for $\ve = -1$, the corresponding imaginary contribution has a positive sign for every other state in the spectrum. However, this cannot change the overall imaginary sign of the spectrum, as the $-i\mcalK_1$ term dominates and the quantum system has resonance properties for all states.

\subsubsection{Anti-resonance solutions} 
As the anti-resonance case is the time-reversal of the resonance one, its physical properties can be inferred by a complex conjugation. In Section~\ref{Section: EWKB_Setup_ITW}, we have already shown that the associated elements in the $T^{(\pm)}$ matrices are linked via the complex conjugation operator and that its effect is expected to appear in the spectrum.

To see how the complex conjugation with the resonance case arises, let us examine the solutions of $T^{(\pm)}_{1,1}= 0$ in Table~\ref{Table: ExactQuantizationCondition_ITW}:
\begin{align}
	 e^{-i\pi\mcalF} 2\cos\left(\pi\mcalF\right) & = i\ve \sqrt{\Pi_{B_2}} \, ,\label{EQC_ARes_1} \\ 
	e^{+i \pi \mcalF} 2\cos\left(\pi \mcalF\right) &= - \Pi_{B_1} + i\ve \sqrt{\Pi_{B_2}} \left(1 + \Pi_{B_1} \right) \, , \label{EQC_ARes_2}
\end{align}
respectively. The effect of the time-reversal appears in the exponential terms in~\eqref{EQC_ARes_1} and \eqref{EQC_ARes_2} on the left-hand side of both equations, which are complex conjugates of their counterparts in~\eqref{EQC_Res1} and \eqref{EQC_Res2}, respectively. Note that the structures of the QCs for $\Im \,\hbar>0$ and $\Im\, \hbar<0$ are also exchanged. This mainly stems from \eqref{ITW_complexConjugate} and indicates the Borel ambiguities of the perturbative expansions resulting from~\eqref{EQC_ARes_1} and \eqref{EQC_ARes_2} have the opposite signs relative to those in~\eqref{EQC_Res1} and \eqref{EQC_Res2}. Then, these two signs change and balance to each other, leading to the same resurgence cancellations and the Borel-summed spectrum with the resonance system up to the imaginary sign.


To get the non-perturbative spectrum and the associated resurgence structure for the anti-resonance case independently, we define $\mcalF = N+ \frac{1}{2}+\d_\mrmAR^{(\pm)}$. Then, by solving~\eqref{EQC_ARes_1} and \eqref{EQC_ARes_2} in the same way as the resonance case, we obtain
\begin{align}
	\d_\mrmAR^{(+)} & \simeq  \ve  \mcalK_2 P_0^{(2)} + \mcalK_2^2 \left(P_0^{(2)} P_1^{(2)} - i \pi \left(P_0^{(2)}\right)^2 \right) \, ,  \label{AntiResonance_NP_Trans-series1} \\
	\d_\mrmAR^{(-)} & \simeq -\ve  \mcalK_2 P_0^{(2)} + \mcalK_2^2 \left(P_0^{(2)} P_1^{(2)} + i \pi \left(P_0^{(2)}\right)^2 \right) {\color{red}+} 2 i \mcalK_1 P_0^{(1)}   \nonumber \\
	& \quad - 4\mcalK_1^2 \left(P_0^{(1)} P_1^{(1)} {\color{red}+} i\pi \left(P_0^{(1)}\right)^2 \right) {\color{red}-} 2i\ve \mcalK_1 \mcalK_2 \left(P_0^{(1)} P_1^{(2)} + P_1^{(1)} P_0^{(2)} \right) \, ,
	\label{AntiResonance_NP_Trans-series2}
\end{align}
where the red-colored signs are different from those for $\d^{(+)}_\mrmR$ in~\eqref{Resonance_NP_Trans-series1}. 

The resurgence cancellations take place in the same manner as the resonance case, up to these sign differences and the ones arising from the Borel summations. For example, at the leading order in $\Im \, \hbar >0$, combining the Borel-summed perturbative series and $[\mcalI\mcalIbar]_\pm$ events leads to
\begin{equation}
	\Im \left(\msfS_+ \left[E_\mrmP\right]\right) + \Im\left([\mcalI\mcalIbar]_+\right)  = i \mcalK_1 P_0^{(1)}\, . 
\end{equation}
For $\Im \, \hbar < 0$, we have $2[\mcalB]_-$ contributing to the trans-series, which leads to
\begin{equation}
	\Im\left(\msfS_-  \left[E_\mrmP\right]\right) + \Im \left([\mcalI\mcalIbar]_-\right) + [\mcalB]_+ = i \mcalK_1 P_0^{(1)}\, . 
\end{equation}
Both results are the same, indicating that there are no ambiguities, and the imaginary sign is fully consistent with the anti-resonance behavior of the quantum system. 

In the same manner as the resonance case, the remaining contributions can be deduced. We do not detail it here because the procedure is the same. As a result, the median-summed non-perturbative part of the anti-resonance case becomes
\begin{align}
	\mcalS_0\left[\d_\mrmAR^{(\pm)}\right] &\simeq -\ve \mcalK_2 \mcalS_0\left[P_0^{(2)}\right] + \mcalK_2^2 \mcalS_0\left[P_0^{(2)} P_1^{(2)}\right] + i\mcalK_1 \mcalS_0\left[P_0^{(1)}\right] \nonumber \\
	&\quad - 2 \mcalK_1^2 \mcalS_0\left[P_0^{(1)}P_1^{(1)} +\pi i \left( P_0^{(1)}\right)^2\right]  - i \ve\mcalK_1\mcalK_2 \mcalS_0\left[ P_0^{(1)} P_1^{(2)} + P_1^{(1)} P_0^{(2)} \right] \, .  \label{AR_NP_BorelSummed}
\end{align}

\paragraph{\underline{Remark}:}As expected, the median summed trans-series solutions for resonance and anti-resonance cases are complex conjugate of each other: 
\begin{equation}
	\mcalS_0\left[\d_\mrmR^{(\pm)}\right] = \mathsf{C}\mcalS_0\left[\d^{(\pm)}_\mrmAR\right]\, .\label{ComplexConjugate_NP}
\end{equation}
The trans-series before the summation, on the other hand, are related via complex conjugation as 
\begin{equation}
	\mathsf{C} \d_\mrmR^{(\pm)} = \d_\mrmAR^{(\mp)} \, .\label{ComplexConjugation_Trans-series}
\end{equation}
Since the Borel summation is single-valued, we also have 
\begin{equation}
	\mfrS_\hbar^{\pm 1} \d_\mrmR^{(\pm)} = \d_\mrmR^{(\mp)} \, , \qquad \mfrS_\hbar^{\pm 1} \d_\mrmAR^{(\pm)} = \d_\mrmAR^{(\mp)} \, . \label{StokesAuto_Trans-series}
\end{equation}
Then, we simply deduce that
\begin{equation}
	\mathsf{C} \mfrS_\hbar \d_\mrmR^{(\pm)} = \mathsf{C} \d_\mrmR^{(\mp)} = \d_\mrmAR^{(\pm)} \, .\label{ComplexConjugation_StokesAuto_Trans-series}
\end{equation}
This is equivalent to the relationship between the respective exact QCs in Table~\ref{Table: ExactQuantizationCondition_ITW}, which is
\begin{equation}
	\mathsf{C} \tilde{\mfrS} T_{1,1}^{(\pm)} = \mathsf{C} T_{1,1}^{(\mp)} = T_{2,2}^{(\pm)} \, , \label{ComplexConjugate_QC}
\end{equation}
where $\tilde{\mfrS}$ corresponds to the combined action of $\mfrS_A$ and $\S$ as in \eqref{StokesAuto_Combined}. 

Finally, we note that in some cases, e.g., in unbroken PT-symmetric or other stable systems, the trans-series for $\Im \,\hbar > 0$ and $\Im \,\hbar <0$ become complex conjugate to each other. This is not the case for (anti-)resonance systems, which is obvious from~\eqref{ComplexConjugation_Trans-series}. Instead, the correct relationship between the two sides of the analytic continuation is given by the Stokes automorphism as in~\eqref{StokesAuto_Trans-series}, which appears to be equivalent to $\mathsf{C}$ if the system is stable. The last point is, in fact, an indication of the resurgence cancellations, which leads to real spectrum in such cases.

\paragraph{\underline{Remark}:} The fractional combinations of $\mcalK_1$ and $\mcalK_2$ in~\eqref{PT_NP_Trans-series_Unbroken} and \eqref{PT_NP_Trans-series_Broken}, on the other hand, do not exist in the resonance or anti-resonance spectrum. This indicates that such effects are closely related to the PT-symmetric boundary conditions, which may probably be understood better if complex boundary conditions are used. 

\subsection{Comparisons between different physical cases}
 

Throughout this section, handling the PT-symmetric and (anti-)resonance cases on an equal footing, we computed the first few non-perturbative corrections to their respective spectra. We discussed the resurgence structures and showed their significance for obtaining the physically viable spectrum. Let us now summarize and compare how resurgence relations work in different cases.  

For the PT-symmetric system, as we showed in Section~\ref{Section: PT-symmetric_Direct}, the main source of having both unbroken and broken phases is the discriminant term $\left[\Disc_\mrmPT\left(\Pi_{B_1},\Pi_{B_2}\right)\right]^{1/2}$ appearing in the exact quantization condition in \eqref{EQC_PT_both} and the different expansions in $\mcalK_1 \ll \mcalK_2$ and $\mcalK_2 \ll \mcalK_1$ limits. In the unbroken phase ($\mcalK_1 \ll \mcalK_2$), we observed that the perturbative Borel ambiguity is canceled against the total contribution arising from bounce and bion configurations at the leading order, i.e.
\begin{equation}
	\Im \msfS_{\pm}\left[E_\mrmP\right] +\Im[\mcalB]_\pm + \Im[\mcalI\mcalIbar]_\pm \sim 0 \label{LeadingOrderResurgence}
\end{equation} up to order $O(\mcalK_1) + O(\mcalK_2^2)$, which we numerically established in Section~\ref{Section: semi-classical_ITW}. 

In the broken case ($\mcalK_2 \ll \mcalK_1$), we still have the resurgence relationship \eqref{LeadingOrderResurgence} between the perturbative sector and bounce+bion configurations. However, the corresponding trans-series solution \eqref{PT_NP_Trans-series_Broken2} has a different structure than the trans-series solution for the unbroken phase in \eqref{PT_NP_Trans-series_Unbroken2}. From the resurgence perspective, the main difference is the additional imaginary contributions with $\ve=\pm 1$ factors at order $\mcalK_1$ and $\mcalK_2^2$. Then, combining them with the perturbative sector leads to
\begin{equation}
	\Im \msfS_{\pm}\left[E_\mrmP\right] + (1+\ve)\Im[\mcalB]_\pm + (1-\ve) \Im[\mcalI\mcalIbar]_\pm =\ve \left(\Im[\mcalB]_\pm  +\Im[\mcalI\mcalIbar]_\pm\right)\, , \label{brokenPT_remainderLeading}
\end{equation}
which contributes to the spectrum and breaks the PT-symmetry non-perturbatively.

Note that when combined with other imaginary contributions, i.e., $\ve\Im[\mcalI\mcalIbar \mcalB^{-1}]_\pm \sim  \pm i \ve \frac{\mcalK^2}{\mcalK_1}$ and $\ve\Im [\mcalB^2]_\pm\sim \pm i \ve \mcalK_1^2 $, the cancellation \eqref{brokenPT_remainderLeading} leads to complex conjugate contributions to the Borel spectrum. This is consistent with the expectations from the numerical solutions in Fig.~\ref{Figure: PTsymmetric_Eigenvalues}. On the other hand, Fig.~\ref{Figure: PTsymmetric_Eigenvalues} indicates no splitting between the real eigenvalues of these states. This is possible via cancellations between real parts as
\begin{align}
	\ve \Re\Big( \msfS_\pm [\mcalB]_\pm+ \msfS_\pm [\mcalI\mcalIbar\mcalB^{-1}]_\pm\Big)   + (1+\ve) \Re\Big( [\mcalB^2]_\pm +[\mcalI\mcalIbar]_\pm\Big) 
	\sim  \Re\Big([\mcalI\mcalIbar]_\pm + [\mcalB^2]_\pm\Big)\, , \label{brokenPT_realRemnant}
\end{align}
where, as shown in \eqref{Cancellation_singleSign} and \eqref{Cancellation_singleSign2}, the Borel summations lead to non-ambiguous real contributions as  $[\mcalB]_\pm$ and $[\mcalI\mcalIbar\mcalB^{-1}]_\pm$ originally have imaginary ambiguities. Both cancellations with imaginary and real remnants in \eqref{brokenPT_remainderLeading} and \eqref{brokenPT_realRemnant} distinguish from the resurgence cancellation in the unbroken phase and the Hermitian models in the literature.

When we consider the resonance boundary conditions with $\Im\hbar > 0$ deformation, the resurgence structure resembles that of the broken PT-symmetric case as the complex spectrum arises from the remainders of the resurgence cancellations. We schematically express the imaginary contributions as
\begin{equation}
	\Im\left(\msfS_+ [E_\mrmP]+ \msfS_+[\mcalI]\right) + \Im[\mcalI\mcalIbar]_+  + 2[\mcalB]_+ + 2\Im[\mcalB^2]_+ + 2[\mcalB\mcalI]_+ \sim [\mcalB]_+ + \Im[\mcalB^2]_+ + [\mcalB\mcalI]_+\, . \label{resonanceResurgence1}
\end{equation}
Note that, as explained in \eqref{resonance_LeadingOrder_Imaginary1}, \eqref{resonance_Imaginary_NLO1} and \eqref{minimalR_NLO}, these remaining contributions are due to the pre-factors 2 in \eqref{resonanceResurgence1}; therefore, their sources are originally non-perturbative contributions in the trans-series solution in \eqref{Resonance_NP_Trans-series1}.

For $\Im\hbar<0$, however, the spectrum arises in a different way since there is no bounce contribution to the trans-series in \eqref{Resonance_NP_Trans-series2}. Then, $\Im[\mcalI\mcalIbar]_\pm$ cancels the perturbative Borel ambiguity only partially, and the imaginary contributions remain via the Borel summations rather than the non-perturbative configurations in the trans-series. In this particular case, it reads as
\begin{equation}
	\Im \msfS_-[E_\mrmP] + \Im \msfS_-[\mcalI] + \Im[\mcalI\mcalIbar]_- \sim  [\mcalB]_+ + \Im[\mcalB^2]_+ + [\mcalB\mcalI]_+ \label{resonanceResurgence2}
\end{equation}
where the terms on the right-hand side arise solely from the Borel summations. 

Although their sources are different, the results in \eqref{resonanceResurgence1} and \eqref{resonanceResurgence2} are exactly the same, leading to a unique contribution to the spectrum. The complex conjugation relation between the imaginary parts in $\Im\hbar =0^\pm$ limits reads as
\begin{equation}
	\Im\left( \msfS_\pm[E_\mrmP] + \msfS_\pm[\mcalI]\right) =  [\mcalB]_\mp + \Im[\mcalI\mcalIbar]_\mp+  \Im[\mcalB^2]_\mp + [\mcalB\mcalI]_\mp \dots \, ,
\end{equation}
where $\dots$ corresponds to terms arising from higher order singularities in the Borel planes of $\msfS_\pm[E_\mrmP]$ and $\msfS_\pm[\mcalI]$. 

Overall, our discussion in this section establishes that the resurgence relations between Borel summations of a divergent expansion and higher order non-perturbative contributions play a crucial role in the physical spectrum regardless of the intrinsic characters of the quantum system. We showed that although numerical resurgence relations remain the same in all cases, the spectrum differs with the boundary conditions imposed on the system. In this sense, on the one hand, the boundary conditions pick the exact quantization conditions. On the other hand, the resurgence relations ensure the required cancellations upon Borel summations, leading to spectra perfectly consistent with the physical properties of the system under consideration. In the following section, we will generalize this framework beyond the asymptotic limits of $\mcalK_1\ll\mcalK_2$ and $\mcalK_2\ll \mcalK_1$ and establish its generalization to all parameter regimes.

\section{Exact solutions from median quantization conditions 
}\label{Section: ExactSolutions}

In this section, we will focus on the median QCs for the PT-symmetric, resonance, and anti-resonance systems. This allows us to work directly on the ray that the Stokes discontinuities (for both $A$-cycle and $\hbar$-plane) lie, which is our main interest as the physical quantities are associated with them. At the level of the QCs, this is achieved by an action of half-Stokes automorphism on the QCs, which we obtained via Airy-type EWKB analysis. 

Formally, these Stokes automorphisms are represented by
\begin{equation}
	D^{\med}_{i,j} = \mfrS^{\pm 1/2}_A T_{i,j}^{(\pm)}\, , \label{MedianQC_general}
\end{equation}
which are the QCs that the Stokes discontinuities are removed from the original version $T_{i,j}^{(\pm)}$. Note that this removal is ensured via the relations\footnote{To be more precise, we should also consider the Stokes automorphism $\S$ in \eqref{StokesAutomorphism_WaveFunction}. However, since we are mainly interested in the QCs of the form $T_{i,j}^{(\pm)}=0$, the contributions of $\S$, which affect only the pre-factors, do not alter the general picture. Therefore, we omit them from our discussion for simplifications.}~\eqref{StokesAuto_ITW_PT}, \eqref{StokesAuto_ITW_Res} and \eqref{StokesAuto_ITW_AntiRes}. Then, the solutions to $D^\med_{i,j}=0$ contain no direct information about the resurgence cancellations, and they yield the Borel-summed trans-series for the systems under consideration.

Compared to the analysis in Section~\ref{Section: Trans-seriesDirect}, considering the median QC provides us a direct route to the \textit{exact} non-perturbative corrections to the spectrum. In this way, we can uncover the physical properties of the ITW system across all parameter regimes, including the fugacities (or actions) of the bounce and bion configurations. This is especially crucial for the PT-symmetric case, as there is an exceptional point. In the following, using the median QC, we will determine the exact transition point and the physical properties at the exceptional point and its immediate proximity in the language of semi-classical building blocks, i.e., a bounce and a (half-)bion. We also provide an equivalent analysis for the resonance and anti-resonance cases. 

In addition to that, we will also revisit the resurgence cancellation using the Alien calculus techniques. In this way, we will be able to specify the resurgence cancellation from the first principles and improve the analysis in Section~\ref{Section: semi-classical_ITW} and Section~\ref{Section: Trans-seriesDirect}, where we deduce the cancellations using the numerical analysis and consistency relationships based on it. Finally, since the Alien calculus does not need quantitative output of the perturbative expansions, we will also extend confirmation of the resurgence cancellations beyond the leading order.

\subsection{PT-symmetric case}\label{Section: PTsymmetric_median}

In the PT-symmetric case, we consider the median QC $D_\mrmPT = D^\med_{1,2} = \mfrS_A^{\pm 1/2} T_{1,2}^{(\pm)}$ and express it as\footnote{Throughout this Section, the actions $\Pi_{A,B}$ are in fact different from their original definitions in~\eqref{Dictionary_Perturbative} and \eqref{Dictionary_NonPerturbative} by a half-Stokes automorphism, and they should be written as $\msfS_\pm \left[\mfrS_A^{\pm 1/2} \Pi_{A}\right]$ and $\msfS \left[\Pi_{B_{1,2}}\right]$. However, to simplify the expressions, we continue to use the notation introduced in Section~\ref{Section: BriefIntro_EWKB}.}
\begin{equation}
	D_\mrmPT =\O \left\{\sqrt{1+\Pi_{B_1}}\sqrt{1+\Pi_{B_2}}\left(1+\Pi_A^2\right) + \Pi_A \left(2 + \Pi_{B_1}\right) \right\} =0 \label{MedianQC_PT}\, . 
\end{equation}
Then, we find the solution of the quadratic equation for $\Pi_A$ as
\begin{equation}
	\Pi_A = -\frac{1+ \frac{1}{2}\Pi_{B_1} +  \ve \sqrt{\Disc_{\mrmPT}}}{ \sqrt{1+\Pi_{B_1}}\sqrt{1+\Pi_{B_2}}}\, , \label{ExactSoln_PT}
\end{equation}
where $\ve = \pm 1$ indicates the breaking of the degenerate perturbative states as before, and 
\begin{equation*}
	\Disc_{\mrmPT} = \frac{\Pi_{B_1}^2}{4} - \Pi_{B_2} \left(1 + \Pi_{B_1}\right)\, ,
\end{equation*}
which was originally defined in~\eqref{discriminantPT}. 

To relate \eqref{ExactSoln_PT}, we utilize the Weber-type expressions as before. We first start by identifying the perturbative action as 
\begin{equation}
	\Pi_A = e^{-i J_0}\, , 
\end{equation}
where $J_0$ in the exponent is related to the standard Weber-type perturbative action as
\begin{equation}
	J_0 = 2\pi \mcalF_0 \, , \label{medianAction}
\end{equation}
where $\mcalF_0 = \msfS_\pm \left[\mfrS^{\pm 1/2}_\hbar \mcalF\right]$ as introduced in Section~\ref{Section: Spectrum_EWKB}. This means that the action $J_0$ is also a quantity without a Stokes discontinuity. The associated resurgence terms are related by the Alien calculus.  Then, separating $J_0$ into perturbative and non-perturbative parts as $J_0 = J_\mrmP + J_\mrmNP$, we re-write \eqref{ExactSoln_PT} as
\begin{equation}
	e^{-i J_\mrmNP} = \frac{1+ \frac{1}{2}\Pi_{B_1} +  \ve \sqrt{\Disc_{\mrmPT}}}{ \sqrt{1+\Pi_{B_1}}\sqrt{1+\Pi_{B_2}}} \, , \label{ExactSoln_PT_NP}
\end{equation}
where we use $J_\mrmP = 2\pi \left(N+\frac{1}{2}\right)$ via the perturbative quantization.

A more convenient way to express the solution is
\begin{equation}
	J_\mrmNP = i \log \mcalY\Big[\Pi_{B_1}(J),\Pi_{B_2}(J)\Big]\, ,  \label{ExactNP_PT}
\end{equation}
with 
\begin{equation}
	\mcalY\left(\Pi_{B_1},\Pi_{B_2}\right) = \frac{1+ \frac{1}{2}\Pi_{B_1} +  \ve \sqrt{\Disc_{\mrmPT}}}{ \sqrt{1+\Pi_{B_1}}\sqrt{1+\Pi_{B_2}}} \label{ExactNP_PT_func} \, . 
\end{equation}
The compact expression in~\eqref{ExactNP_PT} involves the \textit{entire} information about the non-perturbative spectrum after the Borel summations throughout the trans-series and associated resurgence cancellations. This is a very strong statement that demonstrates the capacity of the current approach with respect to the asymptotic version in Section~\ref{Section: Trans-seriesDirect}. If one wants to obtain the semi-classical structure\footnote{We note that the connection to the semi-classical structure in the language of bounce and bion events is more apparent in Section~\ref{Section: Trans-seriesDirect} as the ambiguities arise in the same way with a corresponding path integral analysis. This explicit link becomes obscure in the median QC and its Alien derivatives. However, the semi-classical information is traded with the manifestation of median summed spectrum and the resurgence structure.} in the asymptotic limits, it is possible to tackle \eqref{ExactNP_PT} in the same way we explain in Section~\ref{Section: Trans-seriesDirect}. Then, the result would correspond to the Borel-summed spectrum of the PT-symmetric system, and we would recover the expressions~\eqref{PT_NP_BorelSummed_Unbroken} and \eqref{PT_NP_BorelSummed_Broken}. 

Later in this subsection, we will revisit the asymptotic limits of the median summed spectrum and the corresponding resurgence structure. Before that, we will unlock the full capability of the solution~\eqref{ExactNP_PT} by analyzing the PT-symmetry breaking without any approximation and proving the exact conditions for the broken and unbroken phases. Then, having the exact conditions will allow us to probe the spectrum at the exceptional points, again without any approximation.


\subsubsection{Exact conditions for broken and unbroken phases}\label{Section: Condition_PT_Breaking}
To investigate the transition point, we first recall that the median-summed trans-series of the spectrum~\eqref{Spectrum_TransSeries_Alien} at $\nu = 0$. Then, after setting $2\pi \mcalF_0 = J_0=J_\mrmP+ J_\mrmNP$, we can introduce a Taylor expansion to $E^{(0)}(J_0)$ as
\begin{equation}
	E^{(0)}(J_0) = E^{(0)}(J_\mrmP) + \sum_{k=1}^{\infty} \frac{J^k_\mrmNP}{k!} \left(\frac{\mrmd^k E^{(0)}}{\mrmd J^k}\right)_{J=J_\mrmP} \, , \label{Spectrum_AlienVersion}
\end{equation}
where the first term corresponds to $\mcalS_{0}\left[E_\mrmP\right]$. In some sense, $E^{(0)}(J_\mrmP)$ is a remainder of the Borel-summation process after the resurgence cancellations, and it is a manifestly real quantity, as the original asymptotic series is real. The same also holds for the derivative terms when they are evaluated at $J=J_\mrmP$. Then, the exact spectrum $E^{(0)}(J_0)$ can only acquire imaginary contributions from the $J_\mrmNP$ term.  

Recall that our analysis in Section~\ref{Section: Trans-seriesDirect} showed that the breaking of the PT-symmetry is controlled by the fugacity of bounce and bion configurations, which we defined in~\eqref{ITW_fugacities}. More precisely, we showed that asymptotically $\mcalK_1 \ll \mcalK_2$ and $\mcalK_2 \ll \mcalK_1$ conditions control the sign of the discriminant \eqref{discriminantPT}, which we associated to the PT-symmetry breaking condition. Since now we have a closed median-summed expression for the entire non-perturbative spectrum in \eqref{ExactSoln_PT_NP}, we can examine the reality of the spectrum in all parameter regions, including $\mcalK_1 \sim \mcalK_2$ and even $\mcalK_1 = \mcalK_2$. 

In the following, to uncover the necessary conditions for the reality of the spectrum, we analyze~\eqref{ExactSoln_PT_NP} for three cases: $\Disc_{\mrmPT}<0$, $\Disc_{\mrmPT}>0$, and $\Disc_{\mrmPT}=0$. We prove that the sign of $\Disc_{\mrmPT}$ indeed determines if the PT-symmetry is broken or not, and that its zero gives the condition for the exceptional point. 

\begin{itemize}[wide]
\item \textit{\underline{${ \Disc_\mrmPT < 0}$ (Unbroken phase)}:} Since the square-root in \eqref{ExactSoln_PT_NP} produces an imaginary term, assuming $J_\mrmNP$ is real, we split the real and imaginary parts of the equation as
\begin{align}
	\cos\left(J_\mrmNP\right) &= \frac{1 + \frac{\Pi_{B_1}}{2}}{\sqrt{1+\Pi_{B_1}}\sqrt{1+\Pi_{B_2}}} \, , \label{unbroken_real} \\ 
	\sin\left(J_\mrmNP\right) &= \ve \frac{\sqrt{\Pi_{B_2}\left(1+\Pi_{B_1}\right) - \frac{\Pi_{B_1}^2}{4}}}{\sqrt{1+\Pi_{B_1}}\sqrt{1+\Pi_{B_2}}}\, . \label{unbroken_imaginary}
\end{align}
The first equation simply implies
\begin{equation}
	J_\mrmNP = \arccos\left[\frac{1 + \frac{\Pi_{B_1}}{2}}{\sqrt{1+\Pi_{B_1}}\sqrt{1+\Pi_{B_2}}}\right] \, . 
\end{equation}
In return, using this solution in \eqref{unbroken_imaginary}, we obtain
\begin{equation}
	\sin\big[\arccos\left(\mathcal{X} \right)\big] = \sqrt{1 - \mcalX^2} = \ve \frac{\sqrt{\Pi_{B_2}\left(1+\Pi_{B_1}\right) - \frac{\Pi_{B_1}^2}{4}}}{\sqrt{1+\Pi_{B_1}}\sqrt{1+\Pi_{B_2}}}\, ,
\end{equation}
where we set \[\mcalX = \frac{1 + \frac{\Pi_{B_1}}{2}}{\sqrt{1+\Pi_{B_1}}\sqrt{1+\Pi_{B_2}}}\,. \] This shows~\eqref{unbroken_real} and \eqref{unbroken_imaginary} are consistent with each other. Note that even if we allow $J_\mrmNP$ to be complex such that $e^{-i J_\mrmNP} = e^{-i\Re\left(J_\mrmNP\right)} e^{\Im\left(J_\mrmNP\right)} $, the same procedure yields 
\begin{equation}
	\sin\left[\arccos\left(e^{\Im\left(J_\mrmNP\right)}\mcalX\right)\right] = \sqrt{1 - e^{2\Im J_\mrmNP}\mcalX} = \ve e^{2\Im J_\mrmNP} \frac{\sqrt{\Pi_{B_2}\left(1+\Pi_{B_1}\right) - \frac{\Pi_{B_1}^2}{4}}}{\sqrt{1+\Pi_{B_1}}\sqrt{1+\Pi_{B_2}}}\, ,
\end{equation}
which forces us to set $\Im\left(J_\mrmNP\right)=0$ for consistency. As a result, we conclude that when $\Disc_\mrmPT < 0$, the \textit{exact} spectrum is real and the quantum system still has \textit{unbroken} PT-symmetry.

\item \textit{\underline{${ \Disc_\mrmPT > 0}$  (Broken phase)}:} Let us follow the same procedure when the square-root term in \eqref{ExactSoln_PT_NP} is real. Then, for $J_\mrmNP \in \mbbR$, the real and imaginary parts become
\begin{align}
	\cos\left(J_\mrmNP\right) &= 
	\mcalY\left(\Pi_{B_1},\Pi_{B_2}\right) \, , \label{broken_real} \\ 
	\sin\left(J_\mrmNP\right) &= 0\, . \label{broken_imaginary}
\end{align}
The first equation gives
\begin{equation}
	J_\mrmNP = \arccos\Big[\mcalY\left(\Pi_{B_1},\Pi_{B_2}\right) \Big]\,,
\end{equation}
and inserting it into the second one leads to
\begin{equation}
	\sin\big[\arccos\left(\mcalY \right)\big] = \sqrt{1-\mcalY^2} = \sqrt{2}\sqrt{\frac{\sqrt{\Disc_{\mrmPT}}}{\ve \left(1 + \frac{\Pi_{B_1}}{2}\right) + \sqrt{\Disc_{\mrmPT}}}}\, .
\end{equation}
This is manifestly a non-zero quantity except when $\Disc_{\mrmPT}=0$. Therefore, we conclude that $\Disc_{\mrmPT}>0$ corresponds to a \textit{broken} PT-symmetric phase of the quantum theory and $\Disc_\mrmPT = 0$ gives the \textit{exact} condition for the exceptional point of the PT-symmetric system, which we analyze now.

\item \textit{\underline{${ \Disc_\mrmPT = 0}$ (Exceptional point)}:} In order to uncover the reality of the spectrum for $\Disc_\mrmPT = 0$, we follow the same prescription. In this case, the real and imaginary parts of~\eqref{ExactSoln_PT_NP} are written as
\begin{align}
	\cos\left(J_\mrmNP \right) &= \frac{1 + \frac{\Pi_{B_1}}{2}}{\sqrt{1+\Pi_{B_1}}\sqrt{1+\Pi_{B_2}}} =: \mcalZ\, , \label{exceptional_real} \\ 
	\sin \left(J_\mrmNP\right) &= 0 \label{exceptional_imaginary} \, .
\end{align}
Then, the same procedure yields
\begin{equation}
	\sin\Big[\arccos\left(\mcalZ\right)\Big] = \sqrt{1-\mcalZ^2} = \frac{\sqrt{-\Disc_\mrmPT}}{\sqrt{1+\Pi_{B_1}}\sqrt{1+\Pi_{B_2}}} = 0 \, .
\end{equation}
As a result, we observe that at the transition point, the spectrum is still real. 
Note that $\Disc_\mrmPT = 0$ yields a simple relation between the actions of the $B_1$- and $B_2$-cycles:
\begin{equation}
	\Pi_{B_2} = \frac{\Pi_{B_1}^2}{4\left(1+\Pi_{B_1}\right)}\, . \label{ExceptionalPoint_Condition}
\end{equation}
This is the exact condition for the \textit{exceptional point}. Then, the exact conditions for unbroken and broken phases are written as 
\begin{align}
	\mathrm{\underline{Unbroken\, phase}:}&\qquad \Pi_{B_2}\left(1+\Pi_{B_1}\right) > \frac{\Pi_{B_1}^2}{4} \, , \label{Condition_Unbroken} \\ 
	\mathrm{\underline{Broken\, phase}:}&\qquad \frac{ \Pi_{B_1}^2}{4} > \Pi_{B_2}\left(1+\Pi_{B_1}\right) \, , \label{Condition_Broken}
\end{align}
which are perfectly aligned with our numerical and approximate analytic analysis in Section~\ref{Section: Trans-seriesDirect}.
\end{itemize}

\subsubsection{Borel-summed spectrum}
As we discussed above, the exact solution in~\eqref{ExactNP_PT} is for any $x_0$. 
In the following, we first analyze the spectrum around the exceptional point and examine the transition. Then, we revisit the asymptotic limits and verify our analysis in Section~\ref{Section: Trans-seriesDirect}.

\begin{itemize}[wide]
\item 
\textit{\underline{Around exceptional point}:} Using the condition \eqref{ExceptionalPoint_Condition}, we simplify~\eqref{ExactNP_PT_func} at the exceptional point as
\begin{equation}
	\mcalY\left(\Pi_{B_1},\Pi_{B_2}\right) = 1 \, .
\end{equation}
This points out that the exact median-summed non-perturbative correction to the spectrum, i.e., $E^{(0)}$ in~\eqref{Spectrum_AlienVersion} vanishes exactly as
\begin{equation}
	J_\mrmNP^\mathrm{EP} = 0\, .\label{NP_corrections_EP}
\end{equation}


We emphasize that~\eqref{NP_corrections_EP} should not be interpreted as the disappearance of resurgence itself. What vanishes at the exceptional point is the exact median-summed non-perturbative correction to the spectrum, not the non-perturbative sectors or the resurgent structure encoded in the exact WKB data. Indeed,~\eqref{NP_corrections_EP} does not exclude the presence of bounce and bion configurations, as well as their combinations, which exist for any $x_0$. As we show shortly, the resurgence cancellations occur at all orders in the trans-series in the unbroken and broken phases, as well as at the exceptional point. Recalling our discussion in Section~\ref{Section: semi-classical_ITW}, this is necessary since, for any $x_0$, the perturbative expansions around both wells are divergent and non-Borel summable, and therefore ambiguous unless their ambiguities are canceled by the non-perturbative sectors.

The vanishing $J_\mrmNP^\mathrm{EP}$ in~\eqref{NP_corrections_EP}, on the other hand, shows a remarkable fact: The entire non-perturbative part is organized in such a way that at the exceptional point, the parts that do not participate in the resurgence cancellations cancel each other exactly. We also emphasize that this result needs an exact treatment of the spectrum, and, to our knowledge, the current EWKB approach that utilizes the Alien calculus is the only way to access this information with absolute precision.

Recall that around the two wells of the potential \eqref{ITW_potential}, the perturbative expansions are the same. Therefore, \eqref{NP_corrections_EP} yields coalescence of eigenvalues at $x_0=x_\mrmcr$. Furthermore, as we stated in Section~\ref{Section: Setup_Numerical_ITW}, there is only one independent solution for each eigenvalue, showing that $x_0=x_\mrmcr$ is an exceptional point. Let us finally show that $x_0=x_\mrmcr$ is really a branch point of the eigenvalues, as is expected from an exceptional point.

For this purpose, we utilize the exact exceptional point condition \eqref{ExceptionalPoint_Condition} and the vanishing non-perturbative corrections to analyze the spectrum around the critical point. 
First, let us recall the solution for the median summed non-perturbative correction in \eqref{ExactNP_PT} and \eqref{ExactNP_PT_func}.~Since the critical points $x_0 = x_\mrmcr$ are given by $\Disc_{\mrmPT}=0$, we expand $J_\mrmNP$ for $\Disc_\mrmPT \ll 1$ and for $x_0 - x_\mrmcr \ll 1$ subsequently as
\begin{align}
	J_\mrmNP &\simeq i\ve \frac{\sqrt{\Disc_{\mrmPT}}}{1+\frac{1}{2}\Pi_{B_1}} + O\left(\Disc_{\mrmPT}^{3/2}\right) \nonumber \\
	& \simeq \frac{2i\ve}{2 + \Pi_{B_1}(x_\mrmcr)} \left(  \Disc'_\mrmPT\right)^{1/2}_{x_0=x_\mrmcr} \sqrt{x_0 - x_\mrmcr}+ O\left((x_0 - x_\mrmcr)^{3/2}\right) \, , \label{squareRoot_NP_action}
\end{align}
where $\Disc'_\mrmPT(x) = \frac{\mrmd\, \Disc_\mrmPT}{\mrmd x}$ and we used $\Disc_{\mrmPT}(x_\mrmcr) =0$ at the critical point. Finally, the square-root behavior of the spectrum at the critical points directly follows from the median summed trans-series expression in \eqref{Spectrum_AlienVersion}, which becomes
\begin{equation}
	E^{(0)}(J_0) \simeq E^{(0)}(J_\mrmP) + \frac{2i\ve}{2 + \Pi_{B_1}(x_\mrmcr)}\left(\frac{\mrmd E^{(0)}}{\mrmd J}\right)_{J=J_\mrmP} \left(\Disc'_\mrmPT \right)^{1/2}_{x_0=x_\mrmcr} \sqrt{x_0 - x_\mrmcr} \, + \, O\left((x_0-x_\mrmcr)^{3/2}\right) \, .  \label{squareRoot_Spectrum}
\end{equation}
Provided that $\Disc'_\mrmPT$ doesn't vanish, \eqref{squareRoot_Spectrum} shows the double-valued branch-cut structure around $x_0=x_\mrmcr$ as predicted\footnote{Let us also comment on the possibility that $\Disc'_\mrmPT$ vanishes. By expanding to higher orders, we observe that all the non-vanishing corrections in \eqref{squareRoot_NP_action} appear at half-integer orders, i.e. $(x_0-x_\mrmcr)^{\frac{2n+1}{2}}$. Therefore, even if the derivative vanishes, the two-sheeted branching around the exceptional point would remain. In such a case, the first non-vanishing half-integer term would govern the eigenvalue splitting and determine how fast the eigenvalues approach each other.} in Fig.~\ref{Figure: PTsymmetric_Eigenvalues}; thus, verifies their exceptional point character. 


To further analyse the PT-symmetry breaking quantitatively, we now focus on the below-barrier-top sector. We will first estimate the numerical values of the critical points for individual states in this sector. Then, we will verify that they correspond to the simple zeros of the discriminant $\Disc_\mrmPT$, verifying the exceptional point character of the PT-symmetry breaking points.

Recall that in the below-barrier-top sector, the $B_1$- and $B_2$-cycles are represented by \eqref{Dictionary_NonPerturbative}. Then, we rewrite \eqref{ExceptionalPoint_Condition} as
\begin{equation}
	\frac{e^{-\frac{S_1}{\hbar}}}{4} = e^{-\frac{S_2}{\hbar}}\left[ 1 + \frac{\sqrt{2\pi} e^{-\frac{S_1}{\hbar}}}{\Gamma\left(1+ N\right)} \left(\frac{\mcalC}{\hbar}\right)^{\frac{1}{2} + N} \right]\, , \label{ExceptionalPoints_ActionCondition}
\end{equation}
where $S_{1,2}$ and $\mcalC$ are controlled by the parameter $x_0$ and are given in \eqref{BounceAction_ITW}, \eqref{BionAction_ITW}, and \eqref{One_LoopDeterminant}, respectively. 

The first two numerical solutions of~\eqref{ExceptionalPoints_ActionCondition} are 
\begin{equation}
	x_\mrmcr = \left\{1.08213 , 1.17943 \right\}\, . \label{EP_SemiNumerical}
\end{equation} 
These are the critical points for the pairs which correspond to the perturbative ground states and perturbative first excited states around the two wells of the ITW potential~\eqref{ITW_potential}. As we discussed in Section~\ref{Section: Trans-seriesDirect}, the non-perturbative effects break this degeneracy in the unbroken phase. Then, the eigenvalues of these states coalesce at their respective critical points in~\eqref{EP_SemiNumerical} and the real parts of their eigenvalues become degenerate for $x_0>x_0^\mrmcr$ in the broken phase. This picture and the solutions in \eqref{EP_SemiNumerical} perfectly match the numerical analysis depicted in Fig.~\ref{Figure: PTsymmetric_Eigenvalues}.

For the excited states for $N \geq 2$, their PT-symmetries are broken when they lie above the barrier top. Let us verify this is indeed the case by investigating the transition of each state through the barrier top region individually: 

Since we are still at the exceptional point, \eqref{ExactSoln_PT} still simplifies to $\Pi_A=-1$ via~\eqref{ExceptionalPoint_Condition}. However, around the barrier top, i.e., $u\simeq u^\mrmBT$, both $A$- and $B$-cycles have perturbative characters, meaning that $\Pi_A \sim \Pi_B \sim O(1)$. At the leading order, the action of the $A$-cycle reads
\begin{equation}
	a(u\sim u^\mrmBT) \simeq  \frac{S_3}{\hbar}\, , \label{actionBarrierTop}
\end{equation}
where 
\begin{equation}
	S_3(x_0) = \frac{\left(2+x_0^2\right)\sqrt{\left(8+x_0^2\right)\left(4-x_0^2\right)}}{2\sqrt{3}} + 4\left(x_0^2 - 1\right)\log\left[\frac{\sqrt{8+x_0^2} - \sqrt{3}\sqrt{4-x_0^2}}{2\sqrt{x_0^2 - 1}} \right] \label{ITWdual_Bounce}
\end{equation} 
is the bounce action\footnote{Note that we still use the name bounce since the $A$-cycle at $u=u_\mrmBT$ is intimately related to the bounce configuration for the tilted-double well potential, which is obtained via $V_\mathrm{TTW} = u_\mrmBT -V_\mrmITW$. This correspondence is simply due to the duality between two potentials and their WKB-cycles. We refer~\cite{Misumi:2024gtf,Misumi:2025ijd} for detailed discussion and several other examples.}
at the barrier top. As a result, the leading-order solution to~\eqref{ExactSoln_PT} at $u = u_\mrmBT $ becomes
\begin{equation}
	S_3(x_0^\star)  \simeq N + \frac{1}{2}\, . \label{EP_BarrierTop}
\end{equation}
This is the approximate condition for the quantized states $N=0,1,\dots$ to lie around the barrier top region. 

Since at the exceptional point, $\Pi_A = -1$ is the exact QC, $a = N+\frac{1}{2}$ becomes an exact relation at $x_0 = x_\mrmcr$. Then, comparing $x_\mrmcr$ satisfying \eqref{ExceptionalPoints_ActionCondition} and $x_0^\star$ satisfying \eqref{EP_BarrierTop}, we can determine which PT-symmetry-breaking transition takes place while the states lie below the barrier-top sector. More specifically, we compare the solutions to~\eqref{ExceptionalPoints_ActionCondition} with the curve of $S_3(x_0)$ for various values of $N$. If a point $(x_\mrmcr,N)$ lies below the curve, it means the transition occurs in the below-barrier top sector. Otherwise, those solutions for~\eqref{ExceptionalPoints_ActionCondition} become invalid.


In Fig.~\ref{Figure: ExceptionalPoints_BarrierTop}, we compare \eqref{ITWdual_Bounce} with the critical points in \eqref{EP_SemiNumerical} and next two solutions to \eqref{ExceptionalPoints_ActionCondition}, i.e., $\{1.26167,1.32118\}$. It shows only the first two PT-symmetry breakings occur below the barrier top, where~\eqref{ExceptionalPoints_ActionCondition} is applicable. Despite this limitation, \eqref{ExceptionalPoints_ActionCondition} stands as a remarkably simple condition for the exceptional point in terms of the bion and bounce actions only.
\begin{figure}
	\centering
	\includegraphics[width=0.6\textwidth]{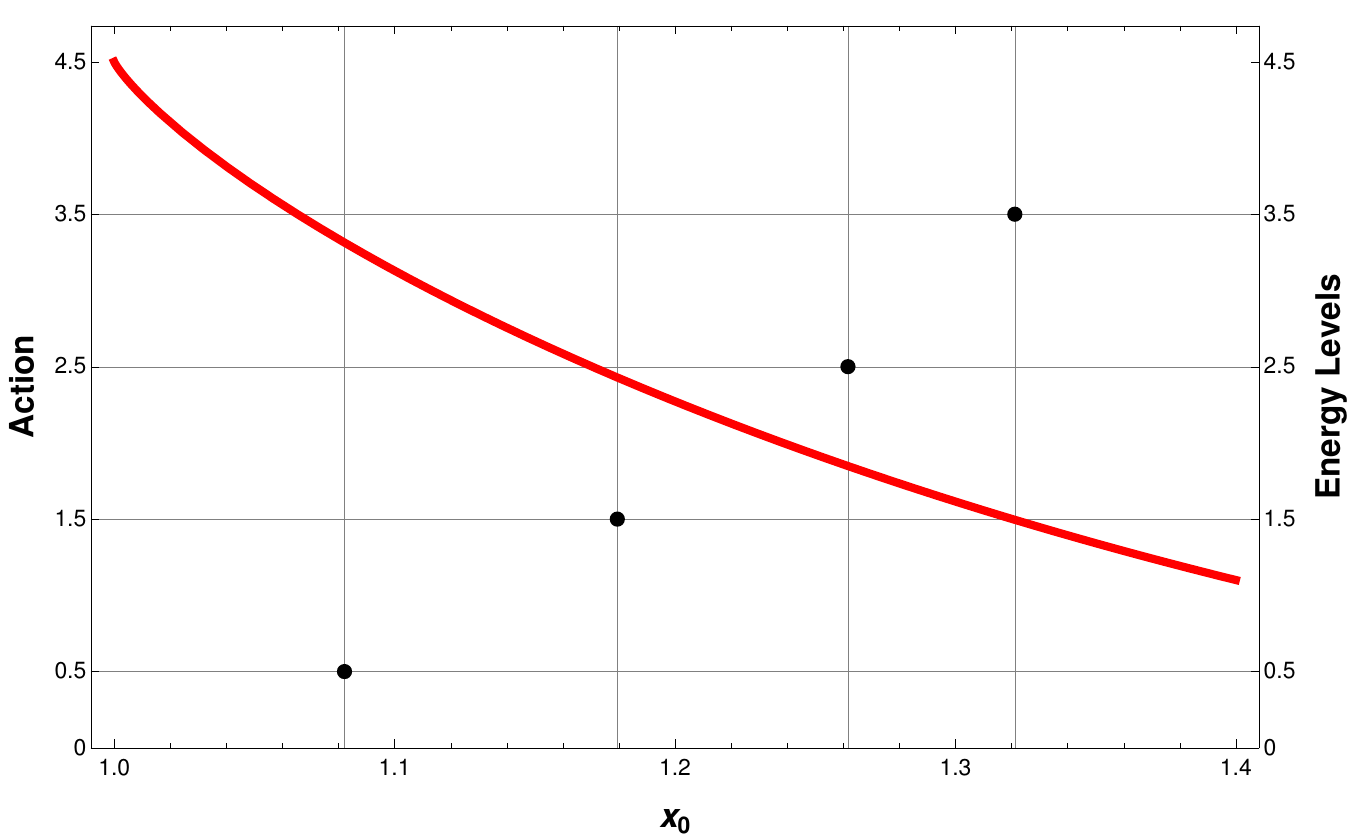}
	\caption{Comparison between the barrier-top condition $S_3(x_0)=N+\tfrac{1}{2}$ and the exceptional-point condition obtained from~\eqref{ExceptionalPoints_ActionCondition}. The {\color{red} red} curve shows $S_3(x_0)$, the horizontal lines denote $N+\tfrac{1}{2}$ for energy levels $N=0,1,2,3$, and the black dots mark the corresponding critical values $x_0^{\rm cr}$. Only the first two intersections lie in the below-barrier regime where~\eqref{ExceptionalPoints_ActionCondition} can be applied.} \label{Figure: ExceptionalPoints_BarrierTop}
\end{figure}

\paragraph{\underline{Remark 1}:} Note that although the specific quantitative forms for $\Pi_A$ and $\Pi_{B_{1,2}}$ in $\Disc_{\mrmPT}$ changes around the barrier top region and above it, the conditions~\eqref{Condition_Unbroken} and \eqref{Condition_Broken} are in fact valid for any parameter region and any classical energy values. This is because of the fact that the transition across a barrier top is a continuous one and that the median QC in both below and above barrier-top sectors is identical (see~\cite{Misumi:2025ijd} for a similar discussion for triple-well potential). Therefore, the proof we provide in Section~\ref{Section: Condition_PT_Breaking} valid for any $x_0\in\left(-2,2\right)$ and $E$ values.

\paragraph{\underline{Remark 2}:} Our analysis in this subsection and Section~\ref{Section: Condition_PT_Breaking} provides an excellent display of the power of the EWKB analysis, particularly with the proper utilization of the median QC via Stokes automorphisms. On the other hand, it can also be interpreted as a possibly unreasonable power of the semi-classical analysis. After all, we showed that the exact exceptional point is simply given by the fugacities of the semi-classical bounce and bion configurations. Moreover, in both exact and semi-classical analysis, we presented that the imaginary contributions arising from the same square-root factor, i.e.~$\sqrt{\Disc_{\mrmPT}}$, which indicates that the PT-symmetry breaking for the ITW potential is due to the competition between the classical actions of bounce and half-bion configurations. This means that the higher-order corrections to the trans-series do not alter this picture at all. 

A possible general reason behind this is the cluster expansion in the semi-classical picture, which was analyzed thoroughly in various setups~\cite{Behtash:2018voa,Pazarbasi:2021ifb,Pazarbasi:2021fey} and briefly discussed in the EWKB framework recently~\cite{Misumi:2025ijd}. In a non-Hermitian setup, such as that we discuss in this paper, the cluster expansion of the semi-classical building block has never been investigated. It would be interesting to uncover if it exists or not and pursue its implications for the PT-symmetry breaking.

\item \textit{\underline{Asymptotic limits}:} Let us now revisit the limiting cases of $\Pi_{B_2}\ll \Pi_{B_1}^2$ and $\Pi_{B_1}^2 \ll \Pi_{B_2}$ in a comparison with the analysis in Section~\ref{Section: Trans-seriesDirect}. Note that since we work on the Stokes rays, we can directly uncover the spectrum that survived after the Borel summation.

As in Section~\ref{Section: Trans-seriesDirect}, we expanded around the perturbative states. Similarly, here we set $J = J_\mrmP + J_\mrmNP$ and assume $J_\mrmNP \ll 1$ due to the exponential suppressions. Since $J_\mrmNP$ is the actual small parameter, we need to expand \eqref{ExactNP_PT} in $J_\mrmNP$ rather than $\Pi_{B_{1,2}}$, which depend on the parameter $J$ themselves, as in Section~\ref{Section: Trans-seriesDirect}. The Alien derivatives to the median-summed contributions~\eqref{ExactNP_PT} will involve the derivatives of $\Pi_{B_{1,2}}$. Therefore, foreseeing such contributions, which we will discuss in Section~\ref{Section: MedianResurgenceStructure_PT}, we define the following expansions:

\begin{itemize}
	\item For bounce configurations: 
	\begin{equation}
		\frac{1}{2}\frac{\mrmd^m \Pi_{B_1}}{\mrmd J^m} = 2\pi \mcalK_1 \sum_{k=0}^{\infty} Q^{(1)}_{k,m}(J_\mrmP)\, J^k_\mrmNP \label{DerivativeNP_expansionB1} \, .
	\end{equation}
	\item For half-bion configurations:
	\begin{align}
		\frac{\mrmd^m \left(\Pi_{B_2}\right)^{1/2}}{\mrmd J^m} &=2\pi  \mcalK_2 \sum_{k=0}^{\infty}  Q^{(2)}_{k,m}(J_\mrmP)\, J^k_\mrmNP\, .\label{DerivativeNP_expansionB2}	
	\end{align}	
\end{itemize}
where $\mcalK_1$ and $\mcalK_2$ are the fugacity terms defined in \eqref{ITW_fugacities}. Note that expansion terms $Q^{(1,2)}_{k,m}$ are simply related to $P_k^{(1,2)}$ by 
\begin{equation}
	Q^{1,2}_{k,m}= \frac{\G(k+m+1)}{\G(k+1)} Q^{(1,2)}_{k+m,0} = \frac{\G(k+m+1)}{\G(k+1)} P_{k+m}^{(1,2)}\, .  \label{relationCoefficients}
\end{equation}
With these relations, it is manifest that $Q_{k,m}^{(1,2)}$, so that the series expansions~\eqref{DerivativeNP_expansionB1} and~\eqref{DerivativeNP_expansionB2} are related to unambiguous parts of the bounce and bion configurations in~\eqref{b1_fugacity} and~\eqref{b2_fugacity}, as expected by construction.

Now, inserting the expansions~\eqref{DerivativeNP_expansionB1} and~\eqref{DerivativeNP_expansionB2} in~\eqref{ExactNP_PT} via \eqref{ExactNP_PT_func}, we can expand~\eqref{ExactNP_PT} for first $J_\mrmNP \ll 1$ and then, for $\mcalK_1\ll \mcalK_2$ or $\mcalK_2\ll\mcalK_1$, depending on the choice. Then, we can solve the resulting expansions order by order for $J_\mrmNP$. We point out that this procedure is similar to the solution scheme in Section~\ref{Section: Trans-seriesDirect}. In the current version, however, the left-hand side involves only one term, i.e.~$J_\mrmNP$, which simplifies the procedure significantly.

Following this procedure for $\mcalK_1 \ll \mcalK_2$ and we obtain the solution up to $O(\mcalK_1^2)$ and $O(\mcalK_2^2)$ order as
\begin{align}
 \frac{1}{2 \pi} J_\mrmPT^\mrmUBr &\simeq - \ve \mcalK_2 \mcalS_0\left[Q_{0,0}^{(2)}\right] + \ve \frac{\mcalK_1^2}{\mcalK_2}\mcalS_0\left[\mathcal{Q}_{\mcalB^2\mcalI^{-1}}\right] + \mcalK_2^2 \mcalS_0\left[Q_{0,0}^{(2)} Q_{1,0}^{(2)}\right] - 2 \mcalK_1^2 \mcalS_0\left[Q_{0,0}^{(1)} Q_{1,0}^{(1)}\right] \nonumber   \\
	& =  - \ve \mcalK_2 \mcalS_0\left[P_0^{(2)}\right] + \ve \frac{\mcalK_1^2}{\mcalK_2}\mcalS_0\left[\mcalP_{\mcalB^2\mcalI^{-1}} \right] + \mcalK_2^2 \mcalS_0\left[P_0^{(2)} P_1^{(2)}\right] - 2 \mcalK_1^2 \mcalS_0\left[P_0^{(1)} P_1^{(1)}\right] 
	 \label{PTmedian_unbroken}
\end{align}
for the unbroken phase. Similarly, same procedure for $\mcalK_2 \ll \mcalK_1$ case, we obtain
\begin{align}
\frac{1}{2 \pi} J_\mrmPT^\mrmBr & \simeq -i\ve \mcalK_1 	\mcalS_0\left[Q^{(1)}_{0,0}\right] + \frac{i\ve}{2} \frac{\mcalK_2^2}{\mcalK_1}\mcalS_0\left[\mathcal{Q}_{\mcalI\mcalIbar\mcalB^{-1}}\right] -2\mcalK_1^2 \mcalS_0\left[Q^{(1)}_{0,0} Q^{(1)}_{1,0} + i \pi \ve \left(Q^{(1)}_{0,0}\right)^2\right] \nonumber \\
& \quad +\mcalK_2^2 \mcalS_0\left[Q^{(2)}_{0,0} Q^{(2)}_{1,0}-i \pi  \ve \left(Q^{(2)}_{0,0}\right)^2\right]\nonumber \\ 
& = i\ve \mcalK_1 	\mcalS_0\left[P^{(1)}_0\right] -\frac{i \ve \mcalK_2^2 }{2 \mcalK_1 }\mcalS_0 \left[\mcalP_{\mcalI\mcalIbar\mcalB^{-1}}\right] -2\mcalK_1^2 \mcalS_0\left[P^{(1)}_0 P^{(1)}_1+ i \pi \ve \left(P^{(1)}_0\right)^2\right] \nonumber \\
&\quad  +\mcalK_2^2 \mcalS_0\left[P^{(2)}_0 P^{(2)}_1-i \pi  \ve 	\left(P^{(2)}_0\right)^2\right]
\label{PTmedian_broken}
\end{align}
for the broken phase. Note that we defined  $\mathcal{Q}_{\mcalB^2\mcalI} = \frac{\left(Q_{0,0}^{(1)}\right)^2}{Q_{0,0}^{(2)}}$ and $\mathcal{Q}_{\mcalI\mcalIbar\mcalB^{-1}} = \frac{\left(Q^{(2)}_{0,0}\right)^2}{Q^{(1)}_{0,0}}$. We also introduced $\frac{1}{2\pi}$ factors as they appear in the final contribution to the spectrum via~\eqref{Spectrum_AlienVersion} and to make the comparison with the results in Section~\ref{Section: Trans-seriesDirect} directly. In this sense, the results in~\eqref{PTmedian_unbroken} and~\eqref{PTmedian_broken} recover the same Borel summed contributions in~\eqref{PT_NP_BorelSummed_Unbroken} and~\eqref{PT_NP_BorelSummed_Broken}, and verify our semi-classical analysis in relation to the bion and bounce configurations.

\end{itemize}

\subsubsection{Resurgence structure (revisited)}\label{Section: MedianResurgenceStructure_PT}
We now turn our focus to the resurgence structure of the spectrum. As discussed in Section~\ref{Section: EWKB_review}, the Alien calculus encodes the ambiguous non-perturbative corrections, which are canceled against the Borel-summed series (both the perturbative sector and instanton fluctuations). For the median QC~\eqref{MedianQC_PT}, using~\eqref{AlienDerivative_Action}, we find the action of the Alien derivative,
\begin{equation}
	D_\mrmPT^{(k \geq 1)}(J) = \ttO(J) \left(\Pi_A(J) + (-1)^k \Pi_A^{(-1)}(J) \right) L^k(J) 
	\, , \label{MedianQC_PT_Alien}
\end{equation}
where 
\begin{equation}
	\ttO(J) = \frac{\sqrt{1+\Pi_{B_1}(J)}\sqrt{1+\Pi_{B_2}(J)}}{\sqrt{\Pi_{B_1}(J)\, \Pi_{B_2}(J)}}\, , \label{MedianQC_PT_prefactor}
\end{equation}
and
\begin{equation}
	L(J) = \log\big[ \left(1+\Pi_{B_1}(J)\right) \left(1+ \Pi_{B_2}(J)\right) \big]\, . \label{functionL}
\end{equation}
Let us also express the median QC \eqref{MedianQC_PT} in this new notation as
\begin{equation}
	D_\mrmPT^{(0)}(J) = \ttO(J) \big(2\cos J + \F(J)  \big)\, ,\label{MedianQC_PT_rearranged}
\end{equation}
where 
\begin{equation}
	\F(J) = \frac{2+\Pi_{B_2}(J)}{\sqrt{1+\Pi_{B_1}(J)}\sqrt{1+\Pi_{B_2}(J)}}\,. 
\end{equation}

In the following, we will only compute the first correction, i.e., $J_1(J_0)$, to the median-summed solution $J_\mrmNP$. Using~\eqref{Alien_EQC_First} and the fact that $D_\mrmPT^{(0)}(J_0) = 0$, the one Alien derivative term is expressed as
\begin{equation}
	J_1(J_0) \left[-2 \sin\left(J_0\right) + \F'(J_0) \right] + 2i \sin(J_0) \, L(J_0) = 0\, ,  \label{PTmedian_firstcorrection}
\end{equation}
where $\F'(J_0) = \frac{\mrmd \F}{\mrmd J}\Big\vert_{J=J_0}$. Then, the formal solution to $J_1(J_0)$ becomes
\begin{equation}
	J_1(J_0) = -i \frac{2\sin (J_0)\,  L(J_0)}{2\sin\left(J_0\right) - \frac{\mrmd \F(J_0)}{\mrmd J}}\, , \label{PTmedian_firstcorrection2}
\end{equation}
provided $2\sin(J_0)\neq \F'(J_0)$.



To relate~\eqref{PTmedian_firstcorrection} to the semi-classical analysis, we expand~\eqref{PTmedian_firstcorrection2} in $J_\mrmNP$ and insert the median-summed non-perturbative corrections~\eqref{PTmedian_unbroken} or~\eqref{PTmedian_broken}. Note that, since the right-hand side of~\eqref{PTmedian_firstcorrection2} is independent of $J_1$, there is no need for an iterative solution procedure. Instead, we directly expand the entire expression for $J=J_\mrmP  + J_\mrmNP$ by using the solutions for $J_\mrmNP^\mrmUBr$ and $J_\mrmNP^\mrmBr$ at the respective limits. 

Up to orders $O(\mcalK_1^2)$ and $O(\mcalK_2^2)$, we obtained
\begin{align}
  \frac{1}{2\cdot2 \pi} J_1^\mrmUBr & \simeq -i \mcalK_1 Q^{(1)}_{0,0}- i \pi \mcalK_2^2 \left(Q^{(2)}_{0,0}\right)^2 + 2\pi  i \mcalK_1^2 \left(Q^{(1)}_{0,0}\right)^2 -i \mcalK_1 \mcalK_2 \left( Q^{(1)}_{1,0} Q^{(2)}_{0,0} + 2\pi  Q^{(1)}_{0,0} Q^{(2)}_{0,1}\right)  \nonumber \\
  & = -i\mcalK_1 P_0^{(1)} -  i\pi \mcalK_2^2 \left(P_0^{(2)}\right)^2 + 2 \pi i \mcalK_1^2 \left(P_0^{(1)}\right)^2 - i \mcalK_1 \mcalK_2 \left(P_1^{(1)} P_0^{(2)} + P_0^{(1)} P_1^{(2)}\right)  \label{PT_J1_unbroken}
\end{align}
for the unbroken phase and 
\begin{align}
\frac{1}{2\cdot 2\pi} J_1^\mrmBr &\simeq  2 \pi  \mcalK_1^2 \left( \ve  Q^{(1)}_{0,0} Q^{(1)}_{0,1}-  i \left(Q^{(1)}_{0,0}\right)^2\right)-i \mcalK_1 Q^{(1)}_{0,0} + \mcalK_2^2 \left(2 \pi \ve  Q^{(2)}_{0,1} Q^{(2)}_{0,0}+3 i \pi \left(Q^{(2)}_{0,0}\right)^2\right)\nonumber \\
&\simeq   2\mcalK_1^2 \left( \ve  P^{(1)}_0 P^{(1)}_1-  \pi i \left(P^{(1)}_0\right)^2\right)-i \mcalK_1 P^{(1)}_0 +  \mcalK_2^2 \left(  \ve  P^{(2)}_{1} P^{(2)}_{0} + 3 i \pi \left(P^{(2)}_{0}\right)^2\right) \nonumber \\  \label{PT_J1_broken} 
\end{align}
for the broken phase. Both expressions recover the parts that appear in the resurgence cancellation in the semi-classical analysis in Section~\ref{Section: Trans-seriesDirect}.

Note that, as we mentioned before, although the general structures of $J_1^\mrmUBr$ and $J_1^\mrmBr$ are quite different, some of their subparts coincide. We have already uncovered this subpart in $\mcalK_1\ll \mcalK_2$ and $\mcalK_2 \ll \mcalK_1$ limits and called it the minimal trans-series [see~\eqref{PT_minimalTransseries}]. Now, let us show its direct link to the perturbative sector.

In the language of Alien calculus, finding the minimal trans-series that emanates from the perturbative sector is quite straightforward. In general, we can turn off the non-perturbative corrections and take the perturbative QC, which is
\begin{equation}
	D_\mathrm{Pert}^{(0)}(J) = 1 + \Pi_A(J) = 0 \label{EQC_perturbative}
\end{equation}
for both wells. Then, the $k$-th Alien derivative becomes
\begin{equation}
	D^{(k\geq 1)}_\mathrm{Pert}(J) = i\Pi_A L^k(J) \, ,
\end{equation}
and setting $J_0=J_\mrmP$, we obtained
\begin{equation}
	J_1^\mathrm{minimal}(J_\mrmP) = -i L(J_\mrmP) = 
	-i \log\big[ \left(1+\Pi_{B_1}(J_\mrmP)\right) \left(1+ \Pi_{B_2}(J_\mrmP)\right) \big] \label{PTmedian_perturbative} \, .
\end{equation}
Then, expanding \eqref{PTmedian_perturbative} for $\mcalK_1 \ll 1$ and $\mcalK_2 \ll 1$, we found the corresponding minimal trans-series as
\begin{equation}
	\frac{1}{2\cdot 2\pi} J_1^\mathrm{\mathrm{minimal}} = -i\left(\mcalK_1 P^{(1)}_{0} +   \pi \mcalK_2^2
	\left(P^{(2)}_{0}\right)^2 - 2\pi \mcalK_1^2 \left(P^{(1)}_{0}\right)^2   \right) \,,\label{PTmedian_minimal}
\end{equation}
which recovers the expansion in~\eqref{PT_minimalTransseries}.~Note that the order of expansion in~\eqref{PTmedian_perturbative} is commutable, indicating that the resulting series~\eqref{PTmedian_minimal} is independent of the quantitative dominance of $\mcalK_1$ or $\mcalK_2$ and that it is valid for all parameter regions. 

Finally, we focus on the resurgence structure at the exceptional point. Since all the non-perturbative corrections vanish in the median-summed trans-series, it is straightforward to deduce that its resurgence structure only consists of the series in~\eqref{PTmedian_minimal}. 
Note that at the exceptional point, 
since $J_\mrmNP=0$, $\Pi_{B_{1,2}}(J_\mrmP)$ are simply given by the respective fugacity terms,
\begin{equation}
	\Pi_{B_i}(J_\mrmP) = \mcalK_i \, .
\end{equation} For completeness, let us compute it as a limit of the general expression of $J_1(J_0)$ for the ITW potential in~\eqref{PTmedian_firstcorrection2}: At the exceptional point, the condition~\eqref{ExceptionalPoint_Condition} leads to $\F=1$. Note that this is independent of the quantization process, regardless of the fact that $J_\mrmNP^\mathrm{EP} =0$. Then, the derivative term in~\eqref{PTmedian_firstcorrection2} vanishes at the exceptional point. Then, using the fact $J_\mrmNP^\mathrm{EP} = 0$, \eqref{PTmedian_firstcorrection2} simplifies to
\begin{equation}
	J_1^\mathrm{EP}(J_0) = -i L(J_0)  \label{PTmedian_correctionEP_1} \, ,
\end{equation}
which induces the minimal series after expanding for $\mcalK_1 \ll 1$ and $\mcalK_2\ll 1$. 


\subsection{Resonance and anti-resonance cases}\label{Section: R_AR_median}

Finally, considering the resonance and anti-resonance cases, we will solve them with the median QC exactly to obtain the median-summed spectrum. Then, we also revisit the semi-classical structure of the spectrum and the associated structure. Our discussion follows the same route as the previous analysis of the PT-symmetric system in Section~\ref{Section: PTsymmetric_median}. Therefore, in the following, we do not elaborate on the details, unless it is different from Section~\ref{Section: PTsymmetric_median}.

\paragraph{\underline{Exact solutions}:}As we discussed at the level of QCs in Section~\ref{Section: EWKB_Setup_ITW} and the trans-series solutions in Section~\ref{Section: R_AR_Direct}, the resonance and anti-resonance systems are time-reversed partners of each other, which reveals itself via complex conjugation. The complex conjugation also appears for their median QCs, which are defined as
\begin{equation}
	D_\med^{(\mrmR)} = \mfrS_A^{\pm 1/2} T_{2,2}^{(\pm)}\, , \qquad D_\med^{(\mrmAR)} = \mfrS_A^{\pm 1/2} T_{1,1}^{(\mrmR)}\, .\label{medianQC_R_AR_definitions}
\end{equation}
Using the expressions in Table~\ref{Table: ExactQuantizationCondition_ITW}, we obtain the median QCs as
\begin{align}
	D_\med^{(\mrmR)} & = \O\left\{\left(1+\Pi_{B_1}\right)\sqrt{1+\Pi_{B_2}} + \Pi_{A}^2 \sqrt{1+\Pi_{B_2}} + 2\Pi_A \sqrt{1+\Pi_{B_1}} \right\} =0\, , \label{MedianQC_Res}
\end{align}
and 
\begin{align}
	D_\med^{(\mrmAR)} =  \O\left\{\sqrt{1+ \Pi_{B_2}} + 2\Pi_A \sqrt{1+\Pi_{B_1}} + \Pi_A^2 \sqrt{1+\Pi_{B_2}}\left(1+\Pi_{B_1}\right) \right\} =0\, .\label{MedianQC_AR}
\end{align}
Then, it is manifest that $D_\med^{(\mrmAR)} = \mathsf{C} D_\med^{(\mrmR)}$. 

To see how this relation is carried to the entire spectrum, let us rewrite~\eqref{MedianQC_Res} as
\begin{equation}
D_\med^{(\mrmR)} = \O\Pi_A^2 \sqrt{1+\Pi_{B_2}}\left(1+\Pi_{B_1}+ 2\Pi_A^{-1} \sqrt{1+\Pi_{B_1}} +\left(\Pi_A^{-2}(1+ \Pi_{B_1})\sqrt{1+ \Pi_{B_2}}  \right)\right\}\, .
\end{equation}
Then, defining $D_\med^{(+)} = D_\med^{(\mrmR)}$ and $D_\med^{(-)} = D_\med^{(\mrmAR)}$, we express both the median QCs in a compact form as
\begin{equation}
	D_\med^{(\pm)} = \Pi_A^{\mp 2} \left(1 + \Pi_{B_1}\right)\sqrt{1+\Pi_{B_2}} + 2 \Pi_A^{\mp 1}\sqrt{1+\Pi_{B_1}} + \sqrt{1+ \Pi_{B_2}} = 0 \, . \label{median_R_AR_both}
\end{equation}

As in Section~\ref{Section: PTsymmetric_median}, these quadratic equations can be solved exactly, and we get
\begin{equation}
	\Pi_{A}^{\mp 1} = \frac{-1 + i \ve \sqrt{\Pi_{B_2}}}{\sqrt{1+ \Pi_{B_1}}  \sqrt{1 + \Pi_{B_2}} }\, , \label{ExactSoln_R_AR}
\end{equation}
where the power $\mp 1$ on the left-hand side refers to the solutions for $D_\med^{(\pm)} =0$, respectively. Finally, recalling the definition $\Pi_A = e^{-i J_0}$, where $J_0 = 2\pi \mcalF_0 = 2\pi \msfS_\pm\left[\mfrS_{\hbar}^{\pm1/2} \mcalF\right]$, we found 
\begin{equation}
	J_0^{(\pm)} = \mp i \log\left[\frac{-1 + i \ve \sqrt{\Pi_{B_2}}}{\sqrt{1+ \Pi_{B_1}} \sqrt{1+ \Pi_{B_2}}}\right] \, ,\label{ExactSoln_R_AR2}
\end{equation}
which can be used to compute the exact trans-series via~\eqref{Spectrum_TransSeries_Alien}. Now, from~\eqref{ExactSoln_R_AR2}, it is manifest that the entire median-summed spectra of the resonance and anti-resonance systems are complex conjugate to each other. We note that $\ve=\pm 1$ in~\eqref{ExactSoln_R_AR2} does not affect this property, as it leads to complex conjugate contributions to $J_0^{(\pm)}$ in both cases.

Since the perturbative spectrum for any boundary condition for the ITW potential~\eqref{ITW_potential} is the same, the real difference takes place at the non-perturbative level. Therefore, it is also appropriate to uncover the exact non-perturbative solution by separating $J_0 = J_\mrmP + J_\mrmNP$. Then, using the perturbative QC, i.e., $e^{\pm i J_\mrmP} = -1$, we find the exact non-perturbative part as
\begin{equation}
	J_\mrmNP^{(\pm)} = \pm i \log\left[\frac{1 - i \ve\sqrt{\Pi_{B_2}}}{\sqrt{1+\Pi_{B_1}}\sqrt{1+\Pi_{B_2}} }\right]\, , \label{ExactNP_R_AR}
\end{equation}
which, of course, preserves the complex conjugacy relationship. 

As in the PT-symmetric case, we can use the exact solutions to check the reality of the spectrum. Again, let us assume $J_\mrmNP^{(\pm)} \in \mbbR$. Then, \eqref{ExactNP_R_AR} can be re-written into two parts as
\begin{align}
	\cos\left(J_\mrmNP^{(\pm)}\right)	&= \frac{1}{\sqrt{1+\Pi_{B_1}}\sqrt{1+\Pi_{B_2}} } \, , \label{ExactReal_R_AR} \\
	\sin\left(J_\mrmNP^{(\pm)}\right) & = \frac{\mp \ve \sqrt{\Pi_{B_2}}}{\sqrt{1+\Pi_{B_1}}\sqrt{1+\Pi_{B_2}} } \, , \label{ExactIm_R_AR}
\end{align}
and setting $\mcalX = \left[\left(1+\Pi_{B_1}\right)\left(1+\Pi_{B_2}\right)\right]^{-\frac{1}{2}}$, \eqref{ExactReal_R_AR} becomes
\begin{equation}
	J_\mrmNP^{(\pm)} = \arccos\left[\mcalX\right]\, .
\end{equation}
In turn, using this solution in \eqref{ExactIm_R_AR}, we obtain
\begin{align}
	\sin\left[\arccos\left(\mcalX\right)\right] = \sqrt{1- \mcalX^2} = \left(1 - \frac{1}{\sqrt{1+\Pi_{B_1}}\sqrt{1+\Pi_{B_2}} }\right)\, ,
\end{align}
which is not equal to~\eqref{ExactIm_R_AR}. This concludes $J_\mrmNP^{(\pm)}\notin \mbbR$. Therefore, the spectrum of both resonance and anti-resonance cases cannot be real.

We note that the above discussion is valid for all parameter regions and that there is no exceptional point for the (anti-)resonance sector for the ITW potential. This is actually manifest already in~\eqref{ExactSoln_R_AR} as there is no competition between $\Pi_{B_1}$ and $\Pi_{B_2}$ actions, which would lead to different behavior of a square-root term as in the PT-symmetric case in Section~\ref{Section: PTsymmetric_median}.

\paragraph{\underline{Semi-classical expansion (revisited)}}
Finally, we conclude our discussion by revisiting the semi-classical resurgence structure that we analyzed in Section~\ref{Section: R_AR_Direct}. Thanks to the elegance of the median QC approach and the exact non-perturbative solution in \eqref{ExactNP_R_AR}, obtaining the median summed trans-series is very straightforward:

Recalling the series definitions in \eqref{DerivativeNP_expansionB1} and \eqref{DerivativeNP_expansionB2}, and expanding for $J_\mrmNP$ for each case in \eqref{ExactNP_R_AR}, we obtained the median summed trans-series for both resonance and anti-resonance cases as
\begin{align}
	\frac{1}{2\pi} J_\mrmNP^{(\pm)} &\simeq  \ve\mcalK_2  P^{(2)}_0 +\mcalK_2^2 P^{(2)}_0 P^{(2)}_1 - 2 \mcalK_1^2 P_0^{(1)} P_1^{(1)} \nonumber \\
	& \qquad \mp i \left(\mcalK_1 P^{(1)}_0 - 2 \pi \mcalK_1^2 \left(P^{(1)}_0\right)^2+ \ve\mcalK_1 \mcalK_2 \left(  P^{(1)}_1 P^{(2)}_0+  P^{(1)}_0 P^{(2)}_1\right) \right)	\, ,\label{R_AR_medianSpectrum}
\end{align}
where we also used \eqref{relationCoefficients} in the last step. This recovers the median summed trans-series in \eqref{R_NP_BorelSummed} and \eqref{AR_NP_BorelSummed}. We also note that two solutions in \eqref{R_AR_medianSpectrum} are complex conjugate to each other, i.e., $J_\mrmNP^{(+)} = \mathsf{C}J_\mrmNP^{(-)}$. This is another manifestation of the time-reversal relationship between the resonance and anti-resonance systems.

Similarly, for the parts involving the resurgence cancellations, we utilize the Alien derivative construction in Section~\ref{Section: Spectrum_EWKB}. For this reason, we first re-organize the median QC \eqref{median_R_AR_both} as
\begin{align}
	D^{(\pm)}_\med &= \left(\Pi_A + \Pi_A^{(-1)}\right) + \Pi_A^{(\pm 1)} \frac{\Pi_{B_1}}{\sqrt{1 + \Pi_{B_2}}} + 2 \frac{\sqrt{1+ \Pi_{B_1}}}{\sqrt{1 + \Pi_{B_2}}}\, .
\end{align}
Then, recalling the notation that we introduced in the Alien-calculus discussion, we express the median QC as
\begin{align} 
	D^{(0)}_\pm & = 2\cos(J) + e^{\pm i J} \F_1(J) + \F_2(J) \, , \label{median_R_AR_organized}
\end{align}
where we put $\pm$ in the subscript to refer to resonance and anti-resonance cases and also defined
\begin{equation}
	\F_1(J) = \frac{\Pi_{B_1}(J)}{\sqrt{1 + \Pi_{B_2}(J)}} \, , \qquad \F_2(J) = 2 \frac{\sqrt{1+ \Pi_{B_1}(J)}}{\sqrt{1 + \Pi_{B_2}(J)}} \,.
\end{equation}

The first Alien derivative of the median QC in \eqref{median_R_AR_organized} can be computed via~\eqref{AlienDerivative_Action} as
\begin{equation}
	D^{(1)}_\pm  = \left(2 i \sin J \pm e^{\pm i J} \F_1(J)\right) J(J)\, ,
\end{equation}
where $L(J)$ is first defined in~\eqref{functionL}. Then, recalling~\eqref{Alien_FirstResurgence}, we obtain
\begin{equation}
	J_1^{(\pm)}(J_0) = \frac{\left(2 i \sin(J_0) \pm e^{\pm i J_0} \F_1(J_0) \right) L(J_0)}{2\sin(J_0) - e^{\pm i J_0} \left(\pm i \F_1(J_0) + \F_1'(J_0)\right) - \F_2'(J_0)}\, . \label{R_AR_firstAlien_action}
\end{equation}
In order to relate $J_1^{(\pm)}$ to the semi-classical resurgence structure, we set $J_0^{\pm} = J_\mrmP^{(\pm)} + J_\mrmNP^{(\pm)}$ and expanded~\eqref{R_AR_firstAlien_action} for $\mcalK_1\ll 1$ and $\mcalK_2 \ll 1$ after inserting~\eqref{R_AR_medianSpectrum} in \eqref{R_AR_firstAlien_action}. As a result, we obtained
\begin{align}
	\frac{1}{2\cdot 2 \pi}J_1^{(\pm)} &\simeq  \mp i\mcalK_1 P_0^{(1)} - 2 \mcalK_1^2 \left(P_0^{(1)}P_1^{(1)} \mp \pi i (P_0^{(1)})^2\right) - i\pi\mcalK_2^2\left(P_0^{(2)}\right)^2  \nonumber \\
	& \quad  \mp i \ve\mcalK_1 \mcalK_2 \left( P^{(1)}_1 P^{(2)}_0+  P^{(1)}_0 P^{(2)}_1\right)  \, , \label{J1_R_AR}
\end{align}
which enters the resurgence cancellations discussed in Section~\ref{Section: R_AR_Direct}. We note that $J_1^{(\pm)}$ are not complex conjugate to each other, as the $O(\mcalK_2^2)$ term has the same sign in both cases. This is because it emanates from the one-bion event $[\mcalI\mcalIbar]_\pm$ and is responsible for canceling the imaginary ambiguity arising from the perturbative sector, which is the same for both the resonance and anti-resonance systems.

Finally, combining the series in \eqref{R_AR_medianSpectrum} and \eqref{J1_R_AR}, we recover the semi-classical trans-series:
\begin{equation}
	J^{(\pm)}(\nu) \simeq J_0^{(\pm)} + \nu  \frac{1}{2 \pi} J_1^{(\pm)}\, . \label{R_AR_transSeries_recover}
\end{equation} 
To prevent confusion, we emphasize that $(\pm)$ in~\eqref{J1_R_AR} and~\eqref{R_AR_transSeries_recover} does not refer to the non-perturbative ambiguities. More precisely, $J^{(+)}$ in \eqref{R_AR_transSeries_recover} is linked to the resonance-system trans-series~\eqref{Resonance_NP_Trans-series1} and \eqref{Resonance_NP_Trans-series2}, i.e., $\mcalS_{\pm}\left[\d_\mrmR^{(\pm)}\right] = J_\med^{(+)}$, and $J^{(-)}$ is linked to the anti-resonance system, i.e., $\mcalS_{\pm}\left[\d_\mrmAR^{(\pm)}\right] = J_\med^{(-)}$ in \eqref{R_NP_BorelSummed}. The non-perturbative ambiguities, on the other hand, are represented by $\nu = \pm \frac{1}{2}$ in \eqref{R_AR_transSeries_recover} as first introduced in Section~\ref{Section: Spectrum_EWKB}.

\section{Discussion and outlook} \label{sec:summary}


Despite its relatively simple definition, the inverted triple-well (ITW) potential provides a concrete example of the semi-classical and exact quantization of non-Hermitian quantum mechanics. Our analysis particularly highlights the effectiveness of resurgence methods in this setting and shows that the exact-WKB (EWKB) framework yields physically consistent and quantitatively precise results. Although it is primarily based on the technical capabilities of the EWKB framework, by a rigorous analysis of the resurgent structure of PT-symmetric and (anti-)resonance systems, this paper also presents possible exploitations for the resurgence theory of general non-Hermitian systems. 

Our analysis shows that the semi-classical analysis of the PT-symmetric and (anti-)resonance systems based on the same building blocks, namely bion $[\mcalI\mcalIbar]$ and bounce $[\mcalB]$ configurations, simply stems from the underlying classical potential. At the quantum level, however, these building blocks enter in different combinations depending on the systems (boundary conditions), and their physical roles change accordingly. For example, we have discovered novel fractional configurations such as $[\mathcal{B}^2 \mathcal{I}^{-1}]$ and $[\mathcal{I}\bar{\mathcal{I}}\mathcal{B}^{-1}]$ in unbroken and broken PT-symmetric phases, respectively. These configurations have no direct analog in conventional Hermitian systems and do not appear in the resonance or anti-resonance ITW problem.

More interestingly, we have shown that a one-bounce configuration $[\mcalB]_\pm$ appears in qualitatively different ways in the unbroken and broken PT-symmetric phases, as well as in resonance and anti-resonance systems. In the unbroken PT-symmetric phase, $[\mcalB]_\pm$ participates in the resurgence cancellations and disappears from the exact spectrum. In the broken phase, in contrast, it appears with an extra factor of $(1+\ve)[\mcalB]_\pm$ and contributes to the imaginary splitting of the eigenvalues. In the resonance system, on the other hand, $[\mcalB]_\pm$ governs the leading instability but does not induce any splitting. 
This demonstrates that the role played by a given semi-classical configuration is determined primarily by the boundary conditions and the exact quantization procedure, rather than solely by the shape of the classical potential or by the existence of the bounce and bion saddles.

One of the highlights of our results is the exact characterization of the PT-symmetry breaking. We have proven that the unbroken and broken phases of the PT-symmetric system are determined directly by the signs of $\Disc_{\mrmPT}$, i.e., $\Disc_\mrmPT <0$ and $>0$ indicate the unbroken and broken phases, respectively. The transition takes place at $\mathrm{Disc}_{\rm PT} = 0$, which yields the exact exceptional-point condition as a strikingly simple algebraic relation: $\Pi_{B_2} = \Pi_{B_1}^2 / 4(1+\Pi_{B_1})$. Very interestingly, the Weber-type expression~\eqref{Dictionary_NonPerturbative} reduces this equation to a simple relationship between the classical actions of the bion $[\mcalI\mcalIbar]_\pm$ and bounce $[\mcalB]_\pm$ configurations, along with the one-loop-determinant prefactor of the latter. (See~\eqref{ExceptionalPoints_ActionCondition}). This implies that the non-Hermitian phase-transition can be probed exactly by semi-classical action data only.


Another particularly interesting characteristic of the exceptional point is the vanishing of the non-perturbative corrections. At the level of the exact median quantization condition (QC), we showed that this follows from the exact cancellations of the median-summed non-perturbative contributions. We also showed that, despite the vanishing of the net non-perturbative correction to the exact spectrum, the resurgent structure originating from the perturbative sector does not disappear. This strikingly beautiful resurgent structure at the exceptional point resembles the Cheshire cat resurgence observed in supersymmetric (SUSY) systems~\cite{Kozcaz:2016wvy,Dorigoni:2017smz,Dunne:2016jsr}. 
In the Cheshire cat resurgence, the perturbative contribution vanishes together with its associated resurgent structure, while the exact answer is maintained by non-perturbative contributions in a manner consistent with supersymmetry. (See also~\cite{Behtash:2015loa,Behtash:2015zha}). 
In the present ITW problem at the exceptional point, the situation is reversed: the perturbative series remains divergent and non-Borel summable, and the minimal resurgent structure survives, whereas the remaining non-perturbative spectral corrections cancel in the exact median-summed spectrum.
We note that this structure also provides consistency with the coalescence of the real eigenvalues as in Fig.~\ref{Figure: PTsymmetric_Real}. 
In the ITW system, such a coalescence of real eigenvalues is possible only if the real non-perturbative corrections vanish, since otherwise they would induce a level splitting, as in the unbroken PT-symmetric phase.


Our analysis also highlights a distinctive feature of the non-perturbative structure of the symmetric inverted double-well (IDW) case. In the PT-symmetric boundary condition, the non-perturbative contribution drops out of the median-summed QC for IDW~\cite{Kamata:2023opn}, whereas in the ITW case it still survives and leads to a much richer non-perturbative sector including the PT-symmetry breaking. This difference also precludes a direct ITW analog of the ABS conjecture, i.e., a relation between the PT-symmetric and (anti-)resonance QCs originally proposed for IDW in~\cite{Ai:2022csx}. Unlike in the IDW case~\cite{Kamata:2023opn}, PT-symmetric and (anti-)resonance systems are not connected by an analytic continuation. Accordingly, the PT-symmetric and (anti-)resonance QCs cannot be connected by a Stokes automorphism, which is mainly due to the presence of the inner-bion cycle in the ITW system. 



There are several natural directions for future work. One immediate extension is to analyze systematically the barrier-top and above-barrier regions, where the hierarchy among WKB actions changes, altering the semi-classical structure. It would also be interesting to clarify the relation between the present EWKB description and a complexified path-integral formulation in which bounce and bion sectors arise directly from complex saddles. A further question is how broadly the structures found here persist in more general non-Hermitian polynomial potentials, especially in systems with richer Stokes geometry or multiple exceptional points. In general, the present results suggest that EWKB and resurgence provide a powerful and flexible framework for the non-perturbative analysis of open quantum systems and non-Hermitian spectral problems beyond the specific ITW example studied here.

\section*{Acknowledgments}
This work is supported by the Japan Society for the Promotion of Science Grant-in-Aid for Scientific Research (JSPS KAKENHI) Grant Nos.~22H05118 and 25K07298 (S. K.), Grant Nos.~23K03425 and 22H05118 (T. M.), No.~24K17058 and the RIKEN TRIP initiative (RIKEN Quantum) (H. T.). C.P. thanks Keio University and the University of Tokyo for the hospitality, where the initial steps of the project took place. 

\bibliographystyle{JHEP}
\bibliography{EWKB.bib}

\end{document}